%% file: main.tex
\newif \iftheory 

\theorytrue 

\newif \ifblind

\blindtrue \blindfalse

\newif \ifdraft

\drafttrue \draftfalse

\newcommand{\authorlist}{
Megumi Ando\iftheory\thanks{MITRE, {\tt mando@mitre.org}}\fi
\and 
Anna Lysyanskaya\iftheory\thanks{Computer Science Department, Brown University, {\tt anna@cs.brown.edu}}\fi
\and 
Eli Upfal\iftheory\thanks{Computer Science Department, Brown University, {\tt eli@cs.brown.edu}}\fi
}

\newcommand{\institutelist}{
{MITRE, Bedford, MA 01730 USA}
\and
{Brown University, Providence, RI 02912 USA}}

\newcommand{\titlelist}{\iftheory\begin{bf}\fi
On the Complexity of Anonymous Communication Through Public Networks
\iftheory\end{bf}\fi}

\newcommand{\pathstyles}{template/stylefiles}

\iftheory
\documentclass[11pt]{article}

\else
\documentclass[envcountsame, 10pt]{\pathstyles/llncs}
\ifblind 
\institute{}
\else
\institute{\institutelist}
\fi
\fi

\ifblind 
\author{}
\else
\author{\authorlist} 
\fi
\title{\titlelist}

\input{\pathpreambles/preamble}

\input{preamble}
\usepackage{multirow}

\iftheory
\else

\fi

\begin{document}

\maketitle
\iftheory
\thispagestyle{empty}
\fi

\begin{abstract}
Onion routing is the most widely used approach to anonymous communication online.  The idea is that Alice wraps her message to Bob in layers of encryption to form an ``onion,'' and routes it through a series of intermediaries. Each intermediary's job is to decrypt (``peel") the onion it receives to obtain instructions for where to send it next, and what to send.  The intuition is that, by the time it gets to Bob, the onion will have mixed with so many other onions, that its origin will be hard to trace even for an adversary that observes the entire network and controls a fraction of the participants, possibly including Bob.  

\iftheory\else~~~~\fi
In spite of its widespread use in practice, until now no onion routing protocol was known that simultaneously achieved, in the presence of an active adversary that observes all network traffic and controls a constant fraction of the participants, (a) fault-tolerance, where even if a few of the onions are dropped, the protocol still delivers the rest; (b) reasonable communication and computational complexity as a function of the security parameter and the number of participants; and (c) anonymity.

\iftheory\else~~~~\fi
In this paper, we give the first onion routing protocol that meets these goals: our protocol (a) tolerates a polylogarithmic (in the security parameter) number of dropped onions and still delivers the rest; (b) requires a polylogarithmic number of rounds and a polylogarithmic number of onions sent per participant per round; and (c) achieves anonymity.  We also show that to achieve anonymity in a fault-tolerant fashion via onion routing, this number of onions and rounds is necessary.

\iftheory\else~~~~\fi
Of independent interest, our analysis introduces two new security properties of onion routing --- mixing and equalizing --- and we show that together they imply anonymity.

\iftheory
\vfill
\fi
\end{abstract}

\newpage
\setcounter{page}{1}
\section{Introduction}

Suppose that Alice wishes to send a message anonymously to Bob. Informally, by \emph{anonymously}, we mean that no one (not even Bob) can distinguish the scenario in which Alice sends a message to Bob from an alternative scenario in which it is Allison who sends a message to Bob. 
To begin with, Alice can encrypt the message and send the encrypted message to Bob so that only Bob can read the message. However, an eavesdropper observing the sequence of bits coming out of Alice's computer and the sequence of bits going into Bob's computer can still determine that Alice and Bob are communicating with each other if the sequences of bits match. Thus, encryption is not enough.

Onion routing~\cite{Chaum81} is the most promising approach to anonymous channels to date. In onion routing, messages are sent via intermediaries and wrapped in layers of encryption, resulting in so-called onions; each intermediary's task is to ``peel off'' a layer of encryption and send the resulting onion to the next intermediary or its final destination. The onion's layers are unlinkable to each other, and so its route through the network cannot be traced from merely observing the sequences of bits that Alice transmits and Bob receives. However, even with Alice sending her message to Bob encoded as an onion, her communication can still be tracked by a resourceful eavesdropper with an extensive view of the network traffic (e.g., an ISP-level or an AS-level adversary) who can observe all Internet traffic. 

An adversary who can observe all network traffic is called a \emph{network adversary}.  An adversary who, in addition to observing all network traffic, controls a subset of the participants, is called \textit{the passive adversary} if it follows the prescribed protocol, or \textit{active} if it does not.  The three adversary models --- the network adversary, the passive adversary, and the active adversary --- are standard for analyzing cryptographic protocols such as multi-party computation (MPC)~\cite{STOC:GolMicWig87}.  The most desirable goal is to achieve security in the presence of the most powerful of these three adversaries, i.e., the active adversary, corrupting as large a fraction of the participants as possible.

It was known how to construct an onion routing protocol that is both efficient and anonymous from the \emph{passive adversary} who corrupts a constant fraction of the parties; an example of this is the protocol $\Pi_p$~\cite{ICALP:AndLysUpf18}. In $\Pi_p$, each user forms an onion bearing his message to its recipient; the users' onions are routed independently and uniformly at random through a network of servers. $\Pi_p$ is anonymous from the passive adversary provided that the onions travel for a superlogarithmic (in the security parameter) number of rounds, and the average number of onions per server per round is also superlogarithmic~\cite{ICALP:AndLysUpf18} . 
However, $\Pi_p$ isn't anonymous from the \emph{active adversary} who causes the parties under his control to deviate from the protocol. To see why this is the case, consider the following attack: Suppose that the adversary $\adv$ suspects that Alice is communicating with Bob. Because $\adv$ is active, he can disrupt Alice's communication by dropping Alice's outgoing onion in the event that Alice's first intermediary is corrupt (the probability of this event is identical to the fraction of parties that are under the adversary's control). If Bob doesn't receive an onion at the end of the protocol, then $\adv$ can infer that her suspicion was correct: Alice's interlocutor is Bob! 

So what can we do instead? Of course, we could use general-purpose multi-party computation (MPC)~\cite{STOC:GolMicWig87}.  Every party will receive as input a message and its destination, and every party will receive as output the messages that were meant for him/her.  In addition to perfect anonymity, this approach provides fault tolerance: in MPC that is secure against the active adversary, the honest parties are guaranteed to receive their output no matter how much the adversary deviates from the protocol. 
The problem with this approach that relies on \emph{general-purpose} MPC is that it is too inefficient: 
the most efficient general MPC protocol still requires that at least some of the participants send and receive $\Omega(N)$ bits, where $N$ is the number of participants.  (See Cramer, Damg{\aa}rd, and Nielsen~\cite{cramer2015secure}.)

Recently proposed protocols, Stadium~\cite{stadium} and Atom~\cite{atom}, are more efficient. However, they are not fault-tolerant: honest parties will abort the protocol run whenever  even a single message packet is dropped. Thus, while this approach provides anonymity from the active adversary, it is also extremely fragile: if just one message is dropped (which could be the result of an innocuous fault), the entire network suffers a catastrophic failure.  In contrast, we would like to design onion routing protocols that can tolerate faults.
Thus, compared to MPC and Stadium-Atom-type protocols, onion routing appears attractive from the efficiency and fault tolerance points of view.  

In this paper, we answer these fundamental questions: 
(1)~Can an onion routing protocol be simultaneously anonymous, fault-tolerant, and efficient? 
(2)~What is the communication complexity \emph{sufficient} for anonymous and fault-tolerant onion routing?
(3)~What is the communication complexity \emph{required} for this? 
We provide a lower bound and match it with a nearly optimal protocol. 

\iftheory
\subsection{Problem setting} \label{sec:setting}
\else
\subsubsection*{Problem setting.} 
\fi

\sloppy Before describing our results in detail, let us first define our problem setting. 
Let $\parties \myeq \{P_1, P_2, \dots, P_N\}$ denote the set of $N$ parties, participating in an onion routing protocol. 
We assume that the protocol progresses in global rounds and that an onion sent at round~$r$ arrives at its destination prior to round~$r+1$. Moreover, the adversary is modelled with \emph{rushing}, i.e., the adversary receives onions sent in round $r$ instantaneously in round $r$.\footnote{
We do not consider the asynchronous communication model~\cite{canrab93} in which Alice's outgoing onions (including her onion to her recipient Bob) can be delayed indefinitely. In such a case, we cannot even guarantee correctness (i.e., message delivery when no party deviates from the protocol). 
}
We assume that the number~$N$ of participants and every other quantity in the protocol is polynomially bounded in the security parameter~$\secpar$.

\paragraph{Setup.}
We define an \emph{onion routing protocol} to be a protocol in which the honest parties form and process only message packets that are cryptographic onions. To do this, the honest parties use a secure onion encryption scheme, which is a triple of algorithms: $(\gen, \formonion, \proconion)$. 
\ifdraft
\begin{itemize}
\item On input 
the security parameter $\secparam$ (written in unary), 
the public parameters $\pp$, and 
(the name of) a party~$P$, the key generation algorithm $\gen$ generates a public key pair $(\pk(P), \sk(P))$ for $P$. The public parameters generally includes the upper bound on the length of the routing path.

\item \sloppy On input 
a message $m$, 
a routing path $(Q_1, \dots, Q_\depth, R)$, 
the public keys $(\pk(Q_1), \dots, \pk(Q_\depth), \pk(R))$ associated with the parties on the routing path, and
the sequence $(s_1, \dots, s_\depth)$ of nonces (i.e., these are short numbers), the algorithm $\formonion$ forms an onion $O$ for the first party $P_1$ on the routing path. 

\item On input 
an onion $O$, 
a processing party $P$, and 
the secret key $\sk(P)$ belonging to $P$, the algorithm $\proconion$ outputs the processed onion $O'$, the destination $P'$ of $O'$, and a nonce~$s$, or the algorithm outputs the message $m$ if $P$ is the recipient of $O$. 
\end{itemize}
\fi
See Section~\ref{sec:or} for more details and Camenisch and Lysyanskaya's paper~\cite{C:CamLys05} and follow-up papers~\cite{KBS19,AL20} for formal definitions, including security definitions.

During setup of an onion routing protocol, each honest party~$P$ generates a public-key pair $(\pk(P), \sk(P)) \gets \gen(\secparam)$ using the onion encryption scheme's key generation algorithm $\gen$. Each party~$P$ publishes his/her public key $\pk(P)$ to a public directory so that everyone knows everyone else's public keys. 

\paragraph{Inputs: the simple input/output setting.}
Let $\mathcal{M}$ be the space of fixed-length messages. 

An input $\sigma = (\sigma_1, \dots, \sigma_N)$ to the protocol is a vector of inputs, where $\sigma_i$ is a set of message-recipient pairs for party~$P_i$. For $m \in \mathcal{M}$ and $P_j \in \parties$, the inclusion of a message-recipient pair $(m, P_j)$ in input $\sigma_i$ means that party~$P_i$ is instructed to send message~$m$ to recipient~$P_j$. 

In this paper, we consider the following ``benchmark'' input space, dubbed the \emph{simple input/output setting (I/O)}. An input $\sigma = (\sigma_1, \dots, \sigma_N)$ is in the simple I/O setting if there exists a permutation function $\pi : \parties \mapsto \parties$ such that each party $P\in\parties$ is instructed to send a message to party $\pi(P)$ and no other message, i.e., $\forall P\in\parties$, $\exists m_P \in\mathcal{M}$ s.t.\ $\sigma_P = \{(m_P, \pi(P))\}$.\footnote{
Why do we need a benchmark? As the following shows, without appropriately constraining the input space, the onion cost can be arbitrarily high: 
%
Suppose that the adversary observes the traffic (i.e., the onions) on all links but does not corrupt any of the parties. The adversary knows that each party will send a fixed-length message to the ``central hub'' $H\in\parties$ (and no one else) but doesn't know which party is $H$. A protocol that is anonymous in this setting necessarily incurs a large communication overhead. This is because a party who receives fewer than $N$ onions cannot be $H$, and so, all but one party must receive at least $N$ dummy onions. 
}
The simple I/O setting is a superset of the spaces considered in prior works~\cite{vuvuzela,stadium,atom,ICALP:AndLysUpf18}.

\paragraph{Adversary model.} 
\sloppy Unless stated otherwise, the adversary is active and can observe the traffic on all communication channels and, additionally, can non-adaptively corrupt and control a constant fraction of the parties. By \emph{non-adaptively}, we mean that the corruptions are made independently of any protocol run.\footnote{
If we were to allow the adversary to adaptively corrupt parties, then the adversary could easily block all of Alice's onions. For every onion $O$ sent by Alice, the adversary can corrupt the party $P$ who receives $O$ in time to direct $P$ to drop the onion obtained from processing $O$ before the next round. 
}
Without loss of generality, this type of corruption is captured by allowing the adversary to select the set $\Bad$ of corrupted parties prior to the beginning of the protocol.  Once the adversary corrupts a party, the adversary can observe the internal state and computations of the corrupted party and arbitrarily alter the behavior of the party.

\paragraph{Views and outputs.}
By $\view{\Pi, \adv}{\secparam, \sigma}$, we denote the adversary $\adv$'s view from interacting with protocol~$\Pi$ on input the security parameter $\secparam$ and the instructions~$\sigma$. 
The view consists of all the observations that $\adv$ makes during the run: 
the values and positions of every onion at every round, 
the states and computations of every corrupted party between every pair of consecutive rounds, 
the randomness used by $\adv$, and 
the numbers of messages received by the honest parties. The view does not include the honest parties' randomness. 
$\view{\Pi, \adv, \Bad}{\secparam, \sigma}$ denotes $\adv$'s view given its choice $\Bad$ for the corrupted parties. 
At the end of the protocol run, each honest party $P_i$ outputs the set $\oh{\Pi, \adv}{i}{\secparam, \sigma}$ of (non-empty) messages from the message space~$\mathcal{M}$ that $P_i$ receives from interacting with adversary $\adv$ in a run of protocol~$\Pi$ on input $\sigma$. We define the \emph{output} $\oh{\Pi, \adv}{}{\secparam, \sigma} $ of protocol $\Pi$ in an interaction with adversary $\adv$ on input $\sigma$ as the $N$ parties' outputs:\footnote{Technically, the view and the output may depend on other parameters, such as the public parameters (denoted, $\pp$) and the parties' states (denoted, $\states$). Thus, we could be more precise by denoting the view and the output as $\view{\Pi, \adv}{\secparam, \pp, \state, \sigma}$ and $\oh{\Pi, \adv}{}{\secparam, \pp, \states, \sigma}$, but we will use the simpler notation for better readability.}  
\mymathenv{
\oh{\Pi, \adv}{}{\secparam, \sigma} \myeq (\oh{\Pi, \adv}{1}{\secparam, \sigma},\oh{\Pi, \adv}{2}{\secparam, \sigma},\dots,\oh{\Pi, \adv}{N}{\secparam, \sigma}) .
}

\iftheory
\subsection{Our results} 
\else
\subsubsection*{Our results.}
\fi
We now describe our results in more detail. Our construction pertains to the problem setting described in 
\iftheory Section~\ref{sec:setting}. 
\else
\emph{Problem setting}. 
\fi
Our lower bound applies more generally to any arbitrary input set (not necessarily constrained to the simple I/O setting). 

\paragraph{Anonymity, mixing, and equalizing.} 
Following prior work~\cite{vuvuzela,stadium,atom,ICALP:AndLysUpf18}, we use a natural game-based definition of anonymity: A protocol is \emph{anonymous} if the adversary cannot distinguish the scenario in which Alice sends a message to Bob while Carol sends one to David, from one in which Alice's message goes to David while Carol's goes to Bob. (See Definition~\ref{def:anon}.) More precisely, for any pair of inputs $(\sigma^0, \sigma^1)$ that agree on the inputs and outputs for the adversarial participants, $\oh{\Pi, \adv}{}{\secparam, \sigma^0} \approx \oh{\Pi, \adv}{}{\secparam, \sigma^1}$, where ``$\approx$" denotes computational indistinguishability.

We relate anonymity of an onion routing protocol to two new concepts: 
An onion routing protocol \emph{mixes} if it sufficiently shuffles the honest users' onions making it infeasible for the adversary to trace a received message back to its sender. A protocol \emph{equalizes} if the adversary cannot determine the input from the numbers of messages received by the parties; in other words, the number of messages output by each participant --- or the fact that a participant did not receive an output at all --- are random variables that are  computationally unrelated to the input vector $\sigma$. (See Definitions~\ref{def:equal} and \ref{def:mix}.)

We show that in many cases, mixing and equalizing implies anonymity, i.e., an onion routing protocol that mixes and equalizes is anonymous. (See Theorem~\ref{thm:impliedby} for the formal theorem statement.) 
We use this to prove that our protocol is anonymous. 
Anonymity also implies equalizing; this observation is useful for proving a lower bound that (almost) matches our protocol. 

\paragraph{Efficient, robust, and anonymous onion routing.}
As we just explained, our strategy is to construct a protocol that mixes and equalizes.

Intuitively, mixing is the easier one to achieve: the onions need to sufficiently shuffle with other onions traveling over the network to ensure that each of them is hard to trace.  This intuition is essentially correct, with the caveat that an active adversary can strategically interfere with this process by dropping onions.  To ensure that each onion shuffles with a sufficiently large number of onions (formed by an honest party) a sufficiently large number of times, our protocol uses \emph{checkpoint onions}~\cite{ICALP:AndLysUpf18} that each intermediary expects to receive, and if a constant fraction (e.g., one-third) of them don't arrive because the adversary dropped them, the protocol aborts.  

An active adversary who controls a fraction of the participants can try to ``isolate'' an honest party Alice from the rest of the network by dropping all of the messages/onions received directly from Alice. In a fault-tolerant network protocol, the remaining participants may still be able to get their messages through to their destinations. Thus, based on who received an output, an adversary can infer who Alice's intended recipient was.  This attack explains why equalizing is difficult to achieve.

To overcome this attack, we introduce a new type of onions, called \emph{merging onions}.  
When two merging onions belonging to the same pair
arrive at some intermediary $I$, $I$ recognizes that they are from the same pair (although, other than their next layer and destination, $I$ does not learn anything else about them).  The protocol directs $I$ to discard one of them (chosen at random) while sending its mate along.  If only one onion of the pair arrived at $I$ while its mate is missing (i.e. the adversary dropped it some time earlier in the protocol run), then $I$ simply sends along the mate that survived, and there is nothing to discard.

Why does this help?  Suppose that both Alice and Allison created $2^h$ merging onions; at rounds $r_1$, $r_2$, $\ldots$, $r_h$ each of these onions (if it hasn't been deleted yet) will meet a mate.  
Say, exactly one of Alice's onions is dropped by the adversary at some point prior to round $r_1$, so its mate (the onion it was supposed to pair with at round $r_1$) was not dropped.  Also, suppose that none of Allison's onions were dropped.  Then at round $r_1$ all but one of Alice's remaining merging onions will meet a mate, and half of them will be dropped, so exactly $2^{h-1}$ of Alice's onions will remain in the system --- which is exactly how many of Allison's onions remain.  Additional $h-1$ opportunities to merge account for the possibility that the adversary has dropped a larger number of Alice's onions.  Merging onions ensure that the number of Alice's onions that remain in the system at the end of the protocol is the same as the number of Allison's onions, i.e., that the protocol equalizes.  The fact that Alice was targeted and many of her onions had been dropped doesn't matter, because the protocol discards all but one of them anyway!
(See Section~\ref{sec:ingredients} for a more in-depth description of merging onions and how to construct them.)

\underline{\textbf{Positive result:}} 
We construct an onion routing protocol $\PiBfly$, pronounced ``Pi-butterfly," because it uses a butterfly network.  $\PiBfly$ takes advantage of the merging onions technique described above. It is (1) anonymous from the active adversary who can corrupt up to a constant fraction $\kappa < \frac{1}{2}$ of the parties and (2) robust, i.e.\ whenever the adversary drops at most logarithmic (in the security parameter) number of message packets (i.e. onions), $\PiBfly$ delivers the messages from honest senders with overwhelming probability. Moreover, (3) during the execution of the protocol, every honest party transmits up to a polylog (in the security parameter) number of onions: specifically $\gamma_1 \log N \log^{3+\gamma_2} \secpar$ onions, where $N$ is the number of participants, and $\secpar$ is the security parameter. $\gamma_1$ and $\gamma_2$ are parameters that can be set as desired: increasing them increases the rate at which the maximum distance in the adversarial views for any two inputs shrinks.  (See Theorem~\ref{thm:upper} for the precise relationship.)

\paragraph{Matching negative result.} 
Our protocol is essentially optimal as far as both the round complexity and the number of onions each participant sends out are concerned.  For why anonymity requires superlogarithmic round complexity, we refer the reader to prior work~\cite{das2020comprehensive, Miranda}.  
In Section~\ref{sec:lower2}, we explain why a protocol that is robust and anonymous in the presence of an active adversary that corrupts a constant fraction of participants requires a polylogarithmic number of onions sent out per participant.

\iftheory
\subsection{Related work}
\else
\subsubsection*{Related work.}
\fi
Our work is inspired by the fact that Tor~\cite{tor}, the most widely adopted anonymous communication system, is also known to have numerous security flaws~\cite{SP:OveSyv06, CCS:JWJSS13, SEVL+15, WSJCM18}: 
Tor is based on a highly efficient design that favors practicality over security and is not secure even from the passive adversary~\cite{SP:DMMK18}. 
Moreover, it has been shown to be vulnerable to network traffic correlation attacks~\cite{SP:OveSyv06, CCS:JWJSS13, SEVL+15, WSJCM18}.  
Thus, our goal was to design a protocol that was as close to Tor's efficiency and fault tolerance as possible, while also being provably anonymous.

We consider a very specific and narrow problem in the much larger field of anonymous messaging systems. Although our definition of anonymity and adversary models are standard in cryptography, other definitions have been considered~\cite{BKMMM, FC:BerFiaTaS04, ChaPalPan08, EC:DodReySmi04, AlvAndCha+12} and positive results for alternative models are known~\cite{FC:BerFiaTaS04, BGKM, BKMMM}. 

Atom~\cite{atom} is a current state-of-the-art anonymous protocol in the active adversary setting. 
It is similar to an onion-routing protocol\footnote{Like a cryptographic onion, each message packet in Atom is layered encryption object but without a sender-defined routing path.} 
and comes in two variants. Atom uses a known random permutation network (e.g., a square network) to mix the message packets. Each node of the random permutation network is really a sufficiently large random sample of the parties, such that the probability that all parties in the sample are corrupt is very small. In Atom \#1, to shuffle message packets at a node, every member of the node verifiably shuffles~\cite{C:FurSak01} the packets and broadcasts a proof to every other member of the node. If an honest party detects a discrepancy, the party aborts the protocol. This guarantees anonymity. The downside is that Atom \#1 is highly \emph{fragile}; the honest parties abort the protocol even if only a single packet is dropped. The second variant, Atom \#2, uses threshold cryptography and so can tolerate some dropped messages at a cost in privacy; it only guarantees $k$-anonymity.  

A slightly older system, called Vuvuzela~\cite{vuvuzela}, assumes that all messages travel through the same set of dedicated servers and is therefore impractical compared to Tor. Their solution is also highly fragile and cannot tolerate a single dropped onion. 
Stadium~\cite{stadium} is a distributed solution that uses verifiable shuffling as its underpinning. Unlike Vuvuzela, Stadium is a properly load-balanced solution but, like Vuvuzela, also suffers from fragility. 

Vuvuzela, Stadium, and Atom~\#2 were not shown to be anonymous. In fact, since these protocols are unable to efficiently equalize, from Theorem~\ref{thm:impliedby}, they cannot be anonymous: In Vuvuzela and in Stadium, the numbers of onions received at a dead-drop (an address at one of the servers) is a function of the number of conversing parties. Thus, adding a random (but polynomially-bounded) number of dummy onions can provide differential privacy but not anonymity. In Atom~\#2, the adversary can drop an honest message packet upfront (at its first hop) and know that any message received in the end was not sent by a particular sender. See Table~\ref{table:compare}. 

\begin{table*}[ht]
\centering
\begingroup
\fontsize{9pt}{11pt}\selectfont
\begin{tabular}{ lcccl }
  \hline
  Protocol & Server load & \# of rounds & Fault tolerant? & Security \\ \hline
  $\Pi_p$~\cite{ICALP:AndLysUpf18} & $\mathsf{polylog}(\secpar)$ & $\mathsf{polylog}(\secpar)$ & not applicable & (passive) anonymity \\ 
  Vuvuzela~\cite{vuvuzela} & $N$ & $n$ & no & diff.\ privacy \\
  Stadium~\cite{stadium} & $\bigsmallO{\frac{N}{|G|}}$ & $\bigsmallO{|G|}$ & no & diff.\ privacy \\ \cline{4-5}
   \multirow{2}{*}{Atom~\cite{atom}} & \multirow{2}{*}{$\bigsmallO{\frac{N |G|}{n}}$} & \multirow{2}{*}{$\bigsmallO{|G|}$} & \hspace{-8mm}\#1: no &  anonymity \\
  & & & \hspace{-8mm}\#2: yes & $k$-anonymity \\ \cline{4-5}
  $\Pi_a$~\cite{ICALP:AndLysUpf18} & $\mathsf{polylog}(\secpar)$ & $\mathsf{polylog}(\secpar)$ & yes & diff.\ privacy \\ 
  $\PiBfly$ (this work) & $\mathsf{polylog}(\secpar)$ & $\log N \mathsf{polylog}(\secpar)$ & yes & anonymity \\ \hline
\end{tabular}
\endgroup
\caption
{\footnotesize A comparison of properties of provably secure onion routing (and similar) protocols. Our protocol $\PiBfly$ is the first construction shown to be simultaneously practical, fault tolerant, and anonymous in the active adversary setting. 
In the table entries: $\mathsf{polylog}(\secpar) \myeq \gamma_1 \log^{1+\gamma_2} \secpar$ for any $\gamma_1, \gamma_2 > 0$; 
$N$ denotes the number of participants; 
$n$ denotes the number of servers; and 
$|G|$ denotes the size of a group of servers. 
In general, we want $n$ and $|G|$ to be at least $\mathsf{polylog}(\secpar)$ to ensure that at least one server in each group is honest.  
}
\label{table:compare}
\end{table*}

Other provably anonymous systems exist~\cite{JC:Chaum88, STOC:RacSim93, SP:CooBir95, SP:CorBonMaz15}, but they are not nearly as efficient. 
Achieving anonymous channels using heavier cryptographic machinery has been considered also. One of the earliest examples is Chaum's dining cryptographer's protocol~\cite{JC:Chaum88}. Rackoff and Simon~\cite{STOC:RacSim93} use secure multiparty computation for providing security from active adversaries. Other cryptographic tools used in constructing anonymity protocols include oblivious RAM (ORAM) and private information retrieval (PIR)~\cite{SP:CooBir95, SP:CorBonMaz15}. Corrigan-Gibbs et al.'s Riposte solution makes use of a global bulletin board with a latency of days~\cite{SP:CorBonMaz15}.

We are not the first to look into lower bounds on the complexity of anonymous messaging protocols (e.g., \cite{SP:DMMK18}). However, all other lower bounds are for the setting where every participant is guaranteed to receive an output, and don't apply to protocols that allow aborts or that allow some participants to receive an output while others' output doesn't make it through.

\input{prelims}

\input{lower1}

\input{construction}

\input{upper}

\input{lower2}

\section{Conclusion and future work}
Here, we mention a few extensions of our results: 
We proved that the required onion cost for an onion routing protocol to provide robustness and (computational) anonymity from the active adversary is polylogarithmic in the security parameter. Our proof for the lower bound can be used to prove the stronger result that polylogarithmic onion cost is required even when (1) the adversary observes the traffic on only $\bigsmallO{1}$ fraction of the links and or when (2) the security definition is weakened to (computational) differential privacy. (3) Also, while we explicitly showed this to be the case for the simple I/O setting, the result holds more generally whenever any party can send a message to any other party. 

We also proved the existence of a robust and anonymous onion routing protocol with polylogarithmic (in the security parameter) onion cost. (4) This result also extends beyond the simple I/O setting; our onion routing protocol is anonymous w.r.t.\ any input set where the size of each party's input is fixed. 

There is a small gap between our lower and upper bounds. A natural direction for future work is to close this gap. 
\iftheory\else
\fi
\iftheory
\bibliographystyle{is-alpha}
\else
\bibliographystyle{plain}
\fi
\bibliography{main} 

\appendix 

\clearpage

\section{Proof that equalizing and mixing $\implies$ anonymity} \label{app:impliedby}
\input{thmimpliedby2}

\section{Proofs of Lemmas~\ref{lem:mix}, \ref{lem:equalizes}, and \ref{lem:first}, and \ref{lem:comm}}

This section contains the full proofs of the lemmas used to prove that $\PiBfly$ is anonymous (Theorem~\ref{thm:upper}). 

We prove that $\PiBfly$ mixes in Appendix~\ref{app:mix} by proving that the protocol mixes \comm\ onions during the last epoch of the execution phase. 
In Appendix~\ref{app:comm}, we define what it means for an onion routing protocol to equalize from \comm\ onions (Definition~\ref{def:equal2}) and show that equalizing from \comm\ onions implies equalizing (Lemma~\ref{lem:comm}). 
Appendices~\ref{sec:first} and \ref{sec:rest} contain the full proofs for Lemmas~\ref{lem:first} and \ref{lem:equalizes}: these are used to support our claim that $\PiBfly$ equalizes from \comm\ onions. 

\subsection{Proof that $\PiBT$ mixes} \label{app:mix}
\input{proof-mix}

\subsection{Proof that equalizing from \comm\ onions $\implies$ equalizing} \label{app:comm}
\input{anon2}

\ifdraft
\subsection{Lemma~\ref{lem:part1}} \label{app:pitree}

\input{proof-pitree}

\input{proof-pibfly-main-lemmas}
\input{proof-pibfly}
\else

\input{proof-pibfly-main-lemmas}
\input{proof-pibfly}
\input{proof2}
\subsection{Proof of Lemma~\ref{lem:part1}} \label{app:pitree}
\input{proof-pitree}
\fi

\input{lowerproof}

\end{document}

%% file: template/preambles/preamble.tex
\usepackage[utf8]{inputenc}
\usepackage{amsfonts}
\iftheory
\usepackage{amsmath}
\fi
\usepackage{amssymb}
\usepackage{cite}
\usepackage[
	n,
	operators,
	advantage,
	sets,
	adversary,
	landau,
	probability,
	notions,	
	logic,
	ff,
	mm,
	primitives,
	events,
	complexity,
	asymptotics,
	keys]{cryptocode}
\usepackage{enumitem}
\setlist{nosep}
\iftheory
\usepackage[margin=1in]{geometry}
\fi
\usepackage[hidelinks]{hyperref}
\usepackage{todonotes}

\usepackage{lscape}

\iftheory
\usepackage{amsthm}

\newtheorem{lemma}{Lemma}
\newtheorem{theorem}{Theorem}

\newtheorem{claim}{Claim}
\newtheorem{definition}{Definition}

\newtheoremstyle{TheoremNum}
        {\topsep}{\topsep}    
        {\itshape}                 
        {}                              
        {\bfseries}                
        {.}                             
        { }                             
        {\thmname{#1}\thmnote{ \bfseries #3}}
\else
\fi

\iftheory
\else

\fi

\renewcommand{\secpar}{\lambda}
\renewcommand{\secparam}{1^\lambda}


%% file: preamble.tex
\usepackage{dashbox}
\usepackage{dashrule}

\usepackage{float}
\floatstyle{boxed} 
\restylefloat{figure}

\usepackage{fancybox}
\usepackage{tikz}
\usetikzlibrary{calc,shapes}
\newcommand{\tikzmark}[1]{\tikz[overlay,remember picture] \node (#1) {};}

\usepackage{xcolor}
\renewcommand\fbox{\fcolorbox{blue}{white}}

\newcommand{\mymathenv}[1]{\ifdraft\[#1\]\else$#1$\fi}

\newcommand{\parties}{\mathcal{P}}

\newcommand{\gen}{\mathsf{Gen}}

\newcommand{\formonion}{\mathsf{FormOnion}}

\newcommand{\proconion}{\mathsf{ProcOnion}}

\renewcommand{\pp}{\mathsf{pp}}

\newcommand{\view}[2]{\mathsf{V}^{#1}{(#2)}}

\newcommand{\Bad}{\mathsf{Bad}}

\newcommand{\states}{\mathsf{states}}

\newcommand{\oh}[3]{\mathsf{O}^{#1}_{#2}{(#3)}}

\newcommand\myeq{\mathrel{\stackrel{\makebox[0pt]{\mbox{\normalfont\tiny def}}}{=}}}

\newcommand{\depth}{d}

\newcommand{\corruptions}{\kappa}

\newcommand{\regOR}{indifferent}

\newcommand{\Pilong}{\Pi(\secparam, \mathsf{pp}, \mathsf{states}, \$, \sigma)}

\newcommand{\senders}{\mathsf{S}}

\newcommand{\receivers}{\mathsf{R}}

\newcommand{\PiTree}{\Pi_{\triangle}}

\newcommand{\PiBfly}{\Pi_{\bowtie}}

\newcommand{\numOnion}{x}

\newcommand{\hTree}{h}

\newcommand{\HOnions}{\mathcal{O}} 

\newcommand{\numHOnions}{X} 

\newcommand{\polylog}{\log^{1+\epsilon}}

\newcommand{\recipient}{r}

\newcommand{\hops}[3]{\#\mathsf{onions}^{#1}_{#2}{(#3)}}

\newcommand{\sio}{\mathsf{SimpleIO}}

\newcommand{\outof}[3]{\mathsf{out}^{#1}_{#2}{(#3)}}

\newcommand{\nonnegl}{\mathsf{nonnegl}(\secpar)}

\newcommand{\volOnions}{\mathsf{vol}}

\newcommand{\PiBT}{\ifdraft\PiTree\else\PiBfly\fi}

\newcommand{\PiTB}{\ifdraft\PiBfly\else\PiTree\fi}

\newcommand{\comm}{commutable}

\newcommand{\start}{\mathsf{start}}

\newcommand{\partway}{\mathsf{partway}}

%% file: prelims.tex
\section{Preliminaries: Onion routing protocols} \label{sec:prelims}

For a set $\mathcal{S}$, we denote the cardinality of $\mathcal{S}$ by $|\mathcal{S}|$, and $s \sample \mathcal{S}$ denotes that $s$ is chosen from $\mathcal{S}$ uniformly at random. For an algorithm $A(x)$, $y \gets A(x)$ is the (possibly probabilistic) output $y$ from running $A$ on the input~$x$. In this paper, $\log(x)$ is the logarithm of $x$ base $2$.  

We say that a function $f: \mathbb{N} \mapsto \mathbb{R}$ is negligible in the parameter~$\secpar$, written $f(\secpar) = \negl$, if for a sufficiently large $\secpar$, $f(\secpar)$ decays faster than any inverse polynomial in $\secpar$. When $\secpar$ is the security parameter, an event $E_\secpar$ is said to occur with (non-)negligible probability if the probability of $E_\secpar$ can(not) be bounded above by a function negligible in $\secpar$. An event occurs with overwhelming probability (abbreviated, w.o.p.) 
if its complement occurs with negligible probability. We use the standard notion of a pseudorandom function~\cite[Chapter~3.6]{Goldreich01}.

\subsection{Onion encryption schemes} \label{sec:or}
Our work on onion routing builds upon a secure onion encryption scheme~\cite{C:CamLys05,KBS19,AL20}. 
Recall that an onion encryption scheme is a triple: $(\gen, \formonion, \proconion)$. The algorithm~$\gen$ generates a participant key pair, i.e., a public key and a secret key. The algorithm $\formonion$ forms onions, and the algorithm $\proconion$ processes onions.  

Let $\parties$ be a set of participants, and let $\Bad\subseteq\parties$ be the set of corrupt parties. 
For every honest $P\in\parties\setminus\Bad$, let $(\pk(P), \sk(P)) \gets \gen(\secparam, \pp, P)$ be the key pair generated for party~$P$, where $\secpar$ is the security parameter, and $\pp$, the public parameters. 
For every corrupt party $P\in\Bad$, let $\pk(P)$ denote $P$'s public key. 

Let $\mathcal{M}$ be the message space consisting of messages of the same fixed length, and let the nonce space $\mathcal{S}$ consist of nonces of the same fixed length.  These lengths may be a function of the security parameter $\secpar$. Here, a \emph{nonce} is really any metadata associated with an onion layer. 

The algorithm $\formonion$ takes as input 
a message~$m\in\mathcal{M}$, 
an ordered list~$(Q_1, \dots, Q_{\depth-1}, R)$ of parties from~$\parties$, 
the public keys~$(\pk(Q_1), \dots, \pk(Q_{\depth-1}), \pk(R))$ associated with these parties, and 
a list~$(s_1, \dots, s_{\depth-1})$ of (possibly empty) nonces from $\mathcal{S}$ associated with the layers of the onion.\footnote{Technically, the input/output syntax and constructions of~\cite{C:CamLys05,AL20} do not include the sequence $(s_1, \dots, s_\depth)$ of nonces but can easily be extended to do so; if we use layered CCA2-secure encryption instead of onion encryption --- which is fine for this application --- then incorporating the nonces is trivial.}
The party~$R$ is interpreted as the \emph{recipient} of the message, and the list~$(Q_1, \dots, Q_{\depth-1}, R)$ is the \emph{routing path}. The output of $\formonion$ is a sequence $(O_1, \dots, O_{\depth})$ of onions. Such a sequence is referred to as an \emph{evolution}, but every $O_i$ in the sequence is an onion. Because it is convenient to think of an onion as a layered encryption object where processing an onion $O_i$ produces the next onion $O_{i+1}$, we sometimes refer to the process of revealing the next onion as ``decrypting the onion'' or ``peeling the onion.'' 

For every $i \in [\depth-1]$, only intermediary party~$Q_i$ can peel onion~$O_i$ to reveal the next layer,  
\mymathenv{
(O_{i+1}, Q_{i+1}, s_{i+1}) \gets \proconion(\sk(Q_i), O_i, Q_i) ,
}
which contains the \emph{peeled} onion~$O_{i+1}$, the \emph{next destination}~$Q_{i+1}$ of the onion, and the nonce~$s_{i+1}$. Only the recipient $R$ can peel the innermost onion~$O_{d}$ to reveal the message,
\mymathenv{
m \gets \proconion(\sk(R), O_{\depth}, R) .
}

In our constructions, a sender of a message~$m$ to a recipient~$R$ ``forms an onion'' by generating nonces and running the $\formonion$ algorithm on the message~$m$, a routing path $(Q_1, \dots, Q_{\depth-1}, R)$, the keys $(\pk(Q_1), \dots, \pk(Q_{\depth-1}), \pk(R))$ associated with the parties on the routing path, and the generated nonces; the \emph{formed onion} is the first onion~$O_1$ from the list of outputted onions. The sender (i.e., the party who formed the onion) can send $O_1$ to the first party $Q_1$ on the routing path, who can process $O_1$ and send the peeled onion~$O_2$ to the next destination $Q_2$, and so on. When the last onion~$O_{\depth}$ is received by the recipient~$R$, $R$ can processes it to obtain the message~$m$. 

\paragraph{Secure onion encryption.} 
Suppose that (honest) Alice generates an onion carrying a message $m$ for Bob. That is, she generates a string of nonces and runs the algorithm $\formonion$ on the inputs: the message $m$, the routing path $(Q_1, \dots, Q_{i-1}, I, Q_{i+1}, \dots, Q_{\depth-1}, \text{Bob})$, the public keys associated with the routing path, and the nonces. Let $O$ denote the onion for intermediary party $I$, i.e., $O$ is the $i^\textit{th}$ onion in the outputted evolution. 

Suppose that (honest) Carol runs the algorithm $\formonion$ on the inputs: the message $m'$, the routing path $(Q'_1, \dots, Q'_{j-1}, I, Q'_{j+1}, \dots, Q'_{e-1}, \text{David})$, the public keys associated with the routing path, and some nonces. Let $O'$ denote the onion for intermediary party $I$. 

Provided that the onion encryption scheme is secure, if party $I$ receives onions $O$ and $O'$ in the same round and consequently processes the two onions in the same batch, then the adversary cannot tell which processed onion resulted from processing $O$, and which resulted from processing $O'$. In other words, onions formed by honest parties ``mix'' at honest parties. For a precise, cryptographic definition of secure onion encryption, see the recent paper by Ando and Lysyanskaya~\cite{AL20}.

\paragraph{Remark.} Note that, in our protocol, the adversary already knows how many layers each onion has to begin with and how many remain at each round.  Thus, the secure onion routing definitions~\cite{C:CamLys05,KBS19,AL20} give us even more security than we need. Onion encryption that satisfies them is good for our purposes, but a simpler and potentially more efficient construction will work too. Forming $O_i$ by encrypting $(O_{i+1},Q_{i+1},s_{i+1})$ under the public key of $Q_{i+1}$ using a CCA2-secure cryptosystem will also work for our purposes. 

\subsection{Formal definition of an onion routing protocol}
In an onion routing protocol, all the packets sent between protocol participants are treated as onions; i.e., upon receipt, they are fed to $\proconion$.  Moreover, internally, there are type checks that ensure that these onions are processed properly.

There are two cases for processing an honestly formed onion properly: the case where peeling the onion reveals its next layer and destination, and the case where it reveals the message of which the processing party is the destination.

In an onion routing protocol, if $Q_i$ runs $\proconion$ and  outputs the next layer of the onion $O_{i+1}$ (together with its destination $Q_{i+i}$ and nonce $s_{i+1}$), then the only two options for what an onion routing protocol permits $Q_i$ to do with $O_{i+1}$ is either send it to $Q_{i+1}$, or drop it (if $Q_{i+1} = Q_i$ then this send step is internal to $Q_i$).   Which of these actions are taken depends on the specifics of the algorithm, and also on the values $(Q_{i+1},s_{i+1})$, but those are the only options.  In other words, the protocol for an onion routing scheme cannot have an onion sent to incorrect destinations or fed as input to another algorithm.  

Further, if $Q_i$ runs $\proconion$ and outputs a message $m\neq \bot$, then this message becomes (ultimately, at the end of the protocol) part of its output, i.e., it will be on the list of messages that have been sent to $Q_i$.   In other words, the message $m$ that is the output of $\proconion$ cannot be internal to the protocol, it must be the message that someone sent to $Q_i$ via the protocol.
Conversely, in an onion routing protocol, the only way that a message $m$ can be on the list of messages received by $Q_i$ is if $Q_i$ obtained it by peeling one of the onions it received.  

These restrictions on protocol design are natural. Indeed, any implementation of onion routing would ensure that it is adhered to by using type checking of the objects created, sent, and processed by the algorithm.  Without such a restriction, any protocol can be thought of as an instance of onion routing protocol, so limiting our attention in this way is meaningful.

Note that this places restrictions just on the protocol that the honest parties are executing; the adversary is still free to do anything he wishes: to mismatch types, to route onions incorrectly, to try to rewrap onions, to form and process onions adversarially, etc.

\paragraph{Correct and indifferent onion routing.}
Onion routing serves a purpose: to route messages from senders to recipients.  Therefore, it needs to satisfy \emph{correctness}:
\begin{definition}
A messaging protocol $\Pilong$ is correct if in an interaction with a passive adversary (i.e., when the adversary doesn't deviate from the protocol), it delivers all the messages with overwhelming probability.
\end{definition} 

In this paper, we will consider only correct onion routing protocols, but we will analyze their interactions with active adversaries. 
Further, the protocols we design in this paper have an additional attractive property of being \emph{indifferent}: 

\begin{definition} [Indifference] \label{def:indiff}
An onion routing protocol $\Pilong$ is \regOR\ if two properties hold: (1) The routing path corresponding to each honestly formed onion is of a fixed length. 
(2) The sequence of intermediaries, including the recipients of dummy onions, and the sequence of nonces corresponding to each honestly formed onion do not depend on the input. (An onion is a dummy if it reveals the empty message ``$\bot$'' when it is peeled all the way.)
\end{definition}

The intuition behind this notion is that the contents of the messages sent and received between parties have no bearing on how the messages are routed and transmitted.  For protocol design, indifference is an attractive property that allows components of an onion to be in place (and possibly the bulk of the cryptographic computation finished) before the message contents even becomes known.  Another attractive feature of indifferent protocols is that their security properties are easier to analyze, as we will explore in the next section.  

Our negative results apply to all onion routing schemes, indifferent or not. 

%% file: lower1.tex
\section{Security definitions: anonymity, equalizing, and mixing} \label{sec:defs}

\paragraph{A motivating example.}
Consider Ando, Lysyanskaya, and Upfal's very simple protocol $\Pi_p$ ($p$, for passive) in the passive adversary setting~\cite{ICALP:AndLysUpf18}. Recall that corrupted parties also follow the protocol in this setting.

Let $\mathsf{Servers} \subseteq \parties$ be the set of servers which is a subset of $\parties$. 

During the \emph{onion-forming phase}, every party $P$ generates an onion from the message-recipient pair $(m, R)$ in $P$'s input by first choosing $\depth-1$ servers $(S_1, \dots, S_{\depth-1})$, each chosen independently and uniformly at random from $\mathsf{Servers}$. Next, $P$ forms an onion $O$ by running $\formonion$ on the message~$m$, the routing path $P^\rightarrow = (S_1, \dots, S_{\depth-1}, R)$, the public keys associated with $P^\rightarrow$, and the sequence of empty nonces. 
At the first round of the \emph{execution phase}, each party~$P$ sends its formed onion $O$ to the first server~$S_1$ on the routing path. For every round $r\in[\depth]$, each server $S$ does the following:
\begin{itemize}
\item Between the $r^\textit{th}$ and $(r+1)^\textit{st}$ rounds, $S$ processes all the onions it received at the $r^\textit{th}$ round. 
\item At the $(r+1)^\textit{st}$ round, $S$ sends the processed onions to their respective next destinations. 
\end{itemize}
At the $\depth^\textit{th}$ round, each party receives an onion that, once processed, reveals a message $m$ for the party. 

$\Pi_p$ is anonymous if the protocol sufficiently shuffles the onions during the execution phase. In prior work~\cite{ICALP:AndLysUpf18}, Ando, Lysyanskaya, and Upfal showed that sufficient shuffling occurs when the server load (i.e., the average number of onions received by a server at a round: $\frac{N}{|\mathsf{Servers}|}$) and the number of rounds (i.e., $\depth$) are both superlogarithmic in the security parameter. 

However, there is no parameter setting for which $\Pi_p$ can be anonymous from the \emph{active} adversary.  If $\corruptions N$ out of $N$ participants are corrupted, then with probability $\corruptions$, the adversary can determine the recipient of any honest party, say Alice: Suppose that during the onion-forming phase, Alice picks a routing path that begins with an adversarial party $S_1$. During the execution phase, the adversary can direct $S_1$ to drop Alice's onion before the second round. In this case, the adversary can figure out who Alice's recipient is (say, it's Bob) by observing who does not receive an onion at the end of the protocol run. 

The motivating example illustrates that while  \emph{mixing} (i.e. sufficiently shuffling onions) is helpful for achieving anonymity, it is not enough. To be anonymous, the protocol must also guarantee that the numbers of messages received by the parties don't reveal the input. We call this property, \emph{equalizing}. 

\paragraph{Relating equalizing and mixing to anonymity.}
Here, we provide formal game-based definitions of anonymity (Section~\ref{sec:anon}), equalizing (Section~\ref{sec:equal}), and mixing (Section~\ref{sec:mix}). Given these definitions, it can be shown that for indifferent onion routing protocols, equalizing and mixing imply anonymity:

\begin{theorem} \label{thm:impliedby}
For any adversary class $\mathbb{A}$, an \regOR\ (Definition~\ref{def:indiff}) onion routing protocol that mixes and equalizes for $\mathbb{A}$ in the simple I/O setting is anonymous for $\mathbb{A}$ in the simple I/O setting, provided that the underlying onion encryption is secure (i.e., UC-realizes the ideal functionality for onion encryption~\cite{AL20}). 
\end{theorem}

The proof is by a reduction and can be found in Appendix~\ref{app:impliedby}. We will use Theorem~\ref{thm:impliedby} to prove our upper bound in Section~\ref{sec:proof}. 

\input{anonymity} 

\input{equalizing}

\input{mixing}

%% file: anonymity.tex
\subsection{Anonymity} \label{sec:anon}
Anonymity is a property of a messaging protocol $\Pi$ (i.e., $\Pi$ doesn't have to be an onion routing protocol). 

The definition of anonymity is standard indistinguishability. 
Recall that indistinguishability is defined using a security game in which the adversary chooses any two inputs to the system: $\sigma^0$ and $\sigma^1$. The system is secure if no adversary can distinguish between the two scenarios: running the system on input $\sigma^0$ (scenario~0) and running the system on input~$\sigma^1$ (scenario~1). 

In the anonymity game (for defining anonymity), the adversary necessarily learns the corrupt parties' inputs and received messages. For example, let $N=4$, and let $P_3$ be a corrupt party. 
\begin{itemize}
\item Suppose that the adversary chooses as inputs $\sigma^0 = (\sigma_1^0, \sigma_2^0, \sigma_3^0, \sigma_4^0)$ and $\sigma^1 = (\sigma_1^1, \sigma_2^1, \sigma_3^1, \sigma_4^1)$ such that $\sigma_3^0 \neq \sigma_3^1$. Then, the adversary can determine the input from $P_3$'s input. 

\item Suppose that the adversary chooses as inputs $\sigma^0$ and $\sigma^1$ such that $\sigma^0$ contains an instruction to  send message $m^0$ to $P_3$, whereas $\sigma^1$ contains an instruction to send message $m^1 \neq m^0$ to $P_3$. Then, the adversary can determine the input from $P_3$'s received message. 
\end{itemize}

Thus, the adversary's choice for $(\sigma^0, \sigma^1)$ is constrained to pairs of inputs that differ only in the honest parties' inputs and ``outputs.'' We define this formally by first defining equivalence classes for inputs as follows: 

\iftheory
\paragraph{Input equivalence classes.}
\fi
Let $\Sigma$ be a set of input vectors. Let $\adv$ be the adversary, and let $\Bad$ be the set of parties controlled by $\adv$. 
Fixing $\Bad$ imposes an equivalence class on $\Sigma$. Each equivalence class is defined by a vector $(e_1, e_2, \dots, e_N)$. For each corrupted party $P_i \in \Bad$, $e_i = (\sigma_i, \mathcal{M}_i)$ ``fixes'' the input $\sigma_i$ for $P_i$ and also, the set $\mathcal{M}_i$ of messages instructed to be sent from honest parties to $P_i$. For each honest party $P_i \in \parties\setminus\Bad$, $e_i = V_i$ ``fixes'' the number $V_i$ of messages instructed to be sent from honest parties to $P_i$.  An input vector belongs to the equivalence class $(e_1, e_2, \dots, e_N)$ if for every $P_i \in \Bad$, the input for $P_i$ is $\sigma_i$, the set of messages from honest parties to $P_i$ is $\mathcal{M}_i$, and $e_i = (\sigma_i, \mathcal{M}_i)$; and if for every $P_i \in \parties \setminus \Bad$, the number of messages from honest parties to $P_i$ is $V_i$, and $e_i = V_i$. 
Two input vectors $\sigma^0$ and $\sigma^1$ are equivalent w.r.t.\ the adversary's choice $\Bad$ for the corrupted parties, denoted $\sigma^0 \equiv_{\Bad} \sigma^1$,  if they belong to the same equivalence class imposed by $\Bad$.

We define anonymity using the anonymity game (below) in which the adversary picks two inputs from the same equivalence class; the protocol is anonymous if this induces indistinguishable adversarial views. 

\paragraph{The anonymity game.}
\sloppy The anonymity game $\mathsf{AnonymityGame}(\secparam, \Pi, \adv, \Sigma)$ is parametrized by the security parameter $\secparam$, a protocol $\Pi$, an adversary $\adv$, and a set~$\Sigma$ of input vectors. 

First, the adversary $\adv$ and the challenger $\cdv$ set up the parties' keys: $\adv$ chooses a subset $\Bad \subseteq \parties$ of the parties to corrupt and sends $\Bad$ to the challenger $\cdv$. For each honest party in $\parties\setminus\Bad$, $\cdv$ generates a key pair for the party; the public keys $\pk(\parties\setminus\Bad)$ of the honest parties are sent to $\adv$. $\adv$ picks the keys for the corrupted parties and sends the corrupted parties' public keys $\pk(\Bad))$ to $\cdv$. 

Next, the input is selected: $\adv$ picks two input vectors $\sigma^0, \sigma^1 \in \Sigma$ such that $\sigma^0 \equiv_{\Bad} \sigma^1$ and sends them to $\cdv$. $\cdv$ chooses a random bit $b \sample \bin$ and interacts with $\adv$ in an execution of protocol $\Pi$ on input $\sigma^b$ with $\cdv$ acting as the honest parties adhering to the protocol and $\adv$ controlling the corrupted parties. 

At the end of the execution, $\adv$ computes a guess $b'$ for $b$ from its view $\view{\Pi, \adv, \Bad}{\secparam, \sigma^b}$ and wins the anonymity game if $b' = b$. 
See Figure~\ref{fig:anon}.   

\input{anonfig}

The standard notion of anonymity is defined as follows:

\begin{definition} [Anonymity] \label{def:anon}
A messaging protocol $\Pilong$ is anonymous from the adversary class $\mathbb{A}$ w.r.t.\ the input set $\Sigma$ if every adversary $\adv \in \mathbb{A}$ wins the anonymity game $\mathsf{AnonymityGame}(\secparam, \Pi, \adv, \Sigma)$ with only negligible advantage, i.e., 
\mymathenv{
\left|\prob{\text{$\adv$ wins $\mathsf{AnonymityGame}(\secparam, \Pi, \adv, \Sigma)$}} - \frac{1}{2} \right| = \negl .
}

The protocol is computationally (resp.\ statistically) anonymous if the adversaries in $\mathbb{A}$ are computationally bounded (resp.\ unbounded). 
\end{definition}

%% file: anonfig.tex
\begin{figure}[ht!]
\centering
\begingroup
\fontsize{9pt}{11pt}\selectfont
\procedure{$\mathsf{AnonymityGame}(\secparam, \Pi, \adv, \Sigma)$}{%
\adv \> \> \cdv \\
\text{pick } \Bad \subseteq \parties  
\> \sendmessageright*[3.0cm]{\Bad} \\
\> \sendmessageleft*[3.0cm]{\pk(\parties\setminus\Bad)} \\
\> \sendmessageright*[3.0cm]{\pk(\Bad)} \\
\text{pick } \sigma^0, \sigma^1 \in \Sigma \text{ s.t.\ } \sigma^0 \equiv_\Bad \sigma^1  
\> \sendmessageright*[3.0cm]{\sigma^0, \sigma^1} \\
\> \> \text{sample } b \sample \bin \pclb
\pcintertext[dotted]{interact in run of protocol $\Pi$ on input $\sigma^b$}
\text{guess and output $b'$} 
}
\endgroup
\caption{\footnotesize Schematic of the anonymity game.}
\label{fig:anon}
\end{figure}

%% file: equalizing.tex
\subsection{Equalizing} \label{sec:equal}
Here, we introduce a new concept called equalizing, which is closely related to anonymity. Like anonymity, equalizing is a property of a messaging protocol $\Pi$. 

Informally, $\Pi$ equalizes if observing how many messages each party received during the protocol run does not reveal whether the protocol ran on $\sigma^0$ or $\sigma^1$. In $\Pi_p$ (in our motivating example), whether Bob receives a message or not exposes who was sending Bob the message: Alice or another party, Allison; so $\Pi_p$ does not equalize. Instead, in an equalizing protocol, the probability that Bob receives a message doesn't depend on the sender's identity. Put another way, Bob is expected to receive the same number of messages in the scenario where Alice is the sender as the one where it is Allison. Formally, equalizing is defined with respect to the equalizing game (below). 

\paragraph{The equalizing game.}
The equalizing game $\mathsf{EqualizingGame}(\secparam, \Pi, \adv, \ddv, \Sigma)$ is parametrized by the security parameter $\secparam$, a protocol $\Pi$, an adversary $\adv$, a distinguisher~$\ddv$, and a set $\Sigma$ of input vectors. 

The challenger for the equalizing game first interacts with the adversary exactly the same way as the challenger for the anonymity game. (See the previous section, Section~\ref{sec:anon}, for the description of the anonymity game.)

Recall that at the end of the anonymity game, each honest party $P_r$ outputs the set $\oh{\Pi, \adv}{r}{\secparam, \sigma^b}$ of (non-empty) messages from the message space $\mathcal{M}$ that it obtained during the execution from processing onions. Let $v_r$ be the \emph{number} of messages that $P_r$ received during the run, i.e., $ v_r \myeq |\oh{\Pi, \adv}{r}{\secparam, \sigma^b}|$. (These statistics are part of the adversary's view in the anonymity game.)

We define the statistics for the corrupt parties differently since $\cdv$ does not get to observe how many messages the corrupt parties output; indeed it is not even clear what it means for a corrupt party to produce an output. For each recipient $P_r \in \Bad$, let $v_r$ correspond to the number of onions that $\cdv$ has routed to an adversarial participant $P'$ such that (1) they had been formed by an honest participant with $P_r$ as the recipient; and (2) all the participants after $P'$ on the remainder of this onion's route are controlled by the adversary.  In other words, $v_r$ is the number of onions from honest participants that $P_r$ would receive if, internal to the adversary, all the onions are processed and delivered to their next destinations.  We define this formally below. 

Let $\mathsf{msPairs}(P_r)$ denote the set of message-sender pairs for $P_r$. That is, for every $(m, P_s) \in \mathsf{msPairs}(P_r)$, the input $\sigma_s$ for $P_s$ includes the message-recipient pair $(m, P_r)$, i.e., $(m, P_r) \in \sigma_s$. Let $\mathsf{receivableOnions}(P_r)$ be the following set of onions: An onion $O$ is in $\mathsf{receivableOnions}(P_r)$ if there exists a message-sender pair $(m, P_s) \in \mathsf{msPairs}(P_r)$ such that 
\begin{enumerate} [label=\roman*.]
\item $O$ was formed by $\cdv$ (on behalf of $P_s$) by running $\formonion$ on input the message $m$, a routing path $P^\rightarrow = (Q_1, \dots, Q_{\depth-1}, P_r)$ ending in $P_r$, the public keys $\pk(P^\rightarrow)$ of the parties on the path, and a sequence $s^\rightarrow$ of nonces, i.e., $O \in \{O_1, \dots, O_{\depth}\}$ where 
\mymathenv{
(O_1, \dots, O_{\depth}) \gets \formonion(m, (P^\rightarrow), \pk(P^\rightarrow), s^\rightarrow) ;
}

\item letting $i$ denote the position of $O$ in the output of the $\formonion$ call, either $i=1$, or the $(i-1)^\textit{st}$ intermediary $Q_{i-1}$ on the routing path is honest; and

\item $O$ is ``peelable all the way'' by $\adv$; i.e., $Q_i, \dots, Q_{\depth-1}, P_r$ are all adversarial.
\end{enumerate}   

For each adversarial recipient $P_r \in \Bad$, we define the statistic $v_r$ to be the number of onions in $\mathsf{receivableOnions}(P_r)$ that the challenger sent out during the execution. 

Let $\mathbf{v} = (v_1, v_2, \dots, v_{N})$. 
$\cdv$ provides these statistics $\mathbf{v}$ alone (and not the rest of the view) to the distinguisher~$\ddv$, who outputs a guess $b'$ for the challenge bit and wins the game if $b' = b$, i.e. if it correctly determines whether the challenger ran the protocol on input $\sigma^0$ or $\sigma^1$.
See Figure~\ref{fig:equal}. 

\input{equalfig}

The definition for equalizing is as follows. 

\begin{definition} [Equalizing] \label{def:equal}
\sloppy A messaging protocol $\Pilong$ equalizes for the adversary class $\mathbb{A}$ w.r.t.\ the input set $\Sigma$ if for every adversary~$\adv \in \mathbb{A}$ and distinguisher $\ddv$, $\ddv$ wins $\mathsf{EqualizingGame}(\secparam, \Pi, \adv, \ddv, \Sigma)$ with negligible advantage, i.e.,
\mymathenv{
\left|\prob{\text{$\ddv$ wins $\mathsf{EqualizingGame}(\secparam, \Pi, \adv, \ddv, \Sigma)$}} - \frac{1}{2} \right| = \negl .
}

The protocol computationally (resp.\ statistically) equalizes if the adversaries and the distinguishers are computationally bounded (resp.\ unbounded).
\end{definition}

Clearly, a protocol that satisfies anonymity must equalize: 
\begin{theorem} \label{thm:implies}
For any adversary class $\mathbb{A}$, a protocol that is anonymous for $\mathbb{A}$ w.r.t.\ the input set $\Sigma$ equalizes for $\mathbb{A}$ w.r.t.\ $\Sigma$.  
\end{theorem}

\underline{Proof:} If $\ddv$ can guess $b$ based on the statistics~$\mathbf{v}$ alone, then the adversary $\adv$ who has access to the entire view of its interaction with $\cdv$ can guess $b$ also. 
It is also easy to see that a protocol need not satisfy anonymity in order to satisfy equalizing.  Thus, equalizing is necessary but not sufficient to achieve anonymity.  

%% file: equalfig.tex
\begin{figure}[ht!]
\centering
\begingroup
\fontsize{9pt}{11pt}\selectfont
\procedure{$\mathsf{EqualizingGame}(\secparam, \Pi, \adv, \ddv, \Sigma)$}{%
   \adv \> \> \cdv \> \ddv \\
 \begin{subprocedure}%
 \dbox{\procedure{$\mathsf{AnonymityGame}$}{%
  	 \\
	 }}
  \end{subprocedure} \\ 
   \> \> \> \sendmessageright*[3.0cm]{\mathbf{v}} \\
  \> \> \> \> \text{guess and output $b'$} }
\caption{\footnotesize Schematic of the equalizing game. }
\endgroup
\label{fig:equal}
\end{figure}

%% file: mixing.tex
\subsection{Mixing  in the simple I/O setting} \label{sec:mix}
Mixing is a property of \emph{onion routing} protocols. 
Informally, an onion routing protocol mixes if the protocol sufficiently shuffles the honest parties' ``message-bearing'' onions. That is, once an honestly generated onion has traveled far enough, getting peeled at every intermediary, the adversary cannot trace it to the original sender.  If the adversary is the recipient of the message contained in the onion, it should not be able to trace it to the sender provided the message itself does not reveal the sender.

Formally, mixing is defined with respect to the mixing game. To keep things simple, we present the definition in the simple I/O setting. This can be extended to any arbitrary input set. 

\paragraph{The mixing game.} 
Let $\mathcal{OE} = (\gen, \formonion, \proconion)$ be a secure onion encryption scheme. 
The mixing game $\mathsf{MixingGame}(\secparam, \Pi, \adv)$ is parametrized by the security parameter $\secparam$, an onion routing protocol $\Pi$, and an adversary $\adv$. 

First, the adversary $\adv$ and the challenger $\cdv$ set up the parties' keys (exactly as we described above for the anonymity game): $\adv$ chooses a subset $\Bad \subseteq \parties$ of the parties to corrupt and sends $\Bad$ to $\cdv$. For each honest party in $\parties\setminus\Bad$, $\cdv$ generates a key pair for the party by running the onion encryption scheme's key generation algorithm $\gen$ and sends the public keys $\pk(\parties\setminus\Bad)$ of the honest parties to the adversary~$\adv$. $\adv$ picks the keys for the corrupted parties and sends the public-key portions $\pk(\Bad)$ to $\cdv$. 

Next, the input is selected: $\adv$ identifies a set $\senders \subseteq \parties \setminus \Bad$ of honest \emph{target senders} and a set $\receivers \subseteq \parties$, $|\receivers| = |\senders|$ of \emph{target receivers}. In addition to $\senders$ and $\receivers$, 
$\adv$ also decides \emph{part} of the input; for every non-target sender $P_s \in \parties\setminus\senders$, $\adv$ chooses a message $m$ and a unique non-target recipient $P_r \in\parties\setminus\receivers$ such that $P_s$'s input becomes $\sigma_s = \{(m, P_r)\}$; and for every target recipient $P_r \in \receivers$, $\adv$ chooses a message $m_r$ to be sent to $P_r$. 
We call the portion of the input that $\adv$ decides ``\emph{the partial input vector},'' and denote it $\tilde \sigma$. $\adv$ sends $(\senders, \receivers, \tilde\sigma)$ to the challenger $\cdv$. $\cdv$ supplies the rest of the input vector $\sigma = (\sigma_1, \dots, \sigma_N)$ by choosing a random bijection $g$ from $\senders$ to $\receivers$; each $P_s \in \senders$ is instructed to send the message $m_{g(P_s)}$ to $g(P_s) \in \receivers$, i.e., $\sigma_s = \{(m_{g(P_s)}, g(P_s)\}$ where the message $m_{g(P_s)}$ was supplied by $\adv$ as part of the partial input vector. 

Next, $\cdv$ interacts with $\adv$ in an execution of protocol $\Pi$ on input $\sigma$ with $\cdv$ acting as the honest parties adhering to the protocol and $\adv$ controlling the corrupted parties. Whenever the protocol $\Pi$ specifies for an onion to be formed or processed, $\cdv$ runs the onion encryption scheme's onion-forming algorithm $\formonion$ or onion-processing algorithm $\proconion$.

Let $\mathsf{O}_{\receivers}$ be the set of onions received by the parties in $\receivers$. 

At the end of the execution, $\adv$ chooses two onions $O_s, O_{\bar{s}}\in\mathsf{O}_{\receivers}$ and 
a target sender $P_s \in \senders$ and 
outputs $(O_s, O_{\bar{s}}, P_s)$. 

Let an onion $O$ be a ``\emph{valid challenge onion}'' if 
(i)~there exists a message $m_r \in \mathcal{M}$ and a target recipient $P_r \in \receivers$ such that $m_r$ is $\adv$'s choice for the message to be sent to $P_r$, and 
(ii) $O$ is the last onion to be received by the recipient over the network in the onion evolution generated by $\cdv$ on behalf of one of the target senders running $\formonion$ on the message $m_r$ and a routing path ending in $P_r$. 

Let $\mathsf{sender}(O_s)$ be the sender of $O_s$, and let $\mathsf{sender}(O_{\bar{s}})$ be the sender of $O_{\bar{s}}$. To maximize his chances of winning the game, the adversary wants both $O_s$ and $O_{\bar{s}}$ to be valid challenge onions such that $O_s$ was sent by $P_s$, while $O_{\bar{s}}$ was not. Formally, if $\adv$ chose two valid challenge onions, and $\{P_s\} \subset \{\mathsf{sender}(O_s), \mathsf{sender}(O_{\bar{s}})\} \subseteq \senders$, then $\adv$ wins iff $P_s = \mathsf{sender}(O_s)$. Otherwise, if $\adv$ did not choose two valid challenge onions, or if $\{P_s\} \not\subset \{\mathsf{sender}(O_s), \mathsf{sender}(O_{\bar{s}})\}$ or $\{\mathsf{sender}(O_s), \mathsf{sender}(O_{\bar{s}})\} \not\subseteq \senders$, then $\adv$ wins with probability one-half. 
See Figure~\ref{fig:mix} for a quick reference to the mixing game.  

\begin{figure}[ht!]
\centering
\begingroup
\fontsize{9pt}{11pt}\selectfont
\procedure{$\mathsf{MixingGame}(\secparam, \Pi, \adv)$}{%
\adv \> \> \cdv \\
\text{pick } \Bad \subseteq \parties  
\> \sendmessageright*[3.0cm]{\Bad} \\
\> \sendmessageleft*[3.0cm]{\pk(\parties\setminus\Bad)} \\
\> \sendmessageright*[3.0cm]{\pk(\Bad)} \\
\text{pick honest } \senders \subseteq \parties \setminus \Bad \\
\text{pick } \receivers \subseteq \parties \text{ s.t. } |\receivers| = |\senders| \\
\text{pick } \tilde\sigma 
\> \sendmessageright*[3.0cm]{\senders, \receivers, \tilde\sigma} \\
\> \> \text{randomly pick } \sigma \pclb
\pcintertext[dotted]{interact in run of protocol $\Pi$ on input $\sigma$} 
\text{output $O_s, O_{\bar{s}} \in \mathsf{O}_{\receivers}$ and $P_s \in \senders$} 
}
\endgroup
\caption{\footnotesize Schematic of the mixing game.}
\label{fig:mix}
\end{figure}

We now define mixing as follows. 

\begin{definition} [Mixing] \label{def:mix}
\sloppy An onion routing protocol $\Pilong$ mixes conditioned on the event~$E$ for the adversary class $\mathbb{A}$ if given $E$, every adversary $\adv \in \mathbb{A}$ wins $\mathsf{MixingGame}(\secparam, \Pi, \adv)$ with negligible advantage, i.e.,
\mymathenv{
\left|\prob{\text{$\adv$ wins $\mathsf{MixingGame}(\secparam, \Pi, \adv)$} \mid E} - \frac{1}{2} \right| = \negl .
}

The protocol computationally (resp.\ statistically) mixes if the adversaries in $\mathbb{A}$ are computationally bounded (resp.\ unbounded). 
\end{definition} 

Now that we have defined mixing formally, let us walk the reader through our definitional choices.  The starting intuition is that this definition needs to capture that it should be hard for the adversary to pinpoint the origin of an onion received by one of the target recipients.  
This goal comes with a caveat that of course an adversary can determine the sender of an onion that one of the target senders has just created, or, more generally, that hasn't traveled very far and hasn't had a chance to mix with any onions from other target senders.  Hence, we need to restrict the set of onions on which the adversary can win to a set of onions that have traveled far and have already had a chance to mix with other onions.  This is why we have the requirement that the onion be a valid challenge onion.  Intuitively, a valid challenge onion is one that was formed by a target sender and has already arrived at its destination, a target recipient, and now the adversary's job is to figure out where it came from.  

Next, let us explain why, to win the game, the adversary must produce two valid challenge onions, and correctly attribute one of them to a sender $P_s$, while the other must have originated with another target sender.  What does it mean that the adversary cannot trace an onion?  One intuitive approach would be to say: the adversary's chances of winning the game where he picks just one onion and guesses its origin are close to a simulator's chances of winning a game where he just guesses a sender, and the challenger picks the onion uniformly at random and independently of the simulator's guess.  The problem with this approach is that we don't know the best strategy for such a simulator and with what probability it would succeed.  

So our approach is to have the adversary pick a sender and two onions.  ``Mixing" means that, if it so happens that exactly one of them comes from $P_s$ and the other comes from another target sender, then try as he may, the adversary cannot tell which is which any better than by guessing randomly.  And if it doesn't happen that way, then the adversary wins with probability one-half.

%% file: construction.tex

\section{Main tools: checkpoint onions and merging onions} \label{sec:ingredients}

We describe the main ingredients for our constructions: checkpoint onions (a tool that was introduced in prior work~\cite{ICALP:AndLysUpf18}) and a new tool: merging onions. 

\subsection{Checkpoint onions}  
Our goal is to achieve anonymity by ensuring that our protocol mixes and equalizes in the presence of an active adversary that drops onions.  The challenge is: if the adversary drops too many onions, then the remaining ones don't have enough onions to mix with, and so the resulting protocol will not mix. Checkpoint onions give the honest participants a way of checking that there are still enough onions in the system for mixing to be possible.

A \emph{checkpoint onion} $O$ is a dummy onion formed by a party $P$ that travels through the network until, at a pre-determined checkpoint round $r$, it arrives at the intermediary $I$, who is \emph{expecting} it.  If it fails to arrive, then $I$ is alerted to the activity of an active adversary.  

More precisely, let $F_{\cdot}(\cdot, \cdot)$ be a pseudo-random function over two inputs, keyed by $\sk(P, I)$ which is a secret key shared between $P$ and $I$.  Let $b$ be a binary predicate. 
Let $\mathcal{D}$ be the \emph{diagnostic rounds}; the honest parties test whether enough onions remain in the system after these rounds.
For each intermediary $I$ and each round $r \in \mathcal{D}$, $P$ determines whether or not to create a checkpoint onion that will arrive at $I$ at round $r$ by computing $f=F(\sk(P, I), (r, 0))$, and then checking if $b(f) = 1$; if so, $P$ creates this checkpoint onion.  Similarly, the intermediary $I$ will know to expect a checkpoint onion from $P$ at round $r$ by computing $f=F(\sk(P, I), (r, 0))$, and then checking if $b(f) = 1$.

$P$ forms $O$ by running $\formonion$ on input the empty message ``$\bot$,'' a randomly chosen routing path $P^\rightarrow = (I_1, \dots, I_{\depth})$, the public keys associated with parties on $P^\rightarrow$, and a sequence $(s_1, \dots, s_{\depth-1})$ of nonces. The nonce $s_r$ which will be received by $I$, is the value that $I$ will know to expect: $s_r=F(\sk(P, I), (r, 1))$; the rest are random nonces.  The reason that $I$ will know to expect $s_r$ is that $I$ can compute it too, since $\sk(P, I)$ is shared between $P$ and $I$.

Of course, the shared key $\sk(P,I)$ need not be set up in advance: it can be generated from an existing PKI, e.g., using Diffie-Hellman (see prior work for details on checkpoint onions~\cite{ICALP:AndLysUpf18}).

If the adversary drops an onion belonging to the same evolution as $O$ before it reaches $I$, $I$ will detect it: it will detect that no onion with nonce $s_r$ was received in round $r$. (Since $F$ is pseudorandom, it is highly unlikely that another onion peels to the same nonce value.)

\underline{Note:} The number of checkpoint onions that $P$ generates is a pseudorandom variable that depends on the pseudo-random function~$F$ and the binary function $b$. For our mixing mechanism to work, we should choose $F$ and $b$ such that the frequency of forming a checkpoint onion for random party $I$ and a random round $r$ is $\frac{\bigOmega{\text{polylog} \,\secpar}}{N|\mathcal{D}|}$, where by ``$\text{polylog} \,\secpar$,'' we mean polylogarithmic in the the security parameter $\secpar$, and so the expected number of checkpoints that $P$ generates is $\bigOmega{\text{polylog} \,\secpar}$. 
This ensures a strong correlation between the number of missing checkpoints observed by a party and the total number of checkpoint onions that have been dropped before the observation was made.  

\subsection{Merging onions}
Checkpoint onions help with mixing, but not with equalizing.  If our routing protocol just has every sender form one ``message-bearing" onion to its recipient and send it along in addition to a set of checkpoint onions (as in the protocol $\Pi_a$ of Ando, Lysyanskaya and Upfal~\cite{ICALP:AndLysUpf18}), then an adversary who targets the sender Alice can cause Alice's recipient Bob to receive the message with a smaller probability than her alternative recipient, Bill; so this protocol will not equalize, and from Theorem~\ref{thm:implies} has no hope of achieving anonymity.

So how can we design a protocol that equalizes? One approach is to detect when the adversary drops any onions at all (e.g., using verifiable shuffling)~\cite{stadium,atom}, and abort when that happens.  While this approach equalizes, it is not at all fault-tolerant. To achieve fault tolerance and equalizing, the protocol must be able to react to the adversary dropping onions in a way that is less dramatic than total abort.  This can be accomplished by using a new tool: merging onions. 

The idea here is that a sender $P$ can create two onions, $O_1$ and $O_2$ that bear the same message to the same recipient $R$.  Further, they will be routed through the same intermediary $I$, arriving at $I$ at the same round $r$.  Let $O'_1$ (resp.\
$O'_2$) denote the $r^\mathit{th}$ layer of $O_1$ (resp.\ $O_2$) that arrives at $I$ at round $r$.  When $I$ peels both $O'_1$ and $O'_2$, $I$ discovers that they are (essentially) the same onion, and only forwards one of them to the next destination.  If $I$ receives just one of them (because the other one had been dropped by the adversary), then it forwards it to the next destination, too.  

Why does this approach help with equalizing?  Suppose we have a protocol in which every participant creates two message-bearing onions that merge at round $r$. Suppose that the adversary targets the sender Alice and succeeds in dropping one of her two outgoing merging onions.  Since these onions were supposed to merge at round $r$, after round $r$, there are just as many onions for which Alice was the sender (namely, just one onion) as for any other participant.

In general, of course, the adversary may drop more than one onion belonging to Alice.  In fact, in order to guarantee that any of Alice's onions survive with overwhelming probability when the adversary controls  a constant fraction of the network's nodes, Alice needs to send out a superlogarithmic (in the security parameter $\secpar$) number of onions.  In order to equalize the number of onions that make it to each destination, our protocol will have to create not a pair, but $2^h = \bigOmega{\text{polylog} \,\secpar}$ merging onions, organized in a binary tree of height $h$.

We now illustrate how to form $2^h$ merging onions through a toy example for $h=3$. We first construct a binary tree graph of height $3 = \log 8$. We label the root vertex of the tree $v$, and the left-child and right-child of $v$, $v_{0}$ and $v_{1}$. More generally, the left-child of a vertex $v_w$ is $v_{w0}$, and the right-child of $v_w$ is $v_{w1}$, so that the leaf vertices are: $v_{000}$, $v_{001}$, $v_{010}$, $v_{011}$, $v_{100}$, $v_{101}$, $v_{110}$, and $v_{111}$. Each of these leaf vertices corresponds to a separate onion. 

Let $y$ denote a fixed number of rounds; this will later correspond to the length of an ``epoch.'' 
Next, for each vertex $v_i$ of the graph, we choose a random sequence $I^\rightarrow_i = (I_i^1, \dots, I_i^y)$ of $y$ parties and a random sequence $s^\rightarrow_i = (s_i^1, \dots, s_i^y)$ of $y$ nonces, i.e., 
\ifdraft
\begin{align*}
\forall j \in[y] \,,\, I_i^j &\sample \parties \\
\forall j \in[y] \,,\, s_i^j &\sample \mathcal{S} .
\end{align*}
\else
$\forall j \in[y]$, $I_i^j \sample \parties$ and $s_i^j \sample \mathcal{S}$.
\fi

Let the ``\emph{direct path from a leaf vertex $v_\ell$ to the root}'' be the path that begins with $v_\ell$ and recursively moves to its parent vertex until the root vertex $v$ is reached. For example, the direct path from $v_{101}$ to the root is $(v_{101}, v_{10}, v_1, v)$. 

Let the ``\emph{sequence of intermediaries corresponding to leaf vertex $v_\ell$}'' be the sequence of parties corresponding to the parties on the direct path from $v_\ell$ to the root, e.g., for $v_{101}$, it is
\mymathenv{
(I^\rightarrow_{101}, I^\rightarrow_{10}, I^\rightarrow_1, I^\rightarrow) ,
}
where $I^\rightarrow$ is the sequence of parties assigned to the root. 
\ifdraft
See Figure~\ref{fig:G}. 
\input{mergefig}

\else
See Figure~\ref{fig:G}. 
\input{mergefig}
\fi
Let the ``\emph{sequence of nonces corresponding to leaf vertex $v_\ell$}'' be the sequence of nonces corresponding to the parties on the direct path from $v_\ell$ to the root, e.g., for $v_{101}$, it is
\mymathenv{
(s^\rightarrow_{101}, s^\rightarrow_{10}, s^\rightarrow_1, s^\rightarrow) ,
}
where $s^\rightarrow$ is the sequence of nonces assigned to the root. 

For each leaf vertex $v_\ell$, we form an onion $O_\ell^1$ using 
the message $m$ from the input, 
the routing path $(I^\rightarrow_{101}, I^\rightarrow_{10}, I^\rightarrow_1, I^\rightarrow, R)$ where $R$ is the recipient from the input, 
the public key associated with the routing path, 
and the sequence $(s^\rightarrow_{101}, s^\rightarrow_{10}, s^\rightarrow_1, s^\rightarrow)$ of nonces. 
\ifdraft
\input{mergeexample}

\else
\fi
We can generalize this idea to generate an arbitrarily large set of merging onions by using an appropriately large binary tree. 

\section{A stepping stone construction, $\PiTree$} \label{sec:pitree-description}
Let us extend the toy example construction we just saw to a protocol, $\PiTree^{\numOnion, y, t}$, which is a stepping stone for our main construction.
$\PiTree^{\numOnion, y, t}$ is pronounced ``Pi-tree'' from the fact that the onions' routing paths are structured like a binary tree graph and is parametrized by 
the number $\numOnion$ of merging onions per sender (this is also the expected number of checkpoint onions per sender), 
the number $y$ of rounds per epoch, and 
the threshold $t$ for missing checkpoint nonces per diagnostic round. (We will generally omit the superscript for better readability.)

\iftheory
\paragraph{The setup phase.}
\fi
We use a secure onion encryption scheme $(\gen, \formonion, \proconion)$ as a building block (see Section~\ref{sec:prelims} for a description of onion encryption schemes). 
During the setup phase, the participants set up their keys. Every honest party $P$ sets up his/her keys $(\pk(P), \sk(P))$ by running the onion encryption scheme's key generation algorithm $\gen$. 

\paragraph{The onion-forming phase.}
During the onion-forming phase, each honest party~$P$ creates two types of onions: merging onions and checkpoint onions. 
\begin{itemize}
\item On input $\{(m, R)\}$, $P$ forms a set of $\numOnion$ merging onions using 
the number $y$ of rounds in an epoch, 
the message~$m$, and 
the recipient~$R$. 

\item In addition to merging onions, $P$ generates (on average) $\numOnion$ checkpoint onions using the set $\mathcal{D} = \{y, 2y, \dots, (\log \numOnion +1)y\}$ as the diagnostic rounds. (Appropriate functions are chosen for $F_{\cdot}(\cdot, \cdot)$ and $b(\cdot)$ such that $P$ generates $\numOnion$ checkpoint onions in expectation. See \emph{Checkpoint onions} in Section~\ref{sec:ingredients} to recall how these functions are used for generating checkpoint onions.) 
\end{itemize}
For both merging onions and checkpoint onions, the length of the routing path is fixed; it is $(\log x + 1)y + 1$. 
See Figure~\ref{fig:PiTree} for a summary of the setup and onion-forming phases. 

\input{pitreefig}

\underline{Remark:} 
The onion layers are tagged with their respective round number to prevent replay attacks. If by peeling an onion received at round $r$, an honest relaying party observes a round number $r' \neq r$, the party ``drops'' the onion (doesn't relay it to its next destination). 

\paragraph{The execution phase.} 
All onions are created during the onion-forming phase and released simultaneously in the first round of the execution phase.
After each round $r$ of the execution phase, $P$ peels all onions it received at the $r^\textit{th}$ round and merges mergeable onions (i.e., if two onions peel to the same nonce value, drop one of them at random). 

If $r$ is a diagnostic round (i.e., $r\in\mathcal{D}$), $P$ runs the following diagnostic test: Let $\mathsf{Ckpts}(P, r)$ denote the set of checkpoints that $P$ expects to see from peeling the onions between rounds $r$ and $r+1$. $P$ counts how many checkpoints from $\mathsf{Ckpts}(P, r)$ are missing. If the number exceeds a fixed threshold value $t$ (i.e., the onions fail the test), then $P$ aborts. Otherwise (if the number of missing checkpoints is below $t$), $P$ continues for another round by sending the processed onions to their respective next destinations in random order.  

At the end of the execution phase, $P$ peels the onions it received at the last round and outputs the set of (non-empty) messages it received. 

\ifdraft
\subsection{Proof that $\PiTree$ is anonymous} \label{sec:pitree-proof}

For an appropriate choice for 
the number of merging onions formed by each honest sender ($\numOnion$), 
the number of rounds in an epoch ($y$), and 
the threshold ($t$), the protocol $\PiTree$ is anonymous from the active adversary: 

\begin{theorem} \label{thm:PiTree}
Let $N$ be the number of parties and $\secpar$, the security parameter. 
For any constants $\delta, \epsilon > 0$ and $0 \le \corruptions < \frac{1}{2}$, the onion routing protocol $\PiTree$ is anonymous from the adversary who corrupts up to $\corruptions$ fraction of the parties when 
\begin{enumerate} [label=\roman*.]
\item the onion encryption scheme is secure (see~\cite{AL20}),
\item the number $\numOnion$ of onions formed by each (honest) party is $\bigOmega{2^{\ceil{\log(\chi(\log \chi + 1))}}}$ where $\chi = \max(\sqrt{N \log^{2 + \epsilon} \secpar}, \log^{2(1+\epsilon)} \secpar)$, 
\item the number $y$ of rounds per epoch is $\bigOmega{\log^{1+\epsilon} \secpar}$, and 
\item the threshold $t$ is $2 (1-\delta)(1-\corruptions)^3\corruptions \polylog \secpar$. 
\end{enumerate} 
\end{theorem}

\input{proof1}

\fi

\paragraph{Performance of $\PiTree$.} As shown in Appendix~\ref{app:pitree-security}, $\PiTree^{\numOnion, y, t}$ is anonymous from the adversary who corrupts up to $\corruptions$ fraction of the parties when 
(a)~the onion encryption scheme is secure,
(b)~the number $\numOnion$ of onions formed by each (honest) party is $\bigOmega{2^{\ceil{\log(\chi(\log \chi + 1))}}}$ where $\chi = \max(\sqrt{N \log^{2 + \epsilon} \secpar}, \log^{2(1+\epsilon)} \secpar)$, 
(c)~the number $y$ of rounds per epoch is $\bigOmega{\log^{1+\epsilon} \secpar}$, and 
(d)~the threshold $t$ is $2 (1-\delta)(1-\corruptions)^3\corruptions \polylog \secpar$. 

The reason that $\PiTree^{\numOnion, y, t}$ needs so many onions is that the adversary can target Alice and drop a lot of her onions before the honest participants realize (via checkpoint onions) the presence of an attack and abort.  The protocol $\PiBfly$ presented in the next section improves on this by giving the routing paths enough structure that missing onions can be detected sooner.


%% file: mergefig.tex
\definecolor{blue}{RGB}{0, 118, 186}

\begin{figure}[ht!]
\center
\begingroup
\fontsize{9pt}{11pt}\selectfont
\iftheory
\begin{tikzpicture}[scale=0.8,level/.style={sibling distance=250pt/#1}] 
\node [circle,draw,minimum size=30pt,color=blue] (z) {\footnotesize $v$}
  child {node [circle,draw,minimum size=30pt] (a) {\footnotesize $v_0$}
    child {node [circle,draw,minimum size=30pt] (b) {\footnotesize $v_{00}$}
      child {node [circle,draw,minimum size=30pt] (c) {\footnotesize $v_{000}$}} 
      child {node [circle,draw,minimum size=30pt] (d) {\footnotesize $v_{001}$}}
    }
    child {node [circle,draw,minimum size=30pt] (e) {\footnotesize $v_{01}$}
      child {node [circle,draw,minimum size=30pt] (f) {\footnotesize $v_{010}$}}
      child {node [circle,draw,minimum size=30pt] (g) {\footnotesize $v_{011}$}}
    }
  }
  child {node [circle,draw,minimum size=30pt,color=blue] (h) {\footnotesize $v_{1}$}
    child {node [circle,draw,minimum size=30pt,color=blue] (i) {\footnotesize $v_{10}$}
      child {node [circle,draw,minimum size=30pt]  (j) {\footnotesize $v_{100}$}}
      child {node [circle,draw,minimum size=30pt,color=blue] (k) {\footnotesize $v_{101}$}}
    }
  child {node [circle,draw,minimum size=30pt] (l) {$v_{11}$}
    child {node [circle,draw,minimum size=30pt] (m) {\footnotesize $v_{110}$}}
    child {node [circle,draw,minimum size=30pt] (n) {\footnotesize $v_{111}$}}
  }
};
\node [above=of z] (zz) {};
\draw[->,color=blue] (z) edge  node[sloped, above,color=blue] {\footnotesize $I^\rightarrow$} (zz); 
\draw[->] (a) edge  node[sloped, above] {\footnotesize $I^\rightarrow_{0}$} (z); 
\draw[->] (b) edge  node[sloped, above] {\footnotesize $I^\rightarrow_{00}$} (a); 
\draw[->] (c) edge  node[sloped, above] {\footnotesize $I^\rightarrow_{000}$} (b); 
\draw[->] (d) edge  node[sloped, above] {\footnotesize $I^\rightarrow_{001}$} (b); 
\draw[->] (e) edge  node[sloped, above] {\footnotesize $I^\rightarrow_{01}$} (a); 
\draw[->] (f) edge  node[sloped, above] {\footnotesize $I^\rightarrow_{010}$} (e); 
\draw[->] (g) edge  node[sloped, above] {\footnotesize $I^\rightarrow_{011}$} (e); 
\draw[->,color=blue] (h) edge  node[sloped, above,color=blue] {\footnotesize $I^\rightarrow_{1}$} (z); 
\draw[->,color=blue] (i) edge  node[sloped, above,color=blue] {\footnotesize $I^\rightarrow_{10}$} (h); 
\draw[->] (j) edge  node[sloped, above] {\footnotesize $I^\rightarrow_{100}$} (i); 
\draw[->,color=blue] (k) edge  node[sloped, above,,color=blue] {\footnotesize $I^\rightarrow_{101}$} (i); 
\draw[->] (l) edge  node[sloped, above] {\footnotesize $I^\rightarrow_{11}$} (h); 
\draw[->] (m) edge  node[sloped, above] {\footnotesize $I^\rightarrow_{110}$} (l); 
\draw[->] (n) edge  node[sloped, above] {\footnotesize $I^\rightarrow_{111}$} (l); 
\end{tikzpicture}

\else
\begin{tikzpicture}[level/.style={sibling distance=170pt/#1}] 
\node [circle,draw,minimum size=26pt,color=magenta] (z) {$v$}
  child {node [circle,draw,minimum size=26pt] (a) {$v_0$}
    child {node [circle,draw,minimum size=26pt] (b) {$v_{00}$}
      child {node [circle,draw,minimum size=26pt] (c) {$v_{000}$}} 
      child {node [circle,draw,minimum size=26pt] (d) {$v_{001}$}}
    }
    child {node [circle,draw,minimum size=26pt] (e) {$v_{01}$}
      child {node [circle,draw,minimum size=26pt] (f) {$v_{010}$}}
      child {node [circle,draw,minimum size=26pt] (g) {$v_{011}$}}
    }
  }
  child {node [circle,draw,minimum size=26pt,color=magenta] (h) {$v_{1}$}
    child {node [circle,draw,minimum size=26pt,color=magenta] (i) {$v_{10}$}
      child {node [circle,draw,minimum size=26pt]  (j) {$v_{100}$}}
      child {node [circle,draw,minimum size=26pt,color=magenta] (k) {$v_{101}$}}
    }
  child {node [circle,draw,minimum size=26pt] (l) {$v_{11}$}
    child {node [circle,draw,minimum size=26pt] (m) {$v_{110}$}}
    child {node [circle,draw,minimum size=26pt] (n) {$v_{111}$}}
  }
};
\node [above=of z] (zz) {};
\draw[->,color=magenta] (z) edge  node[sloped, above,color=magenta] {$I^\rightarrow$} (zz); 
\draw[->] (a) edge  node[sloped, above] {$I^\rightarrow_{0}$} (z); 
\draw[->] (b) edge  node[sloped, above] {$I^\rightarrow_{00}$} (a); 
\draw[->] (c) edge  node[sloped, above] {$I^\rightarrow_{000}$} (b); 
\draw[->] (d) edge  node[sloped, above] {$I^\rightarrow_{001}$} (b); 
\draw[->] (e) edge  node[sloped, above] {$I^\rightarrow_{01}$} (a); 
\draw[->] (f) edge  node[sloped, above] {$I^\rightarrow_{010}$} (e); 
\draw[->] (g) edge  node[sloped, above] {$I^\rightarrow_{011}$} (e); 
\draw[->,color=magenta] (h) edge  node[sloped, above,color=magenta] {$I^\rightarrow_{1}$} (z); 
\draw[->,color=magenta] (i) edge  node[sloped, above,color=magenta] {$I^\rightarrow_{10}$} (h); 
\draw[->] (j) edge  node[sloped, above] {$I^\rightarrow_{100}$} (i); 
\draw[->,color=magenta] (k) edge  node[sloped, above, color=magenta] {$I^\rightarrow_{101}$} (i); 
\draw[->] (l) edge  node[sloped, above] {$I^\rightarrow_{11}$} (h); 
\draw[->] (m) edge  node[sloped, above] {$I^\rightarrow_{110}$} (l); 
\draw[->] (n) edge  node[sloped, above] {$I^\rightarrow_{111}$} (l); 
\end{tikzpicture}
\fi

\endgroup
\caption{\footnotesize The binary tree $T$ for $\numOnion = 8$. Each leaf node in $T$ corresponds to a merging onion; e.g., onion $O_{101}^1$ is formed using the routing route \textcolor{blue}{$(I^\rightarrow_{101}, I^\rightarrow_{10}, I^\rightarrow_1, I^\rightarrow)$}.}
\label{fig:G}
\end{figure}

%% file: mergeexample.tex
For each onion $O_\ell^1$, let $O_\ell^r$ denote the $r^\textit{th}$ onion layer of $O_\ell^1$. 

Suppose that the onions, $O_{000}^1$, $O_{001}^1$, $O_{010}^1$, $O_{011}^1$, $O_{100}^1$, $O_{101}^1$, $O_{110}^1$, and $O_{111}^1$, are released simultaneously at the first round of the execution phase. For the time being, assume that the adversary is passive; in particular, the adversary does not drop onions. During the first $y$ rounds (i.e., the first epoch), the onions are uncoordinated and move independently through the network. The onions become mergeable for the first time at the $(y+1)^\textit{st}$ round (i.e., the first round of the second epoch); that is, the pair $(O_{000}^{y +1}, O_{001}^{y +1})$ is mergeable, as are the pair $(O_{010}^{y +1}, O_{011}^{y +1})$, the pair $(O_{100}^{y +1}, O_{101}^{y +1})$, and the pair $(O_{110}^{y +1}, O_{111}^{y +1})$. For example, the onions $O_{000}^{y+1}$ and $O_{001}^{y+1}$ meet at party~$I_{00}^1$ (the first party on the path $I^{\rightarrow}_{00}$). When $I_{00}^1$ peels these, they each reveal the nonce value $s_{00}^1$ (the first nonce in $s^{\rightarrow}_{00}$), 
\ifdraft i.e.,
\begin{align*}
(O_{000}^{y+2}, I_{00}^2, s_{000}^{y+1}) &= \proconion(\sk(I_{00}^1), O_{000}^{y+1}, I_{00}^1) \\
(O_{001}^{y+2}, I_{00}^2, s_{001}^{y+1}) &= \proconion(\sk(I_{00}^1), O_{001}^{y+1}, I_{00}^1) ,
\end{align*}
\fi
and so $I_{00}^1$ knows to drop one of them at random. 
For concreteness, let $O_{000}^{y +2}$, $O_{001}^{y +2}$, $O_{101}^{y +2}$, and $O_{110}^{y +2}$ be the onions that remain after the merging step in round $y+1$. These onions become mergeable again at the start of the second epoch; the pairs $(O_{000}^{2y +1}, O_{001}^{2y +1})$ is mergeable, and the pair $(O_{101}^{2y +1}, O_{110}^{2y +1})$ is mergeable. For example, they may merge into the merged onions $O_{000}^{2y +2}$ and $O_{101}^{2y +2}$. At the start of the third (and final) epoch, the merged onions become a mergeable pair and are merged into a single onion bearing the message $m$ for the recipient $R$. 

Let's see what happens when the adversary drops onions $O_{001}^1$, $O_{100}^{1}$, and $O_{101}^{1}$. At the start of the first epoch, there are only two mergeable pairs $(O_{010}^{y+1}, O_{011}^{y+1})$ and $(O_{110}^{y+1}, O_{111}^{y+1})$. These are merged into two onions, for concreteness: $O_{011}^{y+2}$ and $O_{111}^{y+2}$. The onion $O_{000}^{y+1}$ is not part of any mergeable pair, so it continues on as $O_{000}^{y+2}$. At the start of the second round, onions $O_{000}^{2y+1}$ and $O_{011}^{2y+1}$ are merged, and the onion $O_{111}^{2y+1}$ continues on. In the final epoch, these are reduced to a single surviving onion. Thus, even though the adversary dropped many onions upfront, the recipient $R$ still receives the same number of message-bearing onions in the end. 
\ifdraft As we shall see in Section~\ref{sec:rest}, this ``evening out'' of the numbers of remaining merging onions can be guaranteed unless the adversary drops too many onions upfront. 
(Of course, the adversary can also drop merging onions later in the execution, but this doesn't prevent the numbers of onions from evening out; see Section~\ref{sec:rest}.)
\fi

%% file: pitreefig.tex
\begin{figure}[ht!]
\begin{minipage}[t]{0.37\textwidth}
\begingroup
\fontsize{9pt}{11pt}\selectfont
\begin{flushleft}
{Let $\secparam$ denote the security parameter, and let $\Sigma = (\gen, \formonion, \proconion)$ be a secure onion encryption scheme.} 

\bigskip
{Let the public parameters $\pp$ include the participants' public keys, the number $\numOnion$ of merging onions, and the number $y$ of rounds in an epoch.}

\bigskip
\underline{$\mathsf{Setup}(\secparam, \Sigma)$}
\vspace{2mm}
\begin{enumerate} [label=\arabic*:]
\item {Generate a public key pair $(\sk(P), \pk(P))$ by running $\gen(\secparam, \pp, P)$.}  
\end{enumerate}

\bigskip
\underline{$\mathsf{FormOnions}(\secparam, \Sigma, \pp, \{(m, R)\})$}
\vspace{2mm}
\begin{enumerate} [label=\arabic*:]
\item {Form merging onions by running $\mathsf{FormMergingOnions}(\secparam, \Sigma, \pp, m, R)$.} 
\item {Form checkpoint onions by running $\mathsf{FormCheckpointOnions}(\secparam, \Sigma, \pp)$.} 
\item {Return all formed onions.}
\end{enumerate}
\end{flushleft}
\endgroup
\end{minipage}
\begin{minipage}[t]{0.63\textwidth}
\begingroup
\fontsize{9pt}{11pt}\selectfont
\begin{flushleft}
\underline{$\mathsf{FormMergingOnions}(\secparam, \Sigma, \pp, m, R)$}
\vspace{2mm}
\begin{enumerate} [label=\arabic*:]
\item[\tikzmark{bl}1:]  {Generate a binary tree graph $T$ with $\numOnion$ leaf nodes.}
\item[2:] {For each node in $T$, assign $y$ random intermediaries and $y$ random nonces. \hfill} \tikzmark{br} 
\item[3:] {For each leaf node in $T$, form an onion by running $\formonion$ on the message $m$, the routing path consisting of the sequence of intermediaries corresponding to the leaf node and $R$, the public keys of the parties in the path, and the sequence of nonces corresponding to the leaf node.}
\item[4:] {Return all formed onions.}
\end{enumerate}
\tikz[overlay,remember picture]{\draw[blue]
  ($(bl)+(-0.2em,0.9em)$) rectangle
  ($(br)+(0em,-0.3em)$)
  ;}

\bigskip
\underline{$\mathsf{FormCheckpointOnions}(\secparam, \Sigma, \pp)$}
\vspace{2mm}
\begin{enumerate} [label=\arabic*:]
\item {For every round $r \in [(\log \numOnion+1)y ]$ and party $I \in \parties$:}
\begin{itemize}
\item {Let $\mathsf{makeCkpt} \myeq b(F(\sk(P, I), (r, 0))$.}
\item {If $\mathsf{makeCkpt} = 1$:}
\begin{itemize}
\item[\tikzmark{BL}--] {Let $s_r \myeq F(\sk(P, I), (r, 1))$, and for all $i\in[(\log \numOnion+1)y]\setminus\{r\}$, let $s_i \sample \mathcal{S}$ where $\mathcal{S}$ is the nonce space.}
\item {Let $I_r \myeq I$, and for all $i\in[(\log \numOnion+1)y+1]\setminus\{r\}$, let $I_i \sample \parties$.\hfill}\tikzmark{BR}
\end{itemize}
\item {Form onion by running $\formonion$ on the empty message, the routing path $(I_1, \dots, I_{(\log \numOnion+1)y +1})$, the public keys associated with the path, and the sequence $(s_1, \dots, s_{(\log \numOnion+1)y})$ of nonces.}
\end{itemize}
\item {Return all formed onions.}
\end{enumerate}
\tikz[overlay,remember picture]{\draw[blue]
  ($(BL)+(-0.2em,0.9em)$) rectangle
  ($(BR)+(0em,-0.3em)$)
  ;}
\end{flushleft}
\endgroup
\end{minipage}
\caption{\footnotesize Protocol $\PiTree$'s setup and onion-forming algorithms for party $P$ on input $\{(m, R)\}$. The code for generating the intermediaries and nonces are \fbox{boxed}; since these do not depend on the input $\{(m, R)\}$, it is evident that $\PiTree$ is \regOR.}
\label{fig:PiTree}
\end{figure} 

%% file: proof1.tex
\ifdraft\begin{proof}\fi

\begin{lemma}
\label{lemma:ideal}
Let $\Pi$ be onion routing protocol that makes use of an onion encryption scheme that is UC-secure~\cite{FOCS:Canetti01} under a computational assumption $A$.  Let $\Pi^{\mathit{ideal}}$ be the same protocol, but the onion encryption scheme is replaced by the ideal onion encryption functionality of Camenisch and Lysyanskaya~\cite{C:CamLys05}.  If  $\Pi^{\mathit{ideal}}$ is anonymous, then $\Pi$ is anonymous under assumption $A$.
\end{lemma}

\begin{proof}
The Lemma follows by the UC composition theorem of Canetti~\cite{FOCS:Canetti01}. \iftheory\else\qed\fi
\end{proof}

\paragraph{Remark.}  Since CCA2-secure public-key encryption UC-realizes the ideal public-key encryption functionality of Canetti, and in $\PiBfly$ the adversary already knows how many layers of a given onion have already been peeled, forming onions by using CCA2-secure encryption to encrypt each layer will also result in an anonymous $\PiBfly$.

\begin{lemma}
\label{lemma:bfly-indifferent}
$\PiBT^{\mathit{ideal}}$ is indifferent.
\end{lemma}

\renewcommand{\PiBfly}{\Pi_{\bowtie}^{\mathit{ideal}}}
\begin{proof}
In $\PiBT$, the length of each routing path is fixed, and the intermediaries and nonces of honestly formed onion layers do not depend on the input $\sigma$ to the protocol. The procedure for generating intermediaries and nonces (\fbox{boxed} in 
Figure~\ref{fig:PiTree})
takes as input only the values 
\ifdraft
$\numOnion$ and $y$. 
\else
$\numOnion$, $y$, and $z$. 
\fi
Thus, by definition, $\PiBT$ is indifferent. \iftheory\else\qed\fi
\end{proof}

For the subsequent lemmas (Lemmas~\ref{lem:mix}-\ref{lem:first}), we analyze only \emph{\comm} onions, where an onion (layer) is \comm\ if (i) an honest party formed it, and (ii) it is not a checkpoint onion for verification by an adversarial party (more precisely, it does not belong to the same evolution as a checkpoint onion for verification by an adversarial party).

\begin{lemma} \label{lem:mix} 
With parameters $\numOnion$, $y$, $z$, and $t$ defined as above, $\PiBT$ mixes for the adversary who corrupts up to half of the parties.
\end{lemma}

\noindent \emph{Proof sketch.}
If $\PiBfly$ delivers messages in the final round $d$, then w.o.p., the adversary dropped (at most) a constant fraction of the \comm\ checkpoint onions before the last epoch: 
The adversary cannot drop more than a constant fraction of all \comm\ onions without also dropping a proportional number of checkpoint onions.  This is because
if the adversary were to drop more than a constant fraction of all \comm\ onions, then, 
from known probability concentration bounds~\cite{HS05}, w.o.p., the adversary would drop close to a proportional number of checkpoint onions, which, in turn, would cause all honest parties to abort the run. 
Combining this with Chernoff bounds we get: 
during each round of the penultimate epoch $e$, each honest party processed a polylogarithmic (in the security parameter) number of commutable onions. From Chernoff bounds, we also get: during epoch~$e$, each commutable onion went to an honest party a polylogarithmic number of times. Thus, either the $\PiBfly$ aborts, or it sufficiently shuffles the commutable onions during the penultimate epoch since shuffling for a polylogarithmic number of rounds with a polylogarithmic number of other onions is sufficient for mixing. Either way, $\PiBfly$ mixes. 
See Appendix~\ref{app:mix} for the proof of Lemma~\ref{lem:mix}. \hfill\qed

%

\begin{lemma} \label{lem:equalizes} 
With parameters $\numOnion$, $y$, $z$, and $t$ defined as above, $\PiBT$ equalizes for the adversary who corrupts up to half of the parties, who also receives everything about non-\comm\ onions as an auxiliary input.
\end{lemma}

Before proving Lemma~\ref{lem:equalizes}, let us prove the following: 

\begin{lemma}
\label{lem:first}
Let $\PiBT$ run with parameters $\numOnion$, $y$, $z$, and $t$ are as defined above on input $\sigma$, with $\adv$ corrupting up to half of the participants, and receiving an auxiliary input about non-\comm\ onions as an auxiliary input.  
If there is an unaborted honest party at the beginning of 
the equalizing phase, 
then with overwhelming probability for each honest party $P$, at least $\frac{1-\corruptions}{9}$ of $P$'s merging onions remained undropped by the adversary at the end of 
the mixing phase. 
(Recall that $\corruptions$ is the corruption rate.)
\end{lemma}

\ifdraft
\underline{Pf of Lemma~\ref{lem:part1}:} Suppose that the adversary drops all the onions that could have been formed by a target sender $P$ that are routed to an adversarial party during the first few rounds of the execution phase. We can show that (if somehow, the honest parties don't abort the run), w.o.p., a constant fraction of $P$'s onions remain undropped at the end of the epoch since a constant fraction of $P$'s onions initially route through only honest locations. We then argue that the adversary cannot do better than this. If the adversary drops at most the same number of onions, then at least as many of $P$'s onions will remain undropped. Whereas, dropping more onions will cause the honest parties to abort the run. (For the full proof, see in Appendix~\ref{app:pitree}.) 
\else 
\noindent \emph{Proof sketch.}
In the first round, the adversary $\adv$ knows the sender of each commutable onion. As the protocol progresses, $\adv$ loses track of this information. Thus, $\adv$'s optimal tactic is to target Alice upfront by dropping every onion that might have come from Alice that is routed to an adversarial party during the first three rounds of the first epoch (as well as the last round of the epoch). 

In the first round, some of Alice's onions route to a corrupt party; $\adv$ drops all of these. However, from Chernoff bounds, w.o.p., at least a constant fraction of Alice's onion go to an honest party first. Let $O$ be such an onion, and let $P$ be the honest party that receives $O$ in the first round. Recall that during each epoch of the mixing phase, $P$ shuffles onions back and forth with another party $P'$. $\adv$ can attempt to drop $O$ if $P'$ is corrupt. However, even if $P'$ is corrupt, $\adv$ cannot drop $O$ if it arrives at $P$ first and remains at $P$ during rounds $2$ and $3$ (and return to $P$ at round $y$) -- so, using probability concentration bounds,  $\frac{1-\corruptions}{9}$ of the time. Thus, even if $\adv$ employs the optimal tactic for dropping Alice's onions, (at least) $\frac{1-\corruptions}{9}$ of Alice's onions will make it to the equalizing phase. Since $\adv$ cannot do better than this, this proves Lemma~\ref{lem:first}. (See Appendix~\ref{sec:first} for the full proof.) \hfill\qed
\fi


\ifdraft
\underline{Pf of Lemma~\ref{lem:pitree-equalizes}} The proof is essentially the same as proof of a result analogous to Lemma~\ref{lem:pitree-equalizes} for the main construction $\PiBfly$, which we will prove in detail in Section~\ref{sec:rest}. 
\else
\begin{proof}[Proof sketch of Lemma~\ref{lem:equalizes}] 
From Lemma~\ref{lem:first}, if $\PiBfly$ continues into the equalizing phase, then a constant fraction of each honest party's merging onions are still in play at the start of the equalizing phase. However, Lemma~\ref{lem:first} does not guarantee that there will be an epoch $i > nyz$ such that the number of Alice's merging onions at epoch~$i$, $\textsf{numMO}_{\textsf{Alice},i}$, will be close to that of Allison's, $\textsf{numMO}_{\textsf{Allison},i}$. To prove that $\PiBfly$ equalizes, we need to show that there exists an epoch $i \le d$ such that (for any two parties Alice and Allison), $\textsf{numMO}_{\textsf{Alice},i} \approx \textsf{numMO}_{\textsf{Allison},i}$. If $\adv$ doesn't drop any commutable onions during the equalizing phase, then this condition is satisfied by the merging of onions. 

So what can $\adv$ do? The only information that $\adv$ has for guessing where any commutable onion came from is which onions are part of a mergeable pair and which are not; this is because the onions are shuffled during the mixing phase and each epoch of the equalizing phase. Let a \emph{singleton} be a commutable onion that is not part of a mergeable pair; note that it can be either a checkpoint onion or a merging onion. W.l.o.g., suppose that $\adv$ dropped more of Alice's onions upfront (during the mixing phase) than Allison's. Then, at the start of the equalizing phase, it is likely that more singletons are Alice's merging onions than Allison's merging onions. So, $\adv$ can attempt to prevent the numbers of merging onions from evening out by dropping singletons. We can show that the best that $\adv$ can do is to drop as many singletons as possible (without causing the protocol to be aborted) at the beginning of the equalizing phase. (Of course, $\adv$ could also drop onions that belong in a mergeable pair, but this would only help to even out the numbers of merging pairs.) Even if $\adv$ does this, there exists an epoch $i\le d$ such that $\textsf{numMO}_{\textsf{Alice},i} \approx \textsf{numMO}_{\textsf{Allison},i}$. 
(See Appendix~\ref{sec:rest} for the full proof.)
\fi

Armed with 
\ifdraft Lemmas~\ref{lem:part1} and \ref{lem:pitree-equalizes},
\else 
Lemma~\ref{lem:first} and the above analysis,
\fi 
we can prove that $\PiBT$ equalizes. If the adversary drops too many onions during
\ifdraft
the first epoch, 
\else
the mixing phase, 
\fi
then $\PiBT$ equalizes since every honest party stops participating 
\ifdraft (Lemma~\ref{lem:part1}), 
\else (Lemma~\ref{lem:first}), 
\fi 
and so no one receives their message. Otherwise, $\PiBT$ equalizes since enough of each sender's merging onions make it to the 
\ifdraft second epoch (Lemma~\ref{lem:part1}), 
\else equalizing phase (Lemma~\ref{lem:first}), 
\fi and the numbers of merging onions are eventually evened out by the merging of onions
\ifdraft (Lemma~\ref{lem:pitree-equalizes}). 
\else (above). 
\fi
\end{proof}
\ifdraft\end{proof}\fi

%% file: upper.tex
\input{PiBfly}

\subsection{Proof that $\PiBfly$ is efficient, robust, and anonymous} \label{sec:proof}
In this section, we will prove that there exists a parameter setting (for $\numOnion$, $y$, $z$, and $t$) such that $\PiBfly$ is simultaneously efficient, fault-tolerant, and anonymous.  

Our measure of efficiency is \emph{onion cost per user}, which measures how many onions are transmitted by each user in the protocol. This is an appropriate measure when the parties pass primarily onions to each other. 
It is also an attractive measure of complexity because it is algorithm-independent: 
If we measured complexity in bits, it would change depending on which underlying encryption scheme was used. Since an onion contains as many layers as there are intermediaries, its bit complexity scales linearly with the number of intermediaries. (We assume that every message~$m$ can be contained in a single onion.) To translate our lower bound from onion complexity to bits, we will consider onions to be at least as long (in bits) as the message $m$ being transmitted and the routing information.
More formally, 
\begin{definition} [Onion cost] \label{def:oc}
Let $\outof{\Pi, \adv}{i}{\secparam, \sigma}$ denote the number of onions formed by an honest party that party $P_i$ transmits directly to another party in a protocol run of $\Pi$ with adversary $\adv$,   security parameter $\secpar$ and $\sigma$.   

The onion cost of $\Pi$ 
is 
\mymathenv{
\mathsf{OC}^{\Pi, \adv}(\secparam, \Sigma) \myeq \mathbb{E}_{\sigma, i, \$}\left[ \outof{\Pi, \adv}{i}{\secparam, \sigma} \right] .
}
The expectation is taken over the input $\sigma \sample \Sigma$, the party $P_i \sample \parties$, and the randomness $\$$ of the protocol. 

For an adversary class $\mathbb{A}$, the onion cost of $\Pi$ interacting with $\mathbb{A}$ w.r.t.\ $\Sigma$ is the maximum onion cost over the adversaries in $\mathbb{A}$, i.e., $\mathsf{OC}^{\Pi, \mathbb{A}}(\secparam, \Sigma) \myeq \max_{\adv\in\mathbb{A}} \mathsf{OC}^{\Pi, \adv}(\secparam, \Sigma)$.
\end{definition}

Our formal notion of fault tolerance is \emph{robustness}, defined below:
\begin{definition} [Robustness] \label{def:robust}
A messaging protocol $\Pilong$ is robust if in every interaction in which the adversary drops at most a logarithmic (in the security parameter) number of message packets, $\Pi$ delivers all messages sent out by honest participants with overwhelming probability.
\end{definition}

Let $\mathbb{A}_\corruptions$ denote the class of active adversaries who can corrupt up to a constant $\corruptions$ fraction of the participants. 
In this section, we will prove the following upper bound on onion cost: 

\begin{theorem} \label{thm:upper} 
For any constants $\corruptions < \frac{1}{2}$ and $\gamma_1, \gamma_2 > 0$, there is a setting of $\numOnion$, $y$, $z$, and $t$ such that 
$\PiBfly^{\numOnion,y,z,t}$ is robust and anonymous from the adversary class $\mathbb{A}_\corruptions$ with onion cost at most $\gamma_1 \log N \log^{3+\gamma_2} \secpar$ (in the presence of $\mathbb{A}_\corruptions$), where $\secpar$ is the security parameter and $N = \omega(\log \secpar)$ is the number of participants. 
\end{theorem}

\newcommand{\onionCost}{\mathsf{OC}}

\begin{proof}

Recall that the number of corruptions is $\corruptions < \frac{1}{2}$.  Set $\epsilon_1$ such that $\gamma_1 = 6\epsilon_1^3$ and  $\epsilon_2$ such that $\gamma_2 = 3\epsilon_2$.  
Let $\numOnion = y = z = \epsilon_1 \log^{1+\epsilon_2} \secpar$; and let $t = \frac{W}{3}$, where 
$W = \frac{(1-\corruptions)\numOnion}{z\log N + \log \numOnion}$ is the expected number of (commutable) checkpoint nonces at a party at a diagnostic round. 

Having set the parameters, we wish to show that the protocol $\PiBfly^{\numOnion,y,z,t}$  
(a)~is robust; 
(b)~has onion cost $\onionCost\leq\gamma_1 \log N \log^{3+\gamma_2} \secpar$; and 
(c)~is anonymous, provided that the underlying onion encryption scheme is secure. 

Part (a) is true by inspection.

To see why (b) follows, recall that each participant forms $\numOnion$ merging onions, and, on average $\numOnion$ checkpoint onions; let $X$ be the maximum number of onions formed by an honest party.  Each of these onions will need to be processed in each round, so $\onionCost \leq Xd$, where $d$ is the number of rounds.  Using Chernoff bounds, $X<3\numOnion$ with overwhelming probability.  The number of rounds is $d = (nyz + 1) + y \log x + 1$; for our setting of parameters, therefore, $\onionCost \leq 6 \epsilon_1 ^3 \log N \log^{3(1+ \epsilon_2)} \secpar$.

We show part (c) via a series of lemmas that follow.  First, we invoke the UC composition theorem of Canetti~\cite{FOCS:Canetti01} in order to replace cryptographic algorithms for onion encryption with ideal encryption; this allows our further analysis to assume that onions reveal nothing to an intermediary $I$ other than the information that is intended for $I$ (Lemma~\ref{lemma:ideal}). 
Let $\PiBT^{\mathit{ideal}}$ be the resulting protocol.
Next, we argue that $\PiBT^{\mathit{ideal}}$ is an indifferent onion routing protocol (Lemma~\ref{lemma:bfly-indifferent}).  This is helpful because then we will be able to invoke Theorem~\ref{thm:impliedby}.   Third, we discard, for the purposes of analysis, all the checkpoint onions that are checked by the adversary; we show that if a protocol mixes (resp.\ equalizes) in this setting, then it mixes (resp.\ equalizes) (Lemmas~\ref{lem:comm2} and~\ref{lem:comm} in Appendices~\ref{app:mix} and \ref{app:comm}).  Finally, we show that in this setting, $\PiBfly$ mixes (Lemma~\ref{lem:mix}) and equalizes (Lemma~\ref{lem:equalizes}).  Then, putting it all together, we get our desired result.
\iftheory\else\qed\fi
\end{proof}

\ifdraft
\input{proof2}
\else
\input{proof1}
\fi

\ifdraft
\input{proof-pibfly-main-lemmas}
\fi

%% file: PiBfly.tex
\section{Our main construction, $\PiBfly$}
In this section, we present our main construction $\PiBfly$ (pronounced ``Pi-butterfly''). $\PiBfly$ uses a variant of a butterfly graph described below.

\subsection{The butterfly network and variants} \label{sec:butterfly}  Recall~\cite[Chapter 4.5.2]{MU05} that the \emph{butterfly network} $B = (V(B),E(B))$ is a directed graph on $(n+1)2^n$ vertices.  The vertices are organized into $N=2^n$ rows and $n+1$ columns, so each vertex has an address $(r,c)$ where $1\leq r \leq N$ and $0\leq c \leq n$.  Vertices in column $i$ represent potential locations of a data packet (here, an onion) at epoch $i$; each participant $P$ has a dedicated row.  An edge from $(P,i)$ to $(Q,i+1)$ means that an onion can travel from participant $P$ to participant $Q$ in epoch $i$.  
The edges of the specific butterfly network that will be useful for us are
\begin{eqnarray*}
E(B) = \{((P,i),(Q,i+1)) &\mid& \mbox{$P=Q$ or binary representations of $P$ and $Q$} \\ 
&& \mbox{differ in position $i+1$ only}\}.
\end{eqnarray*}

Let $J$ and $J'$ be two participants whose binary representation differs in bit $i+1$ only.  In $\PiBfly$, epoch~$i$ is dedicated to having an onion bounce $y$ times between $J$ and $J'$. This way, by the end of the epoch, the onions that $J$ and $J'$ held at the beginning of the epoch will be mixed together if one of them is honest.  More formally, the onions travel along the edges of a \emph{stretched} butterfly network, defined as follows: its $N(ny+1)$ vertices are organized into $N$ rows and $ny+1$ columns; and its edges are:
\mymathenv{E(\beta) = \{((P,j),(Q,j+1)) \mid \mbox{for $i=\lfloor j/y \rfloor$, $((P,i),(Q,i+1))\in E(B)$} \}.}

However, what if both $J$ and $J'$ are adversarial?  Then sending the onions through the stretched butterfly network just once will result in the adversary  knowing the $i^\mathit{th}$ bit of an onion's destination!  So to prevent this, we will send the onions through the \emph{iterated} stretched butterfly network. 
For an integer $z$, let $\beta^z$ denote the stretched butterfly network iterated $z$ times.  More precisely, $\beta^z$ is a directed graph in which the vertices are organized into $N$ rows and $nyz+1$ columns, i.e., a vertex has an address $(r,c)$ where $1\leq r \leq N$ and $0 \leq c \leq nyz$.  The edges are as follows:
\mymathenv{E(\beta^z) = \{((P,j),(Q,j+1))\mid  \mbox{for $i=j\bmod ny$, $((P,i),(Q,i+1))\in E(\beta)$} \}.}

To summarize, we begin with a butterfly network $B$, then we stretch it by $y$ to get $\beta$, then we iterate it $z$ times to get $\beta^z$; see Figure~\ref{fig:butterfly}.
By a ``\emph{walk through} $\beta^z$'' we mean a sequence $(J_0,\ldots J_{nyz})$ such that, for each $i < nyz$, $((J_i,i),(J_{i+1},i+1)) \in E(\beta^z)$. A random walk from a node $J_0$ is a sequence that begins with $J_0$ such that for $i>0$, each $J_i$ is a walk selected uniformly at random conditioned on the first $i$ elements being $(J_0,\ldots,J_{i-1})$.  A random walk starting at any address can be sampled efficiently. Moreover, we can also efficiently sample a random walk backwards from a given address $J$.  And we can sample a random walk that hits a given address $I$ at round $i$ efficiently as well, by walking backwards on the butterfly network from $(J_i,i)$ to $(J_0,0)$ and forward from $(J_i,i)$ to $(J_{nyz},nyz)$.

\input{butterfly}

\subsection{Description of the construction} \label{sec:pibfly-description}
Like $\PiTree^{\numOnion,y,t}$, $\PiBfly^{\numOnion,y,z,t}$ consists of the setup phase, the onion-forming phase, and the execution phase. It is parameterized by 
the number $\numOnion$ of merging onions per sender, 
the number $y$ of rounds per epoch, 
the number $z$ of iterations of a variant of a butterfly graph, and 
the threshold $t$ for missing checkpoint nonces. (We will generally omit the superscript for better readability.)
The execution phase is divided into the \emph{mixing sub-phase} and the \emph{equalizing sub-phase}.  The iterated stretched butterfly graph determines routing options for the mixing sub-phase. 

\iftheory
\paragraph{The setup phase.}
\fi
Let $\mathcal{OE} = (\gen, \formonion, \proconion)$ be a secure onion encryption scheme.  During setup, each honest participant $P$ generates its public key pair $(\pk(P), \sk(P))$ using $\mathcal{OE}$'s key generation algorithm $\gen$.  (This is the same as the setup phase in $\PiTree$. )

\paragraph{The onion-forming phase.}
On input $\{(m, R)\}$, each honest party $P$ generates exactly $\numOnion$ merging onions and (on average) $\numOnion$ checkpoint onions. 
To form an onion, $P$ first needs to pick a path for it.
Each onion will (potentially) travel to 
$d \myeq (nyz+1) + y \log\numOnion+1$ parties to reach its destination: the first $nyz+1$ steps involve a random walk through the iterated stretched butterfly network (the mixing sub-phase), and the next $y \log\numOnion+1$ steps will take the onion through the equalizing sub-phase and to the recipient.  

To begin with, $P$ generates the $\numOnion$ merging onions as follows: Let $T$ be the binary tree of height $\log \numOnion$.  Let $k$ be an address of a node in $T$ (i.e., $k$ is  a binary string of length at most $\log \numOnion$); let $v_k$ denote this node. I.e., $V(T) = \{v_k\mid\mbox{$k$ is a binary string, } |k| \leq \log \numOnion\}$.  To each non-leaf vertex $v_k$ in $T$, $P$ assigns a sequence of $y$ random parties and $y$ random nonces; let $I^\rightarrow_{v_k} = (I_{v_k,1},\ldots,I_{v_k,y})$ denote the sequence of vertices and $s^\rightarrow_{v_k} = (s_{v_k,1},\ldots,s_{v_k,y})$ denote the sequence of nonces corresponding to vertex $v_k$. (Up until this step, this is exactly how merging onions are formed in $\PiTree$.)  
For each leaf vertex $v_\ell$, $P$ picks a random walk through the iterated stretched butterfly $\beta^z$ and $nyz+1$ random nonces; let  $J^\rightarrow_{v_\ell} = (J_{v_\ell,0},\ldots, J_{v_\ell,nyz})$, denote the random walk, and let $t^\rightarrow_{v_\ell} = (t_{v_\ell,0},\ldots,t_{v_\ell,nyz})$ be the sequence of nonces. 

Let $v_\ell$ be a leaf of $T$.  Let $v_{\ell,i} = v_{k_i}$ where $k_i$ is the $i$-bit prefix of $\ell$.  I.e. $v_{\ell,\ell} = v_\ell$ and $v_{\ell,0} = v_\varepsilon$, and $(v_{\ell,h},v_{\ell,h-1},\ldots,v_{\ell,0})$ is the path from $v_\ell$ to the root of the tree, where $h = \log \numOnion$.

$P$ will create an onion $O_\ell$ for each leaf $v_\ell$.  Its routing path is $I^\rightarrow_\ell = (J^\rightarrow_{v_\ell}, I^\rightarrow_{\ell,2}, \dots, I^\rightarrow_{\ell,h}, R)$ 
where $J^\rightarrow_{v_\ell}$ is as defined above, $I^\rightarrow_{\ell,i} = I^{\rightarrow}_{k_i}$ where $k_i$ is the $i$-bit prefix of $\ell$, and $R$ is the recipient, and such that $|I^\rightarrow_\ell| = d$.  Similarly, let $s^\rightarrow_\ell=(t^\rightarrow_{v_\ell}, s^\rightarrow_{\ell,2}, \dots, s^\rightarrow_{\ell,h})$ 
denote the sequence of nonces corresponding to this path.

To form the onion $O_\ell$ corresponding to $v_\ell$, $P$ runs the algorithm $\formonion$ on the message $m$, the routing path $I^\rightarrow_\ell$, the public keys associated with the routing path, and the nonce sequence $s^\rightarrow_\ell$.

After forming the merging onions, $P$ generates the checkpoint onions.  Just as in $\PiTree$, the execution phase consists of epochs, and the last round of every epoch is a diagnostic round.   Here, each epoch lasts $y$ rounds, thus round $r>0$ is a diagnostic round if $r$ is a multiple of $y$.  For each diagnostic round $r$ and for each intermediary $I$, $P$ uses the pseudorandom function $F_{\sk(P,I)}(r, 0)$ to determine whether to form a checkpoint onion to send to $I$ at round~$r$, and if so, calculates the nonce $s = F_{\sk(P,I)}(r, 1)$.  

When $F_{\sk(P,I)}(r, 0)=1$, $P$ generates a checkpoint onion to be verified by party $I$ in round $r$.  
Recall that $d \myeq (nyz +1) + y\log \numOnion+1$; so round $d$ is the last round of the execution phase.  
Since the checkpoint onion should not be distinguishable from a merging one during the mixing sub-phase, it needs to travel over the edges of the iterated stretched butterfly network for the first $nyz+1$ rounds, and follow a random path through the network during the equalizing sub-phase, all the way until the last round $d$.

As a result, for $r \geq nyz+1$, $P$ generates the routing path by first picking a random walk $J^{0\rightarrow nyz}=(J_{0},\ldots,J_{nyz})$ through the iterated stretched butterfly network starting at a random node $J_0$, and then choosing each participant on the next part of the path $J^{nyz+1\rightarrow r-1}= (J_{nyz+1},\ldots,J_{r-1})$ uniformly at random from $\parties$.  Next, $J_r = I$, and each router on the remaining stretch of the path $J^{r+1\rightarrow d}$ is, again, chosen uniformly at random from $\parties$.  So the resulting routing path is $J^\rightarrow_{I,r} = (J^{0\rightarrow nyz},J^{nyz+1\rightarrow r-1},J_r,J^{r+1\rightarrow d})$.  $P$ chooses the corresponding nonces $\{s_{I,r,j}\}_{j\in\{0,\dots,d-1\}\setminus\{r\}}$ uniformly at random, sets $s_{I,r,r}=s$, and gives the resulting routing path, sequence $(s_{I,r,0},\dots,s_{I,r,d-1})$ of nonces and the empty message to $\formonion$ to obtain checkpoint onion $O_{I,r}$.

If $r \leq nyz$, then round $r$ occurs during the mixing sub-phase, as the onion is making its way through the butterfly network.  So its path has to be formed in such a way that it arrives at $I$ at round $r$; but it needs to be a randomly chosen path conditioned on this event (so that a checkpoint onion's path is distributed the same way as one of a merging onion).  
Let $J^{0\rightarrow nyz}$ be a random walk through $\beta^z$ that is at address $I$ at round $r$ (see Section~\ref{sec:butterfly} for how to sample this walk efficiently).  Let each intermediary in the sequence $J^{nyz+1\rightarrow d}$ be chosen uniformly at random from $\parties$.  Again, for $j \neq r$, $0 \leq j \leq d-1$, the nonce $s_{I,r,j}$ is chosen at random, while $s_{I,r,r} = s$.  Let $J^\rightarrow_{I,r} = (J^{0\rightarrow nyz},J^{nyz+1\rightarrow d})$.  Run $\formonion$ on input the  routing path $J^\rightarrow_{I,r}$, sequence of nonces $s^\rightarrow_{I,r}$ and the empty message to obtain checkpoint onion $O_{I,r}$. 
%
See Figure~\ref{fig:PiBfly}. 

\input{pibflyfig}

\underline{Remark:} As in $\PiTree$, the onion layers are tagged with their respective round number to prevent replay attacks. If by peeling an onion received at round $r$, an honest relaying party observes a round number $r' \neq r$, the party drops the onion. 
(We can, therefore, assume that replay attacks do not happen. We can safely do so since the security of the onion encryption scheme prevents the adversary from modifying the onions formed by honest participants in any meaningful way. See, for example Ando and Lysyanskaya's work on onion encryption~\cite{AL20}, for a sufficiently strong construction.) 

\paragraph{The execution phase.}
At the beginning of the execution phase, each party $P$ is \emph{live}.  $P$'s status will change from live to \emph{aborted} if it ever receives a special abort message from another party.  An aborted party sends the special abort message to a random sample of $\numOnion$ parties.  (A slight technicality is that, since all messages must be onions, the special abort message is a specially formed onion.)

For each $r \in \{0, \dots, d-1\}$, each live honest party $P$ first 
peels all the onions it received at the $r^\textit{th}$ round.  It merges onions that are mergeable: if it received two onions that have the same nonce, then it drops one of them, selected at random, and sends the other one to its next destination.

If $r$ is a diagnostic round (i.e., a multiple of $y$), then $P$ runs the diagnostic test: $P$ compares the number of checkpoint onions it expects to receive with the number it received.   For every participant $Q\in \parties$, if $F_{\sk(Q,P)}(r, 0)=1$, then $P$ expects to receive a checkpoint onion with nonce $s = F_{\sk(Q,P)}(r, 1)$ in this round.  
In the mixing sub-phase, if fewer than $t$ checkpoint onions are missing so far in the protocol run (not just in this round, but cumulatively), then $P$ continues the run by processing all the other onions. Otherwise, $P$'s status changes: it is no longer live but becomes an \emph{aborted} party. 
In the equalizing sub-phase, change status to aborted if there are $t$ or more missing checkpoint onions in this round, else continue.

At the last round (round $d$) of the execution phase, $P$ peels the onions it received and outputs the set of (non-empty) messages it received. 


%% file: butterfly.tex
\begin{figure}[ht!]
\center
\begin{tikzpicture}[scale=1.2]
\iftheory
\draw[->] (-0.5,0) -- (0,0);
\draw[->] (-0.5, 0.5) -- (0,0.5);
\draw[->] (-0.5,1) -- (0,1);
\draw[->] (-0.5,1.5) -- (0,1.5);
\draw[->] (-0.5,2) -- (0,2);
\draw[->] (-0.5,2.5) -- (0,2.5);
\draw[->] (-0.5,3) -- (0,3);
\draw[->] (-0.5,3.5) -- (0,3.5);
\draw[->] (-0.5,0) -- (0,0.5);
\draw[->] (-0.5,0.5) -- (0,0);
\draw[->] (-0.5,1) -- (0,1.5);
\draw[->] (-0.5,1.5) -- (0,1);
\draw[->] (-0.5,2) -- (0,2.5);
\draw[->] (-0.5,2.5) -- (0,2);
\draw[->] (-0.5,3) -- (0,3.5);
\draw[->] (-0.5,3.5) -- (0,3);

\draw[->] (0,0) -- (0.5,0);
\draw[->] (0,0.5) -- (0.5,0.5);
\draw[->] (0,1) -- (0.5,1);
\draw[->] (0,1.5) -- (0.5,1.5);
\draw[->] (0,2) -- (0.5,2);
\draw[->] (0,2.5) -- (0.5,2.5);
\draw[->] (0,3) -- (0.5,3);
\draw[->] (0,3.5) -- (0.5,3.5);
\draw[->] (0,0) -- (0.5,1);
\draw[->] (0,0.5) -- (0.5,1.5);
\draw[->] (0,1) -- (0.5,0);
\draw[->] (0,1.5) -- (0.5,0.5);
\draw[->] (0,2) -- (0.5,3);
\draw[->] (0,2.5) -- (0.5,3.5);
\draw[->] (0,3) -- (0.5,2);
\draw[->] (0,3.5) -- (0.5,2.5);

\draw[->] (0.5,0) -- (1,0);
\draw[->] (0.5,0.5) -- (1,0.5);
\draw[->] (0.5,1) -- (1,1);
\draw[->] (0.5,1.5) -- (1,1.5);
\draw[->] (0.5,2) -- (1,2);
\draw[->] (0.5,2.5) -- (1,2.5);
\draw[->] (0.5,3) -- (1,3);
\draw[->] (0.5,3.5) -- (1,3.5);
\draw[->] (0.5,0) -- (1,2);
\draw[->] (0.5,0.5) -- (1,2.5);
\draw[->] (0.5,1) -- (1,3);
\draw[->] (0.5,1.5) -- (1,3.5);
\draw[->] (0.5,2) -- (1,0);
\draw[->] (0.5,2.5) -- (1,0.5);
\draw[->] (0.5,3) -- (1,1);
\draw[->] (0.5,3.5) -- (1,1.5);

\filldraw[black] (-0.5,-0.5) circle (0pt) node[anchor=west] {$B$};

\draw[->] (2,0) -- (2.5,0);
\draw[->] (2,0.5) -- (2.5,0.5);
\draw[->] (2,1) -- (2.5,1);
\draw[->] (2,1.5) -- (2.5,1.5);
\draw[->] (2,2) -- (2.5,2);
\draw[->] (2,2.5) -- (2.5,2.5);
\draw[->] (2,3) -- (2.5,3);
\draw[->] (2,3.5) -- (2.5,3.5);
\draw[->] (2,0) -- (2.5,0.5);
\draw[->] (2,0.5) -- (2.5,0);
\draw[->] (2,1) -- (2.5,1.5);
\draw[->] (2,1.5) -- (2.5,1);
\draw[->] (2,2) -- (2.5,2.5);
\draw[->] (2,2.5) -- (2.5,2);
\draw[->] (2,3) -- (2.5,3.5);
\draw[->] (2,3.5) -- (2.5,3);
\draw[->] (2.5,0) -- (3,0);
\draw[->] (2.5,0.5) -- (3,0.5);
\draw[->] (2.5,1) -- (3,1);
\draw[->] (2.5,1.5) -- (3,1.5);
\draw[->] (2.5,2) -- (3,2);
\draw[->] (2.5,2.5) -- (3,2.5);
\draw[->] (2.5,3) -- (3,3);
\draw[->] (2.5,3.5) -- (3,3.5);
\draw[->] (2.5,0) -- (3,0.5);
\draw[->] (2.5,0.5) -- (3,0);
\draw[->] (2.5,1) -- (3,1.5);
\draw[->] (2.5,1.5) -- (3,1);
\draw[->] (2.5,2) -- (3,2.5);
\draw[->] (2.5,2.5) -- (3,2);
\draw[->] (2.5,3) -- (3,3.5);
\draw[->] (2.5,3.5) -- (3,3);

\draw[->] (3,0) -- (3.5,0);
\draw[->] (3,0.5) -- (3.5,0.5);
\draw[->] (3,1) -- (3.5,1);
\draw[->] (3,1.5) -- (3.5,1.5);
\draw[->] (3,2) -- (3.5,2);
\draw[->] (3,2.5) -- (3.5,2.5);
\draw[->] (3,3) -- (3.5,3);
\draw[->] (3,3.5) -- (3.5,3.5);
\draw[->] (3,0) -- (3.5,1);
\draw[->] (3,0.5) -- (3.5,1.5);
\draw[->] (3,1) -- (3.5,0);
\draw[->] (3,1.5) -- (3.5,0.5);
\draw[->] (3,2) -- (3.5,3);
\draw[->] (3,2.5) -- (3.5,3.5);
\draw[->] (3,3) -- (3.5,2);
\draw[->] (3,3.5) -- (3.5,2.5);
\draw[->] (3.5,0) -- (4,0);
\draw[->] (3.5,0.5) -- (4,0.5);
\draw[->] (3.5,1) -- (4,1);
\draw[->] (3.5,1.5) -- (4,1.5);
\draw[->] (3.5,2) -- (4,2);
\draw[->] (3.5,2.5) -- (4,2.5);
\draw[->] (3.5,3) -- (4,3);
\draw[->] (3.5,3.5) -- (4,3.5);
\draw[->] (3.5,0) -- (4,1);
\draw[->] (3.5,0.5) -- (4,1.5);
\draw[->] (3.5,1) -- (4,0);
\draw[->] (3.5,1.5) -- (4,0.5);
\draw[->] (3.5,2) -- (4,3);
\draw[->] (3.5,2.5) -- (4,3.5);
\draw[->] (3.5,3) -- (4,2);
\draw[->] (3.5,3.5) -- (4,2.5);

\draw[->] (4,0) -- (4.5,0);
\draw[->] (4,0.5) -- (4.5,0.5);
\draw[->] (4,1) -- (4.5,1);
\draw[->] (4,1.5) -- (4.5,1.5);
\draw[->] (4,2) -- (4.5,2);
\draw[->] (4,2.5) -- (4.5,2.5);
\draw[->] (4,3) -- (4.5,3);
\draw[->] (4,3.5) -- (4.5,3.5);
\draw[->] (4,0) -- (4.5,2);
\draw[->] (4,0.5) -- (4.5,2.5);
\draw[->] (4,1) -- (4.5,3);
\draw[->] (4,1.5) -- (4.5,3.5);
\draw[->] (4,2) -- (4.5,0);
\draw[->] (4,2.5) -- (4.5,0.5);
\draw[->] (4,3) -- (4.5,1);
\draw[->] (4,3.5) -- (4.5,1.5);
\draw[->] (4.5,0) -- (5,0);
\draw[->] (4.5,0.5) -- (5,0.5);
\draw[->] (4.5,1) -- (5,1);
\draw[->] (4.5,1.5) -- (5,1.5);
\draw[->] (4.5,2) -- (5,2);
\draw[->] (4.5,2.5) -- (5,2.5);
\draw[->] (4.5,3) -- (5,3);
\draw[->] (4.5,3.5) -- (5,3.5);
\draw[->] (4.5,0) -- (5,2);
\draw[->] (4.5,0.5) -- (5,2.5);
\draw[->] (4.5,1) -- (5,3);
\draw[->] (4.5,1.5) -- (5,3.5);
\draw[->] (4.5,2) -- (5,0);
\draw[->] (4.5,2.5) -- (5,0.5);
\draw[->] (4.5,3) -- (5,1);
\draw[->] (4.5,3.5) -- (5,1.5);

\filldraw[black] (2,-0.5) circle (0pt) node[anchor=west] {$\beta$};

\else
\draw[->] (0.5,0) -- (1,0);
\draw[->] (0.5, 0.5) -- (1,0.5);
\draw[->] (0.5,1) -- (1,1);
\draw[->] (0.5,1.5) -- (1,1.5);
\draw[->] (0.5,2) -- (1,2);
\draw[->] (0.5,2.5) -- (1,2.5);
\draw[->] (0.5,3) -- (1,3);
\draw[->] (0.5,3.5) -- (1,3.5);
\draw[->] (0.5,0) -- (1,0.5);
\draw[->] (0.5,0.5) -- (1,0);
\draw[->] (0.5,1) -- (1,1.5);
\draw[->] (0.5,1.5) -- (1,1);
\draw[->] (0.5,2) -- (1,2.5);
\draw[->] (0.5,2.5) -- (1,2);
\draw[->] (0.5,3) -- (1,3.5);
\draw[->] (0.5,3.5) -- (1,3);

\draw[->] (1,0) -- (1.5,0);
\draw[->] (1,0.5) -- (1.5,0.5);
\draw[->] (1,1) -- (1.5,1);
\draw[->] (1,1.5) -- (1.5,1.5);
\draw[->] (1,2) -- (1.5,2);
\draw[->] (1,2.5) -- (1.5,2.5);
\draw[->] (1,3) -- (1.5,3);
\draw[->] (1,3.5) -- (1.5,3.5);
\draw[->] (1,0) -- (1.5,1);
\draw[->] (1,0.5) -- (1.5,1.5);
\draw[->] (1,1) -- (1.5,0);
\draw[->] (1,1.5) -- (1.5,0.5);
\draw[->] (1,2) -- (1.5,3);
\draw[->] (1,2.5) -- (1.5,3.5);
\draw[->] (1,3) -- (1.5,2);
\draw[->] (1,3.5) -- (1.5,2.5);

\draw[->] (1.5,0) -- (2,0);
\draw[->] (1.5,0.5) -- (2,0.5);
\draw[->] (1.5,1) -- (2,1);
\draw[->] (1.5,1.5) -- (2,1.5);
\draw[->] (1.5,2) -- (2,2);
\draw[->] (1.5,2.5) -- (2,2.5);
\draw[->] (1.5,3) -- (2,3);
\draw[->] (1.5,3.5) -- (2,3.5);
\draw[->] (1.5,0) -- (2,2);
\draw[->] (1.5,0.5) -- (2,2.5);
\draw[->] (1.5,1) -- (2,3);
\draw[->] (1.5,1.5) -- (2,3.5);
\draw[->] (1.5,2) -- (2,0);
\draw[->] (1.5,2.5) -- (2,0.5);
\draw[->] (1.5,3) -- (2,1);
\draw[->] (1.5,3.5) -- (2,1.5);

\filldraw[black] (0.5,-0.5) circle (0pt) node[anchor=west] {$B$};

\draw[->] (2.5,0) -- (3,0);
\draw[->] (2.5,0.5) -- (3,0.5);
\draw[->] (2.5,1) -- (3,1);
\draw[->] (2.5,1.5) -- (3,1.5);
\draw[->] (2.5,2) -- (3,2);
\draw[->] (2.5,2.5) -- (3,2.5);
\draw[->] (2.5,3) -- (3,3);
\draw[->] (2.5,3.5) -- (3,3.5);
\draw[->] (2.5,0) -- (3,0.5);
\draw[->] (2.5,0.5) -- (3,0);
\draw[->] (2.5,1) -- (3,1.5);
\draw[->] (2.5,1.5) -- (3,1);
\draw[->] (2.5,2) -- (3,2.5);
\draw[->] (2.5,2.5) -- (3,2);
\draw[->] (2.5,3) -- (3,3.5);
\draw[->] (2.5,3.5) -- (3,3);
\draw[->] (3,0) -- (3.5,0);
\draw[->] (3,0.5) -- (3.5,0.5);
\draw[->] (3,1) -- (3.5,1);
\draw[->] (3,1.5) -- (3.5,1.5);
\draw[->] (3,2) -- (3.5,2);
\draw[->] (3,2.5) -- (3.5,2.5);
\draw[->] (3,3) -- (3.5,3);
\draw[->] (3,3.5) -- (3.5,3.5);
\draw[->] (3,0) -- (3.5,0.5);
\draw[->] (3,0.5) -- (3.5,0);
\draw[->] (3,1) -- (3.5,1.5);
\draw[->] (3,1.5) -- (3.5,1);
\draw[->] (3,2) -- (3.5,2.5);
\draw[->] (3,2.5) -- (3.5,2);
\draw[->] (3,3) -- (3.5,3.5);
\draw[->] (3,3.5) -- (3.5,3);

\draw[->] (3.5,0) -- (4,0);
\draw[->] (3.5,0.5) -- (4,0.5);
\draw[->] (3.5,1) -- (4,1);
\draw[->] (3.5,1.5) -- (4,1.5);
\draw[->] (3.5,2) -- (4,2);
\draw[->] (3.5,2.5) -- (4,2.5);
\draw[->] (3.5,3) -- (4,3);
\draw[->] (3.5,3.5) -- (4,3.5);
\draw[->] (3.5,0) -- (4,1);
\draw[->] (3.5,0.5) -- (4,1.5);
\draw[->] (3.5,1) -- (4,0);
\draw[->] (3.5,1.5) -- (4,0.5);
\draw[->] (3.5,2) -- (4,3);
\draw[->] (3.5,2.5) -- (4,3.5);
\draw[->] (3.5,3) -- (4,2);
\draw[->] (3.5,3.5) -- (4,2.5);
\draw[->] (4,0) -- (4.5,0);
\draw[->] (4,0.5) -- (4.5,0.5);
\draw[->] (4,1) -- (4.5,1);
\draw[->] (4,1.5) -- (4.5,1.5);
\draw[->] (4,2) -- (4.5,2);
\draw[->] (4,2.5) -- (4.5,2.5);
\draw[->] (4,3) -- (4.5,3);
\draw[->] (4,3.5) -- (4.5,3.5);
\draw[->] (4,0) -- (4.5,1);
\draw[->] (4,0.5) -- (4.5,1.5);
\draw[->] (4,1) -- (4.5,0);
\draw[->] (4,1.5) -- (4.5,0.5);
\draw[->] (4,2) -- (4.5,3);
\draw[->] (4,2.5) -- (4.5,3.5);
\draw[->] (4,3) -- (4.5,2);
\draw[->] (4,3.5) -- (4.5,2.5);

\draw[->] (4.5,0) -- (5,0);
\draw[->] (4.5,0.5) -- (5,0.5);
\draw[->] (4.5,1) -- (5,1);
\draw[->] (4.5,1.5) -- (5,1.5);
\draw[->] (4.5,2) -- (5,2);
\draw[->] (4.5,2.5) -- (5,2.5);
\draw[->] (4.5,3) -- (5,3);
\draw[->] (4.5,3.5) -- (5,3.5);
\draw[->] (4.5,0) -- (5,2);
\draw[->] (4.5,0.5) -- (5,2.5);
\draw[->] (4.5,1) -- (5,3);
\draw[->] (4.5,1.5) -- (5,3.5);
\draw[->] (4.5,2) -- (5,0);
\draw[->] (4.5,2.5) -- (5,0.5);
\draw[->] (4.5,3) -- (5,1);
\draw[->] (4.5,3.5) -- (5,1.5);
\draw[->] (5,0) -- (5.5,0);
\draw[->] (5,0.5) -- (5.5,0.5);
\draw[->] (5,1) -- (5.5,1);
\draw[->] (5,1.5) -- (5.5,1.5);
\draw[->] (5,2) -- (5.5,2);
\draw[->] (5,2.5) -- (5.5,2.5);
\draw[->] (5,3) -- (5.5,3);
\draw[->] (5,3.5) -- (5.5,3.5);
\draw[->] (5,0) -- (5.5,2);
\draw[->] (5,0.5) -- (5.5,2.5);
\draw[->] (5,1) -- (5.5,3);
\draw[->] (5,1.5) -- (5.5,3.5);
\draw[->] (5,2) -- (5.5,0);
\draw[->] (5,2.5) -- (5.5,0.5);
\draw[->] (5,3) -- (5.5,1);
\draw[->] (5,3.5) -- (5.5,1.5);

\filldraw[black] (2.5,-0.5) circle (0pt) node[anchor=west] {$\beta$};
\fi

\draw[->] (6,0) -- (6.5,0);
\draw[->] (6,0.5) -- (6.5,0.5);
\draw[->] (6,1) -- (6.5,1);
\draw[->] (6,1.5) -- (6.5,1.5);
\draw[->] (6,2) -- (6.5,2);
\draw[->] (6,2.5) -- (6.5,2.5);
\draw[->] (6,3) -- (6.5,3);
\draw[->] (6,3.5) -- (6.5,3.5);
\draw[->] (6,0) -- (6.5,0.5);
\draw[->] (6,0.5) -- (6.5,0);
\draw[->] (6,1) -- (6.5,1.5);
\draw[->] (6,1.5) -- (6.5,1);
\draw[->] (6,2) -- (6.5,2.5);
\draw[->] (6,2.5) -- (6.5,2);
\draw[->] (6,3) -- (6.5,3.5);
\draw[->] (6,3.5) -- (6.5,3);
\draw[->] (6.5,0) -- (7,0);
\draw[->] (6.5,0.5) -- (7,0.5);
\draw[->] (6.5,1) -- (7,1);
\draw[->] (6.5,1.5) -- (7,1.5);
\draw[->] (6.5,2) -- (7,2);
\draw[->] (6.5,2.5) -- (7,2.5);
\draw[->] (6.5,3) -- (7,3);
\draw[->] (6.5,3.5) -- (7,3.5);
\draw[->] (6.5,0) -- (7,0.5);
\draw[->] (6.5,0.5) -- (7,0);
\draw[->] (6.5,1) -- (7,1.5);
\draw[->] (6.5,1.5) -- (7,1);
\draw[->] (6.5,2) -- (7,2.5);
\draw[->] (6.5,2.5) -- (7,2);
\draw[->] (6.5,3) -- (7,3.5);
\draw[->] (6.5,3.5) -- (7,3);
\draw[->] (7,0) -- (7.5,0);
\draw[->] (7,0.5) -- (7.5,0.5);
\draw[->] (7,1) -- (7.5,1);
\draw[->] (7,1.5) -- (7.5,1.5);
\draw[->] (7,2) -- (7.5,2);
\draw[->] (7,2.5) -- (7.5,2.5);
\draw[->] (7,3) -- (7.5,3);
\draw[->] (7,3.5) -- (7.5,3.5);
\draw[->] (7,0) -- (7.5,1);
\draw[->] (7,0.5) -- (7.5,1.5);
\draw[->] (7,1) -- (7.5,0);
\draw[->] (7,1.5) -- (7.5,0.5);
\draw[->] (7,2) -- (7.5,3);
\draw[->] (7,2.5) -- (7.5,3.5);
\draw[->] (7,3) -- (7.5,2);
\draw[->] (7,3.5) -- (7.5,2.5);
\draw[->] (7.5,0) -- (8,0);
\draw[->] (7.5,0.5) -- (8,0.5);
\draw[->] (7.5,1) -- (8,1);
\draw[->] (7.5,1.5) -- (8,1.5);
\draw[->] (7.5,2) -- (8,2);
\draw[->] (7.5,2.5) -- (8,2.5);
\draw[->] (7.5,3) -- (8,3);
\draw[->] (7.5,3.5) -- (8,3.5);
\draw[->] (7.5,0) -- (8,1);
\draw[->] (7.5,0.5) -- (8,1.5);
\draw[->] (7.5,1) -- (8,0);
\draw[->] (7.5,1.5) -- (8,0.5);
\draw[->] (7.5,2) -- (8,3);
\draw[->] (7.5,2.5) -- (8,3.5);
\draw[->] (7.5,3) -- (8,2);
\draw[->] (7.5,3.5) -- (8,2.5);
\draw[->] (8,0) -- (8.5,0);
\draw[->] (8,0.5) -- (8.5,0.5);
\draw[->] (8,1) -- (8.5,1);
\draw[->] (8,1.5) -- (8.5,1.5);
\draw[->] (8,2) -- (8.5,2);
\draw[->] (8,2.5) -- (8.5,2.5);
\draw[->] (8,3) -- (8.5,3);
\draw[->] (8,3.5) -- (8.5,3.5);
\draw[->] (8,0) -- (8.5,2);
\draw[->] (8,0.5) -- (8.5,2.5);
\draw[->] (8,1) -- (8.5,3);
\draw[->] (8,1.5) -- (8.5,3.5);
\draw[->] (8,2) -- (8.5,0);
\draw[->] (8,2.5) -- (8.5,0.5);
\draw[->] (8,3) -- (8.5,1);
\draw[->] (8,3.5) -- (8.5,1.5);
\draw[->] (8.5,0) -- (9,0);
\draw[->] (8.5,0.5) -- (9,0.5);
\draw[->] (8.5,1) -- (9,1);
\draw[->] (8.5,1.5) -- (9,1.5);
\draw[->] (8.5,2) -- (9,2);
\draw[->] (8.5,2.5) -- (9,2.5);
\draw[->] (8.5,3) -- (9,3);
\draw[->] (8.5,3.5) -- (9,3.5);
\draw[->] (8.5,0) -- (9,2);
\draw[->] (8.5,0.5) -- (9,2.5);
\draw[->] (8.5,1) -- (9,3);
\draw[->] (8.5,1.5) -- (9,3.5);
\draw[->] (8.5,2) -- (9,0);
\draw[->] (8.5,2.5) -- (9,0.5);
\draw[->] (8.5,3) -- (9,1);
\draw[->] (8.5,3.5) -- (9,1.5);

\draw[->] (9,0) -- (9.5,0);
\draw[->] (9,0.5) -- (9.5,0.5);
\draw[->] (9,1) -- (9.5,1);
\draw[->] (9,1.5) -- (9.5,1.5);
\draw[->] (9,2) -- (9.5,2);
\draw[->] (9,2.5) -- (9.5,2.5);
\draw[->] (9,3) -- (9.5,3);
\draw[->] (9,3.5) -- (9.5,3.5);
\draw[->] (9,0) -- (9.5,0.5);
\draw[->] (9,0.5) -- (9.5,0);
\draw[->] (9,1) -- (9.5,1.5);
\draw[->] (9,1.5) -- (9.5,1);
\draw[->] (9,2) -- (9.5,2.5);
\draw[->] (9,2.5) -- (9.5,2);
\draw[->] (9,3) -- (9.5,3.5);
\draw[->] (9,3.5) -- (9.5,3);
\draw[->] (9.5,0) -- (10,0);
\draw[->] (9.5,0.5) -- (10,0.5);
\draw[->] (9.5,1) -- (10,1);
\draw[->] (9.5,1.5) -- (10,1.5);
\draw[->] (9.5,2) -- (10,2);
\draw[->] (9.5,2.5) -- (10,2.5);
\draw[->] (9.5,3) -- (10,3);
\draw[->] (9.5,3.5) -- (10,3.5);
\draw[->] (9.5,0) -- (10,0.5);
\draw[->] (9.5,0.5) -- (10,0);
\draw[->] (9.5,1) -- (10,1.5);
\draw[->] (9.5,1.5) -- (10,1);
\draw[->] (9.5,2) -- (10,2.5);
\draw[->] (9.5,2.5) -- (10,2);
\draw[->] (9.5,3) -- (10,3.5);
\draw[->] (9.5,3.5) -- (10,3);
\draw[->] (10,0) -- (10.5,0);
\draw[->] (10,0.5) -- (10.5,0.5);
\draw[->] (10,1) -- (10.5,1);
\draw[->] (10,1.5) -- (10.5,1.5);
\draw[->] (10,2) -- (10.5,2);
\draw[->] (10,2.5) -- (10.5,2.5);
\draw[->] (10,3) -- (10.5,3);
\draw[->] (10,3.5) -- (10.5,3.5);
\draw[->] (10,0) -- (10.5,1);
\draw[->] (10,0.5) -- (10.5,1.5);
\draw[->] (10,1) -- (10.5,0);
\draw[->] (10,1.5) -- (10.5,0.5);
\draw[->] (10,2) -- (10.5,3);
\draw[->] (10,2.5) -- (10.5,3.5);
\draw[->] (10,3) -- (10.5,2);
\draw[->] (10,3.5) -- (10.5,2.5);
\draw[->] (10.5,0) -- (11,0);
\draw[->] (10.5,0.5) -- (11,0.5);
\draw[->] (10.5,1) -- (11,1);
\draw[->] (10.5,1.5) -- (11,1.5);
\draw[->] (10.5,2) -- (11,2);
\draw[->] (10.5,2.5) -- (11,2.5);
\draw[->] (10.5,3) -- (11,3);
\draw[->] (10.5,3.5) -- (11,3.5);
\draw[->] (10.5,0) -- (11,1);
\draw[->] (10.5,0.5) -- (11,1.5);
\draw[->] (10.5,1) -- (11,0);
\draw[->] (10.5,1.5) -- (11,0.5);
\draw[->] (10.5,2) -- (11,3);
\draw[->] (10.5,2.5) -- (11,3.5);
\draw[->] (10.5,3) -- (11,2);
\draw[->] (10.5,3.5) -- (11,2.5);
\draw[->] (11,0) -- (11.5,0);
\draw[->] (11,0.5) -- (11.5,0.5);
\draw[->] (11,1) -- (11.5,1);
\draw[->] (11,1.5) -- (11.5,1.5);
\draw[->] (11,2) -- (11.5,2);
\draw[->] (11,2.5) -- (11.5,2.5);
\draw[->] (11,3) -- (11.5,3);
\draw[->] (11,3.5) -- (11.5,3.5);
\draw[->] (11,0) -- (11.5,2);
\draw[->] (11,0.5) -- (11.5,2.5);
\draw[->] (11,1) -- (11.5,3);
\draw[->] (11,1.5) -- (11.5,3.5);
\draw[->] (11,2) -- (11.5,0);
\draw[->] (11,2.5) -- (11.5,0.5);
\draw[->] (11,3) -- (11.5,1);
\draw[->] (11,3.5) -- (11.5,1.5);
\draw[->] (11.5,0) -- (12,0);
\draw[->] (11.5,0.5) -- (12,0.5);
\draw[->] (11.5,1) -- (12,1);
\draw[->] (11.5,1.5) -- (12,1.5);
\draw[->] (11.5,2) -- (12,2);
\draw[->] (11.5,2.5) -- (12,2.5);
\draw[->] (11.5,3) -- (12,3);
\draw[->] (11.5,3.5) -- (12,3.5);
\draw[->] (11.5,0) -- (12,2);
\draw[->] (11.5,0.5) -- (12,2.5);
\draw[->] (11.5,1) -- (12,3);
\draw[->] (11.5,1.5) -- (12,3.5);
\draw[->] (11.5,2) -- (12,0);
\draw[->] (11.5,2.5) -- (12,0.5);
\draw[->] (11.5,3) -- (12,1);
\draw[->] (11.5,3.5) -- (12,1.5);

\filldraw[black] (6,-0.5) circle (0pt) node[anchor=west] {$\beta^z$};
\end{tikzpicture}

\caption{\footnotesize Diagrams of the butterfly network $B$, the stretched butterfly network $\beta$, and the iterated stretched butterfly network $\beta^z$ for $n = \log(8)=3$, and $y = z = 2$.}
\label{fig:butterfly}
\end{figure}

%% file: pibflyfig.tex
\begin{figure}[ht!]
\begin{minipage}[t]{\textwidth}
\begingroup
\fontsize{9pt}{11pt}\selectfont
\begin{flushleft}
{Let the public parameters $\pp$ include the participants' public keys, the number $\numOnion$ of merging onions, the number $y$ of rounds in an epoch, and $z$ the number of iterations of the iterated stretched butterfly network $\beta^z$.}

\bigskip
\underline{$\mathsf{FormMergingOnions}(\secparam, \Sigma, \pp, m, R)$}
\vspace{2mm}
\begin{enumerate} [label=\arabic*:]
\item[\tikzmark{bl}1:]  {Generate a binary tree graph $T$ with $\numOnion$ leaf vertices. Let $h \myeq \log \numOnion+1$.}
\item[2:] {For each non-leaf vertex, assign $y$ random intermediaries and $y$ random nonces.} 
\item[3:] {For each leaf vertex $v_\ell$, assign a random walk $J^\rightarrow_{v_\ell}$ through $\beta^z$ and a sequence $t^\rightarrow_{v_\ell}$ of random nonces s.t.\ $|t^\rightarrow_{v_\ell}|=|J^\rightarrow_{v_\ell}|$.}
\item[4:] {For each leaf vertex $v_\ell$, let $(J^\rightarrow_{v_\ell}, I^\rightarrow_{\ell, 2}, \dots, I^\rightarrow_{\ell, hy})$ denote the sequence of intermediaries corresponding to $v_\ell$, and let $(t^\rightarrow_{v_\ell}, s^\rightarrow_{\ell, 1}, \dots, s^\rightarrow_{\ell, hy})$ denote the sequence of nonces corresponding to $v_\ell$.\hfill} \tikzmark{br} 
\item[5:] {For each leaf vertex $v_\ell$, form an onion by running $\formonion$ on the message $m$, the routing path $(J^\rightarrow_{v_\ell}, I^\rightarrow_{\ell, 2}, \dots, I^\rightarrow_{\ell, hy}, R)$, the public keys of the parties in the path, and the sequence $(t^\rightarrow_{v_\ell}, s^\rightarrow_{\ell, 2}, \dots, s^\rightarrow_{\ell, hy})$.}
\item[6:] {Return all formed onions.}
\end{enumerate}
\tikz[overlay,remember picture]{\draw[blue]
  ($(bl)+(-0.2em,0.9em)$) rectangle
  ($(br)+(0em,-0.3em)$)
  ;}

\underline{$\mathsf{FormCheckpointOnions}(\secparam, \Sigma, \pp)$}
\vspace{2mm}
\begin{enumerate} [label=\arabic*:]
\item {Let $d \myeq (\numOnion yz+1) + y\log\numOnion+1$. (This is the total number of rounds in the execution phase.)}
\item {For every round $r \in [d]$ and party $I \in \parties$:}
\begin{itemize}
\item {Let $\mathsf{makeCkpt} \myeq b(F(\sk(P, I), (r, 0))$.}
\item {If $\mathsf{makeCkpt} = 1$:} 
\begin{itemize}
\item[\tikzmark{BL}--] {Let $I_r \myeq I$; $s_r \myeq F(\sk(P, I), (r, 1))$.}
\item {If $r > nyz+1$:}
\begin{itemize}
\item {Let $J^\rightarrow$ be a random walk through $\beta^z$.}
\item {Let $I_r \myeq I$. For all $i\in\{\numOnion yz+1, \dots, d\}\setminus\{r\}$, let $I_i \sample \parties$.}
\end{itemize}
\item {Else (if $r \le nyz+1$):}
\begin{itemize}
\item {Let $J^\leftarrow$ be a random walk $(I_0, \dots, I_r)$ from $I_r$ to $I_0$.}
\item {Let $J^\rightarrow$ be a random walk $(I_r, \dots, I_d)$ from $I_r$ to $I_{\numOnion yz}$.}
\item {For all $i\in\{\numOnion yz+1, \dots, d\}$, let $I_i \sample \parties$.}
\end{itemize}
\item {For all $i\in[d-1]\setminus\{r\}$, let $s_i \sample \mathcal{S}$. \hfill} \tikzmark{BR}
\end{itemize}
\item {Form an onion by running $\formonion$ on the empty message, the routing path $(J^\leftarrow, J^\rightarrow, I_{nyz+1}, \dots, I_{d})$, the public keys associated with the path, and the sequence $(s_0, \dots, s_{d-1})$ of nonces.}
\end{itemize}
\item {Return all formed onions.}
\end{enumerate}
\tikz[overlay,remember picture]{\draw[blue]
  ($(BL)+(-0.2em,0.9em)$) rectangle
  ($(BR)+(0em,-0.3em)$)
  ;}
\end{flushleft}
\endgroup
\end{minipage}
\caption{\footnotesize Protocol $\PiBfly$'s onion-forming algorithms for party $P$ on input $\{(m, R)\}$. The code for generating the intermediaries and nonces are \fbox{boxed}; since these do not depend on the input $\{(m, R)\}$, it is evident that $\PiBfly$ is \regOR.}
\label{fig:PiBfly}
\end{figure} 

%% file: proof2.tex
\ifdraft \else
\section{Proof that $\PiTree$ is anonymous} \label{app:pitree-security}
\fi
The proof that $\PiTB$ is anonymous mirrors the proof of \ifdraft
Theorem~\ref{thm:PiTree}). 
\else 
Theorem~\ref{thm:upper}). 
\fi
$\PiTB$ is indifferent; see the \fbox{boxed} areas in 
\ifdraft
Figure~\ref{fig:PiBfly}. 
\else
Figure~\ref{fig:PiTree}. 
\fi
So from Theorem~\ref{thm:impliedby}, $\PiTB$ is anonymous if it mixes and equalizes. 

For the analysis, we assume that replay attacks are not possible, and that the onion encryption is ideal (Simplifications 1 and 2 in the proof of
\ifdraft
Theorem~\ref{thm:PiTree}). 
\else 
Theorem~\ref{thm:upper}). 
\fi
Given these simplifications, we show that for any two inputs $\sigma^0$ and $\sigma^1$ (from the same equivalence class), the adversarial view consisting of just the \comm\ onions on $\sigma^0$ is statistically-close to that on $\sigma^1$. (Recall that an onion is \comm\ if it was generated by an honest party and is not a checkpoint onions to be verified by an adversarial party.)

$\PiTB$ mixes for the same reason that $\PiTree$ mixes; $\PiTB$ mixes because either a polylogarithmic number of \comm\ checkpoint onions shuffle with the remaining \comm\ merging onions for a polylogarithmic number of rounds during the penultimate epoch, or because not enough \comm\ onions remain by the last epoch, and so the protocol is aborted. (For the full proof, see Lemma~\ref{lem:mix} in Appendix~\ref{app:mix}.) 

From Theorem~\ref{thm:impliedby}, it remains to prove that $\PiTB$ equalizes from \comm\ onions. 
\ifdraft 
In the next sections, we will show: 
\fi

\ifdraft
\underline{Informal Lemma~\ref{lem:first}.} 
\else
\begin{lemma} \label{lem:part1}
\fi
If the adversary drops at most a logarithmic (in the security parameter) number onions during 
\ifdraft the mixing phase, 
\else the first epoch, 
\fi then w.o.p., a constant fraction of each honest sender's merging onions remain undropped by the start of the 
\ifdraft equalizing phase. 
\else second epoch. 
\fi Otherwise, if the adversary drops too many onions, then w.o.p., each honest participant detects that the adversary dropped too many onions and aborts the run. (See 
\ifdraft Section~\ref{sec:first} for the formal lemma statement and proof.)
\else Appendix~\ref{app:pitree} for the proof.)
\fi 
\ifdraft
\else
\end{lemma}
\fi

\ifdraft
\underline{Informal Lemma~\ref{lem:rest}.} 
\else 
\begin{lemma}
\fi
If a constant fraction of each honest sender's merging onions remain at the start of the 
\ifdraft equalizing phase, 
\else second epoch, 
\fi 
then w.o.p., the numbers of merging onions will ``even out'' by the end of the execution phase. That is, for any two honest parties $P$ and $Q$, the quantity of $P$'s merging onions will become statistically-close to that of $Q$'s merging onions. 
\ifdraft (See Section~\ref{sec:rest} for the formal lemma statement and proof.)
\else (The proof is essentially the same as that of Lemma~\ref{lem:equalizes}.)
\fi
\ifdraft
\else
\end{lemma}
\fi

If the adversary drops too many onions during the first epoch, then $\PiTB$ equalizes since every honest party stops participating (Lemma~\ref{lem:first}), and so no one receives their message. Otherwise, $\PiTB$ equalizes since enough of each sender's merging onions make it to the second epoch (Lemma~\ref{lem:first}), and the numbers of merging onions are eventually evened out by the merging of onions (Lemma~\ref{lem:equalizes}). 

%% file: proof-pibfly-main-lemmas.tex
\ifdraft
\subsubsection{Proof of Lemma~\ref{lem:first}}\label{sec:first}
\else
\subsection{Proof of Lemma~\ref{lem:first}}\label{sec:first}
\fi
The adversary's goal is to prevent the protocol from equalizing. The adversary can win if for any honest parties $P$ and $Q$, there is a strategy for dropping $P$'s message with probability noticeably higher than that of dropping $Q$'s message. At the first round, the adversary still knows the sender of every onion. As the protocol progresses, the onions get shuffled, and the adversary loses track of who sent which onion. Thus, a conceivable attack by the adversary is to drop onions likely to have originated at $P$ upfront. 

Lemma~\ref{lem:first} states that if the adversary drops at most a logarithmic number of onions during the mixing phase, then a constant fraction of $P$'s onions will remain at the start of the equalizing phase. If the adversary drops many more onions than this, then every honest party will detect the attack and abort the protocol. 

We now formally state and prove Lemma~\ref{lem:first}. 

Let $\start \myeq (\log N + 1) z$ be the number of epochs in the mixing phase, and let ``\emph{the start of the equalizing phase}'' be diagnostic round $\start$. (It is more precisely the end of the mixing phase.)

\ifdraft
\begin{lemma} \label{lem:first} 
\else
\paragraph{Formal statement of Lemma~\ref{lem:first}.}
\fi
In $\PiBfly$, let the onion encryption scheme be perfectly secure; let $\numOnion = y = z = \epsilon_1 \log^{1+\epsilon_2} \secpar$; and let 
$t = \frac{W}{3}$, where $N$ is the number of participants, $\secpar$ is the security parameter, and $W = \frac{(1-\corruptions)\numOnion}{z\log N + \log \numOnion}$. 

Let $0 \le \corruptions < \frac{1}{2}$ be a fixed constant, representing the ``corruption rate.''
Let $\Bad \subseteq \parties$ be any set of participants in $\parties$ such that $|\Bad| \le \corruptions N$, and let $\adv$ be any active adversary who corrupts the set $\Bad$ of parties. 

Let $\sigma$ be any input in the simple I/O setting.

For a party $P$, let $V^P_\start$ denote the number of $P$'s merging onions that remain at the start of the equalizing phase in an interaction between $\PiBfly$ and $\adv$ on input $\sigma$. If there exists an honest party $Q \in \parties\setminus\Bad$ such that $Q$ is unaborted after the start of the equalizing phase, then w.o.p.,
\ifdraft
\begin{align*}
&V^P_\start \ge \frac{(1 - \corruptions)\numOnion}{3} , &\forall P \in\parties\setminus\Bad .
\end{align*}
\else
$V^P_\start \ge \frac{(1 - \corruptions)\numOnion}{9}$ for all $P \in\parties\setminus\Bad$.
\fi
\ifdraft
\end{lemma}
\fi

\begin{proof} 
\iftheory [Proof of Lemma~\ref{lem:first}] 
\else [of Lemma~\ref{lem:first}] 
\fi
Fix any target honest party $P$. 

We will first consider what happens when the adversary $\adv$ employs the following tactic $T$. (Later on, we will analyze that happens when $\adv$ tries a different tactic.) In between the first and second rounds, $\adv$ drops every onion that $P$ sends to an adversarial party. In between the second and third rounds and between the third and fourth rounds, $\adv$ drops every onion that might have come from $P$. For example, suppose that in the first round, $P$ sends an onion to an honest party. Then, $\adv$ drops every onion that the honest party sends to an adversarial party in the second round. 

Let $O$ be any onion that first routes to an honest party, which we will denote $H$. 
W.l.o.g., assume that $O$ is designed to shuffle between $H$ and an \emph{adversarial} intermediary $I$ during the first epoch of the mixing phase. More precisely, the position of the peeled version of $O$ shuffles between $H$ and $I$. (We can assume that $I$ is adversarial since otherwise $I$ is also honest.) By design, all honestly formed onions that first go to $H$ or $I$ shuffle randomly between $H$ and $I$ for the first epoch of the mixing phase. 

By using tactic $T$, the adversary drops too many onions for the protocol to continue. 
If $\adv$ drops every onion that $H$ sends to $I$ in the second and third rounds, then w.o.p., $H$ will detect this and abort the protocol run before the second epoch. (This follows from a known probability concentration bound for the hypergeometric distribution~\cite{HS05}.) $H$ aborting the protocol run, in turn, will cause the network to be flooded with abort messages, and the remaining honest parties to eventually abort. (While at least half of the honest parties are unaborted, the number of aborted honest parties grows super-exponentially w.r.t.\ the number of rounds. This follows from recasting the problem as a martingale problem and applying the Azuma-Hoeffding inequality; see Lemma~\ref{lem:bb} below.) 

At the end of the first epoch, the adversary can drop almost all of the onions he has. So the $P$'s onions that are ``safe'' are those that are routed to an honest party in the first round, and again to the same honest party in the second, third, and final rounds of the first epoch. 

What fraction of $P$'s onions are ``safe?'' For an arbitrarily small positive constant $\delta > 0$, at least $(1-\delta)(1 - \corruptions)\numOnion$ of $P$'s merging onions go to an honest party at the first round (Chernoff bounds), and at least $\frac{1-\delta}{8}$ fraction of these go to an honest party in the second, third, and final rounds (Chernoff bounds). In particular, these bounds hold for any $\delta \le 1 - \frac{2\sqrt{2}}{3}$. 
Thus, the answer is: w.o.p., at least $\frac{(1 - \corruptions)}{9}$. 

We now consider what happens when $\adv$ behaves arbitrarily. 

For every honest party $Q$, let $I(Q)$ denote the party that $Q$ shuffles onions with during the first epoch. Let $D(Q)$ be the number of onions that $\adv$ drops during the first epoch that would have shuffled between $Q$ and $I(Q)$, and let $D_{\max} \myeq \max_{Q\in\parties\setminus\Bad} D(Q)$. 

\underline{Case 1:} If $D_{\max} \ge \frac{(1-\corruptions)\numOnion}{2}$, then w.o.p., the honest parties will abort the protocol for the same reason that they would in an interaction with an adversary employing tactic $T$: $\adv$ drops too many onions. 

\underline{Case 2:} If $D_{\max} \le \frac{(1-\corruptions)\numOnion}{2}$, then $\adv$ either makes fewer targeted drops or delays targeted drops. Either way, $\adv$ cannot eliminate more of $P$'s onions than it would by employing tactic $T$. Thus, in this case, $V^P_\start \ge \frac{(1 - \corruptions)\numOnion}{9}$. 
\end{proof}

\begin{lemma} \label{lem:bb}
Let $\secpar$ be the security parameter, and 
let ${N'} = (1-\corruptions)N \le \poly$ be the number of honest parties (or locations). 
Let $\HOnions$ be the set of \comm\ onions at any round $r$ of the protocol execution.  
If $|\HOnions| = \bigOmega{\log^2 \secpar}$ and 
$|\HOnions| \le ({N'}+2)/2$, then, with overwhelming probability in~$\secpar$, at least $|\HOnions|/\log \secpar$ parties receive at least one onion from $\HOnions$. 
\end{lemma}

\begin{proof}
Let $\numHOnions = |\HOnions|$ be the size of $\HOnions$. 

We recast this problem as a balls-and-bins problems, where the onions in $\HOnions$ are the balls, and the parties (or locations) are the bins. To prove the lemma, we show that when $\numHOnions = |\HOnions|$ balls are thrown independently and uniformly at random into the ${N'}$ bins, with overwhelming probability in~$\secpar$, the number of non-empty bins is at least $\frac{\numHOnions}{\log \secpar}$.

Let $L_1, \dots, L_{N'}$ be the ${N'}$ bins. For each $i\in[{N'}]$, let $X_i$ be the indicator random variable that is one if the $i$-th bin $L_i$ is \emph{empty} (and zero, otherwise). The probability that $L_i$ remains empty is given as
$
\prob{X_i = 1} = \expect{X_i} = \left(1- \frac{1}{{N'}}\right)^\numHOnions .
$
The total number $X = \sum_{i = 1}^{N'} X_i$ of empty bins is the summation of all the $X_i$'s. By the linearity of expectation, 
$
\expect{X} = \sum_{i=1}^{N'} \expect{X_i} = {N'} \left(1- \frac{1}{{N'}}\right)^\numHOnions .
$

Let $W$ be the total number of \emph{non-empty} bins; i.e., $W = {N'} - X$. Again by linearity of expectation, 
\begin{align}
\expect{W} &= {N'} - \expect{X} \nonumber \\
&= {N'} - {N'} \left(1- \frac{1}{{N'}}\right)^\numHOnions \nonumber \\
&> {N'} - {N'} \left(1 - \frac{\numHOnions}{{N'}} + \frac{\numHOnions(\numHOnions-1)}{{N'}^2} \right) \label{eq:laurent} \\
&= \numHOnions \left(1 - \frac{\numHOnions-1}{{N'}} \right) 
\ge \frac{\numHOnions}{2} \label{eq:upper} ,
\end{align}
where \eqref{eq:laurent} holds since $\left(1- \frac{1}{{N'}}\right)^\numHOnions$ is strictly less than the first three terms of its Laurent series, and \eqref{eq:upper} holds since $\numHOnions \le \frac{{N'}+2}{2}$ by the hypothesis. 

For every $i\in[\numHOnions]$, let $Y_i$ be the location of the $i$-th ball, and let
\[
Z_i = \expect{ X | Y_1, \dots, Y_i } .
\]
The sequence $Z_1, \dots, Z_{\numHOnions}$ is a Doob martingale by construction, satisfying the Lipschitz condition with constant bound one, i.e., $| Z_{i-1} - Z_i | \le 1$. Thus, we may apply the Azuma-Hoeffding inequality as follows: For $\delta \ge \frac{\log \secpar-2}{\log \secpar}$, 
\begin{align}
\prob{X - \expect{X} \ge \delta \expect{W}} 
&\le \mathsf{exp} \left( - \frac{\delta^2 \expect{W}^2}{2 \sum_{i=1}^\numHOnions 1} \right) \nonumber \\
&\le \mathsf{exp} \left( - \frac{\delta^2 (\numHOnions/2)^2}{2\numHOnions} \right) \label{eq:upper2} \\
&= \mathsf{exp} \left( - \frac{\delta^2\numHOnions}{8} \right) \nonumber \\
&\le \mathsf{exp} \left( - \frac{(\log \secpar-2)^2 \alpha\log^2 \secpar}{8 \log^2 \secpar} \right) \label{eq:lower} \\
&= \mathsf{exp} \left( - \frac{\alpha(\log \secpar-2)^2}{8} \right) \nonumber \\
&= \negl , \nonumber
\end{align}
where \eqref{eq:upper2} follows directly from \eqref{eq:upper}, and \eqref{eq:lower} holds since  $\delta \ge \frac{\log \secpar-2}{\log \secpar}$ and $\numHOnions \ge \alpha\log^2 \secpar$ from the hypothesis. 

In other words, with overwhelming probability in $\secpar$, 
\[
X - \expect{X} \le \delta \expect{W} .
\]
Thus, it follows that, with overwhelming probability in $\secpar$, 
\begin{align}
W &\ge (1-\delta) \expect{W} \nonumber \\
&= \left(1 - \frac{\log \secpar-2}{\log \secpar}\right) \expect{W} \nonumber \\
&= \frac{2}{\log \secpar} \expect{W} \nonumber \\
&\ge \frac{\numHOnions}{\log \secpar} \label{eq:upper3} ,
\end{align}
where \eqref{eq:upper3} follows directly from \eqref{eq:upper}. 
\end{proof}

\ifdraft
\subsubsection{Proof of Lemma~\ref{lem:rest}} \label{sec:rest}
\else
\subsection{Proof of Lemma~\ref{lem:equalizes}.} \label{sec:rest}
\fi
Let $\epsilon > 0$ be any small constant such that $\frac{1+\epsilon}{1-\epsilon} \le 1 + \epsilon_2$. 

Recall that ``$\start$'' is the start of the equalizing phase. Let the ``\emph{partway point}'' be the diagnostic round in epoch $\partway$, where $\partway \myeq \start+\ceil{\log \numOnion/\epsilon}$.

In this section, we prove the following: provided that there are enough of each honest sender's merging onions at epoch $\start$, there exists an epoch $\start < e \le \partway$ such that w.o.p., for any honest parties $P$ and $Q$, the number of $P$'s merging onions will be statistically-close to $Q$'s by the $e^\textit{th}$ epoch:

\ifdraft
\begin{lemma} \label{lem:rest}
\else
\paragraph{Formal statement of Lemma~\ref{lem:equalizes}.}
\fi
In $\PiBfly$, let the onion encryption scheme be secure; let $\numOnion = y = z = \epsilon_1 \log^{1+\epsilon_2} \secpar$; and let 
$t = \frac{W}{3}$, where $N$ is the number of participants, $\secpar$ is the security parameter, and $W = \frac{(1-\corruptions)\numOnion}{z\log N + \log \numOnion}$. 

Let $0 \le \corruptions < \frac{1}{2}$ be a fixed constant. Let $\Bad \subseteq \parties$ be any set of participants in $\parties$ such that $|\Bad| \le \corruptions N$, and let $\adv$ be any active adversary who corrupts the set $\Bad$ of parties.  

Let $\sigma$ be any input in the simple I/O setting. 

For any party $P$ and epoch $e$, let $V^P_{e}$ be the number of $P$'s merging onions that remain at the $e^\textit{th}$ diagnostic round in an interaction between $\PiBfly$ and $\adv$ on input $\sigma$. 
If for all $P \in\parties\setminus\Bad$, $V^P_\start = \bigsmallO{\numOnion}$, then w.o.p., 
\ifdraft
\begin{align*}
\Delta(V^P_\partway, V^Q_\partway) &\le \negl &\forall P, Q \in \parties\setminus\Bad ,
\end{align*}
\else
$\Delta(V^P_\partway, V^Q_\partway) \le \negl$ for all $P, Q \in \parties\setminus\Bad$,
\fi
where $\Delta(\cdot, \cdot)$ denotes the total variation (a.k.a.\ statistical) distance. 
\ifdraft
\end{lemma}
\fi

\begin{proof} \iftheory[Proof of Lemma~\ref{lem:equalizes}] \else[of Lemma~\ref{lem:equalizes}] \fi
A pair of \comm\ onions is \emph{mergeable} if the onions arrive at the same place and time and produce the same nonce when processed (once). Let an \comm\ onion be a singleton if it does not belong in any mergeable pair.

At every round, $\adv$ observes some statistics on singletons and pairs of mergeable onions. With overwhelming probability, these are the only categories of \comm\ onions in the system; e.g., w.o.p., there cannot be three onions that produce the same nonce when peeled (once). This is because a pair of mergeable onions at the start of an epoch cannot remain unmerged for too long. Either $\adv$ drops one or both of the onions, or w.o.p., within the epoch (lasting $y = \text{polylog} \,\secpar$ rounds), an honest party merges the pair. This last fact follows from Chernoff bounds.

During the mixing phase and during each epoch of the equalizing phase, $\PiBfly$ sufficiently shuffles the \comm\ onions. This produces the effect that the only useful information that $\adv$ has for determining the sender of an onion is whether it is a singleton or a merging onion. 
$\adv$ may know that fewer merging onions from one honest party, say $P$, remain in the system compared with those from another honest party, $Q$; in which case, $\adv$ might bet that more singletons are $P$'s merging onions than $Q$'s. In an attempt to prevent the protocol from equalizing, $\adv$ might try dropping singletons during the equalizing phase. (See Appendix~\ref{app:zetafrac} for a precise definition of what we mean by sufficiently shuffling and for the proof that $\PiBfly$ sufficiently shuffles \comm\ onions.)

To prove Lemma~\ref{lem:equalizes}, we will prove, 

\bigskip
\noindent \emph{Claim.} The adversary cannot prevent $\PiBfly$ from equalizing by attacking singletons. No matter how many singletons the adversary drops and when, w.o.p., the probability that $P$'s recipient receives $P$'s message can be only negligibly different from that of $Q$'s recipient receiving $Q$'s message. 

\bigskip
\noindent 
\emph{Proof of claim.}
Let $\mathcal{I}_{e}$ be the set of \comm\ singletons in the first round of the ${e}^\textit{th}$ epoch, and let $Y$ be the (total) number of \comm\ checkpoint onions that are formed. 

Suppose that for every epoch~${e}$, the adversary drops $\alpha_{e}$ fraction of the onions in $\mathcal{I}_{e}$. We expect that the adversary drops $\alpha_1$ of the $Y$ \comm\ checkpoint onions during epoch $1$, and another $\alpha_2$ of the remaining $(1-\alpha_1)Y$ \comm\ checkpoint onions during epoch $2$, and so on. 
Following this logic, \emph{by} the ${e}^\textit{th}$ epoch, we expect that $\expect{\zeta_{e}}$ fraction of the $Y$ \comm\ checkpoint onions have been dropped, where $\expect{\zeta_{e}}$ is defined recursively as follows: 
\ifdraft
\begin{align*}
\expect{\zeta_1} &= 0 \\
\expect{\zeta_{e}} &= \sum_{\tau = 1}^{{e}-1} (1-\expect{\zeta_{\tau}}) \alpha_{\tau} , & \forall e \ge 2. 
\end{align*}
\else
$\expect{\zeta_1} = 0$, and $\expect{\zeta_{e}} = \sum_{\tau = 1}^{{e}-1} (1-\expect{\zeta_{\tau}}) \alpha_{\tau}$ for all $e \ge 2$.
\fi

\underline{Case 1:} if $\expect{\zeta_\partway} \ge \frac{1}{2}$. In this case, the adversary drops ``many'' onions. 
From repeated applications of probability concentration bounds, we can show that 
(1) the adversary essentially drops a random sample from the remaining singletons, 
(2) w.o.p., the actual fraction~$\zeta$ of dropped \comm\ checkpoint onions is close to $\expect{\zeta}$, and 
(3) w.o.p., the number of missing checkpoint onions at a party and round is strongly correlated with $\zeta$. 
Thus, when $\expect{\zeta_\partway} \ge \frac{1}{2}$, w.o.p., every honest party aborts the protocol run by the partway point, i.e., $V^P_\partway = 0$ for every honest $P$. (See Lemma~\ref{lem:zetafrac} in Appendix~\ref{app:zetafrac} for the formal proofs of these claims.) 

\underline{Case 2:} if $\expect{\zeta_\partway} < \frac{1}{2}$. In this case, the adversary drops relatively few onions. 

Fix any honest party $P \in \parties\setminus\Bad$. 

Let ${e}$ be any epoch between the start of the equalizing phase and the partway point, and 
let $\nu_{{e}} \myeq \frac{V_{{e}}}{U_{{e}}}$ be the ratio between the actual number $V_{{e}}$ of $P$'s onions at the ${e}^\textit{th}$ diagnostic round and its upper bound, $U_{{e}} \myeq \frac{\numOnion}{2^{{e}-\start}}$, which is necessarily superlogarithmic in the security parameter.  

Fix $0 \le \expect{\zeta} \le \frac{1}{2}$, and let $0 \le \alpha \le \expect{\zeta}$ be any fraction between zero and $\expect{\zeta}$. 
We will first analyze what happens when the adversary $\adv$ drops $\alpha$ fraction of the remaining singletons during the $({e}+1)^\textit{st}$ epoch (between the ${e}^\textit{th}$ diagnostic and the $({e}+1)^\textit{st}$ diagnostic) and another $\frac{\expect{\zeta}-\alpha}{1-\alpha}$ fraction of the remaining singletons during the $({e}+2)^\textit{nd}$ epoch. 

At the ${e}^\textit{th}$ diagnostic round, there are an expected (approx.) $\nu_{{e}}^2U_{e}$ paired onions and an expected (approx.) $\nu_{e} (1 - \nu_{e})U_{e}$ singletons. Since the actual quantities of mergeable pairs are close to the expected values (see Lemma~\ref{lem:singletons} in Appendix~\ref{app:singletons}), 
if the adversary $\adv$ drops $\alpha$ fraction of the singletons, then for any small constant $\delta \ge \frac{V_\partway}{V_\partway-1} - 1$, w.o.p., 
\begin{align}
\nu_{{e}+1} 
&\ge \frac{U_{e}}{U_{{e}+1}}  \left( (1-\delta) \frac{\nu_{{e}}^2}{2} + (1-\delta) (1-\alpha) \nu_{e} (1 - \nu_{e}) \right) \nonumber \\
&= 2(1-\delta)  \left( \frac{\nu_{{e}}^2}{2} + (1-\alpha) \nu_{e} (1 - \nu_{e}) \right) \nonumber \\
&= (1-\delta)\nu_{{e}}^2 + 2(1-\delta)\nu_{e} - 2(1-\delta)\nu_{e}^2 - 2(1-\delta)\alpha \nu_{e} + 2(1-\delta)\alpha \nu_{e}^2 \label{eq:fderiv} \\
&\myeq \xi_{{e}+1} . \nonumber 
\end{align}

At the $({e}+1)^\textit{st}$ diagnostic round, there are an expected (approx.) $\nu_{{e}+1}^2U_{{e}+1}$ paired onions and an expected (approx.) $\nu_{{e}+1} (1 - \nu_{{e}+1})U_{{e}+1}$ singletons. So if the adversary $\adv$ drops $\beta \myeq \frac{\expect{\zeta}-\alpha}{1-\alpha}$ fraction of the singletons, then (from Lemma~\ref{lem:singletons} in Appendix~\ref{app:singletons}) w.o.p., 
\begin{align*}
\nu_{{e}+2} 
&\ge \frac{U_{{e}+1}}{U_{{e}+2}}  \left( (1-\delta) \frac{\xi_{{e}+1}^2}{2} + (1-\delta) \left(1 - \beta\right) \xi_{{e}+1} (1 - \xi_{{e}+1}) \right) \\
&= 2(1-\delta) \left( \frac{\xi_{{e}+1}^2}{2} + \left(1 - \beta\right) \xi_{{e}+1} (1 - \xi_{{e}+1}) \right) \\
&= 2 (1-\delta) \left(1-\beta \right) \xi_{{e}+1} + 2 (1-\delta) \left(\beta - \frac{1}{2}\right) \xi_{{e}+1}^2 \\
&\myeq \xi_{{e}+2} .
\end{align*}

Taking the derivative of $\xi_{{e}+2}$ with respect to $\alpha$, we get
\begin{align}
\frac{\partial \xi_{{e}+2}}{\partial \alpha} 
&= \left( \frac{\partial \xi_{{e}+2} }{\partial \xi_{{e}+1} } \right) \left(\frac{\partial \xi_{{e}+1} }{\partial\alpha}\right) \nonumber \\
&= \left( 2 (1-\delta) \left( 1-\beta + (2\beta-1) \xi_{{e}+1} \right) \right) \left( 2 (1-\delta) \left(\nu_{e}^2 - \nu_{e}\right) \right) \nonumber \\
&= 4 (1-\delta)^2 \left(\nu_{e}^2 - \nu_{e}\right) \left( 1-\beta + (2\beta-1) \xi_{{e}+1} \right) \nonumber \\
&\le 0 , \label{eq:dropfirst}
\end{align}
since $\frac{\partial \xi_{{e}+1} }{\partial \alpha} = 2 (1-\delta) \left( 1-\beta + (2\beta-1) \xi_{{e}+1} \right)$ from \eqref{eq:fderiv}.  
This last inequality follows because 
$(1-\delta)^2 \ge 0$, 
$\left(\nu_{e}^2 - \nu_{e}\right) \le 0$ since $\nu_{e} \le 1$, and 
$\xi_{{e}+1} \le \frac{1-\beta}{1-2\beta}$ since $\beta = \frac{\expect{\zeta}-\alpha}{1-\alpha} \le \frac{1}{2}$.

From \eqref{eq:dropfirst}, the best that the adversary can do is to drop at most half of all the singletons upfront, in the $(\start+1)^\textit{st}$ epoch, in which case $V^P_{\partway} = U_{\partway}$ for every honest $P$. 
This last follows from a known concentration bound for the hypergeometric distribution~\cite{HS05}. This concludes our proof of the claim. 

From the analysis above, to complete our proof of Lemma~\ref{lem:equalizes}, it suffices to show that it doesn't help $\adv$ to also drop mergeable pairs. 

Fix an epoch $e$ between the start of the equalizing phase (i.e., $e > \start$) and the partway point (i.e., $e \le \partway$). 

Let $P, Q \in \parties\setminus\Bad$ be any two honest parties. Let $U \myeq \frac{1}{2^{e-\start}}$. Let $V^P \myeq \nu^P U$ be the number of $P$'s merging onions at the first round of the $e^\textit{th}$ epoch so that $\nu^P$ is the ratio between the number of $P$'s merging onions (i.e., $V^P$) and the maximum it could be (i.e., $U$). Likewise, let $V^Q \myeq \nu^Q U$ be the number of $Q$'s merging onions at the first round of the $e^\textit{th}$ epoch. 

The adversary $\adv$ can identify an onion to be part of a mergeable pair only if she observes it to be so. The only mergeable onions that $\adv$ can drop are those that first arrive at adversarial parties (at the first round of the $e^\textit{th}$ epoch). (Those that first arrive at honest parties are merged into singletons before the adversary can drop them.) Thus, $\adv$ can drop up to a random half of all mergeable onions since w.o.p., at least half of all mergeable pairs go to honest parties in the first round the epoch (Chernoff bounds). 

If $\adv$ drops $\alpha \le \frac{1}{2}$ fraction of the mergeable pairs, then the expected number $\mathbb{V}^P$ of $P$'s mergeable onions at the second round is given by
\ifdraft
\begin{align*}
\mathbb{V}^P = V^P - \alpha\nu^P V^P &=  \nu^P U - \alpha(\nu^P)^2 U ,
\intertext{and the expected number $v^Q$ of $Q$'s mergeable onions at the second round is given by }
\mathbb{V}^Q &= \nu^Q U - \alpha(\nu^Q)^2 U .
\intertext{W.l.o.g., let $V^{P} \le V^{Q}$. (That is, we expect there to be fewer mergeable pairs among $P$'s merging onions than among $Q$'s merging onions.)
It follows that $0 \le \mathbb{V}^Q - \mathbb{V}^P \le {V^Q} - {V^P}$ since}
\mathbb{V}^Q - \mathbb{V}^P
&= (\nu^Q U - \nu^P U) -  \alpha((\nu^Q)^2-(\nu^P)^2) U \\
&\le V^Q - V^P 
\intertext{since $\nu^Q U = V^Q$, $\nu^P U = V^P$, and $\alpha((\nu^Q)^2-(\nu^P)^2) U  \ge 0$; and }
\mathbb{V}^Q - \mathbb{V}^P
&= U \left(\nu^Q\left(1 -\alpha\nu^Q\right) - \nu^P \left(1 - \alpha\nu^P \right)\right) \\
&= \frac{U}{\alpha} \left(\alpha\nu^Q\left(1 - \alpha\nu^Q \right) - \alpha\nu^P \left(1 - \alpha\nu^P \right)\right) \\
&\ge 0 ,
\end{align*}
since $U > 0$, and $\alpha\nu^P \le \alpha\nu^Q \le \frac{1}{2}$. 

\else
$\mathbb{V}^P = V^P - \alpha\nu^P V^P = \nu^P U - \alpha(\nu^P)^2 U$, and the expected number $v^Q$ of $Q$'s mergeable onions at the second round is given by $\mathbb{V}^Q = \nu^Q U - \alpha(\nu^Q)^2 U$. 
W.l.o.g., let $V^{P} \le V^{Q}$. (That is, we expect there to be fewer mergeable pairs among $P$'s merging onions than among $Q$'s merging onions.)
It can easily be checked that $0 \le \mathbb{V}^Q - \mathbb{V}^P \le {V^Q} - {V^P}$.
\fi

Since the actual quantities of mergeable pairs are close to the expected values (Lemma~\ref{lem:singletons} in \emph{Supplementary materials}), this implies that the protocol equalizes faster when $\adv$ also drops mergeable pairs compared to the scenario in which the adversary drops only singletons.
\end{proof}

This completes our proof of our upper bound, Theorem~\ref{thm:upper}.

%% file: lower2.tex
\section{Our lower bound: polylog onion cost is required} \label{sec:lower2}
In this section, we present our lower bound: an onion routing protocol can be anonymous from the active adversary only if the onion cost is superlogarithmic in the security parameter. 
Our lower bound holds for protocols that are minimally functional for the active adversary. We call this notion weakly robustness, defined below.  The reason this definition is weaker than robustness (Definition~\ref{def:robust}) is that here we only insist that the protocol guarantee delivery for senders whose onions are never dropped.

\begin{definition} [Weakly robust]
\label{def:weakrobust}
Let $\Pilong$ be an onion routing protocol and let $\adv$ be an adversary attacking $\Pi$ that drops at most $\bigO{\log(\secpar)}$ onions. $\Pi$ is weakly robust if whenever $\adv$ doesn't drop any onions sent by honest party $P$, $P$'s message will be delivered to its recipient with overehlming probability. 
\end{definition}

\begin{theorem} \label{thm:lower}
If the onion routing protocol $\Pilong$ is weakly robust and (computationally) anonymous from the adversary $\adv$ who corrupts up to a constant fraction of the parties and drops at most $f(\secpar) = \bigO{\log(\secpar)}$ onions, then the onion cost of $\Pi$ interacting with $\adv$ is $\smallOmega{f(\secpar)}$. 
\end{theorem} 

Let us give the intuition for the proof of this theorem. If an honest $P_i$ sends out only $\bigO{\log(\lambda)}$ onions, then an adversary that chooses which participants to corrupt uniformly at random has a $1/\lambda^{\bigO{1}}$ chance of controlling each and every participant that ever receives an onion directly from $P_i$.  (This is because $\bigO{\log(\lambda)} = \bigO{\log(N)}$, since $\lambda$ and $N$ are polynomially related.)  Thus with non-negligible probability it can cut off $P_i$ entirely by dropping all of the onions it sends out, guaranteeing that the intended recipient of $P_i$'s message never receives the message; yet, by weak robustness (Definition~\ref{def:weakrobust}), we can show that there will be some recipient whose probability of receiving his message is high.  Therefore, $\Pi$ will not equalize (Definition~\ref{def:equal}): based on who failed to receive the message, it is possible to determine whether $P_i$'s intended recipient was Bob or Bill.  Since it does not equalize, by Theorem~\ref{thm:implies}, it is not anonymous.  The full proof is in Section~\ref{sec:lowerproof}.

%% file: thmimpliedby2.tex
\begin{proof} 
\iftheory[Proof of Theorem~\ref{thm:impliedby}] 
\else [of Theorem~\ref{thm:impliedby}]
\fi
Suppose that $\Pi$ equalizes.  We wish to show that in this case, mixing implies anonymity.  To show this, we will provide a reduction~$\bdv$ that wins the mixing game with non-negligible advantage using an adversary $\adv$ that wins the anonymity game with non-negligible advantage.  

Without loss of generality we will only consider $\adv$ that always chooses swap-neighboring input vectors for the challenge input vectors $\sigma^0$ and $\sigma^1$, where
\emph{swap-neighboring} means that  there exist parties $P_i, P_j \in \parties$ such that (i) $\sigma^0_k = \sigma^1_k$ for every party $P_k \notin \{P_i, P_j\}$, (ii) $\sigma^0_i = \sigma^1_j$, and (iii) $\sigma^0_j = \sigma^1_i$. 

By a straightforward hybrid argument, it follows that:
The onion routing protocol $\Pi$ anonymizes from the adversary class $\mathbb{A}$  in the simply I/O setting if no every adversary $\adv$ that sets  $\sigma^0$, $\sigma^1$ to be swap-neighboring can win the anonymity game with non-negligible advantage.

So our goal is to construct a reduction~$\bdv$ that wins the mixing game $\mathsf{MixingGame}(\secparam, \Pi, \adv)$ with nonnegligible advantage from an adversary $\adv$ that wins the anonymity game with swap-neighboring $\sigma^0,\sigma^1$ with nonnegligible advantage. 

\begin{enumerate}
\item \underline{Key setup:} 
\begin{itemize}
\item The adversary $\adv$ picks the set $\Bad$ of corrupt parties. This information is relayed to the reduction~$\bdv$ who forwards it to the challenger $\cdv$ of the mixing game. 

\item $\cdv$ replies to $\bdv$ with the public keys $\pk(\parties\setminus\Bad)$ of the honest parties. $\bdv$ forwards these public keys to $\adv$. 

\item $\adv$ then picks the keys for the adversarial parties and sends the public-key portions $\pk(\Bad)$ to $\bdv$, who relays them to $\cdv$. 
\end{itemize}

\item \underline{Input selection:}
\begin{itemize}
\item $\adv$ picks swap-neighboring input vectors $\sigma^0$ and $\sigma^1$ such that $\sigma^0 \equiv_{\Bad} \sigma^1$. Let $P_i$ and $P_j$ be the (necessarily) honest parties whose inputs are swapped in $\sigma^0$ and $\sigma^1$. 
$\adv$ sends the input vectors $(\sigma^0, \sigma^1)$ to $\bdv$. 

\item Let $R_i^0$ be the recipient of $P_i$ in $\sigma^0$, and let $R_i^1$ be the recipient of $P_i$ in $\sigma^1$. $\bdv$ converts the information $(\sigma^0, \sigma^1)$ to the corresponding partial input vector $\tilde \sigma$, i.e., in $\tilde \sigma$, $\sigma_k = \sigma_k^0 = \sigma_k^1$ for $k\notin\{i, j\}$. $\bdv$ sends the partial input vector $\tilde \sigma$ to $\cdv$. 

\item $\cdv$ picks a random bijection from the set $\senders = \{P_i, P_j\}$ to the set $\receivers = \{R_i^0, R_i^1\}$ to complete the input vector $\sigma$.   
\end{itemize}

\item $\adv$, $\bdv$, and $\cdv$ interact in an execution of protocol $\Pi$ on input $\sigma$ (kept secret by $\cdv$) with $\cdv$ acting as the honest parties adhering to the protocol and $\adv$ (via $\bdv$) controlling the corrupted parties. 

\item Let $\mathsf{O}_i^0$ be the set of honest onions received by $R_i^0$, and 
let $\mathsf{O}_i^1$ be the set of honest onions received by $R_i^1$. 
Upon receiving a guess~$b'$ from $\adv$,  
$\bdv$ picks onions $O_s$ and $O_{\overline{s}}$ and a sender $P_s$ as follows: 
\begin{align*}
P_s &\myeq P_i \\
O^0 &\sample \mathsf{O}_i^0 \\
O^1 &\sample \mathsf{O}_i^1 
\end{align*}

$\bdv$ guesses that onion $O_s \myeq O^{b'}$ was sent by $P_s$ and $O_{\overline{s}} \myeq O^{\overline{b'}}$ was not. 
\end{enumerate}

We now explain why the reduction works. 
\newcommand{\Good}{\mathsf{Good}}
\newcommand{\Zero}{\mathsf{Zero}}
\newcommand{\Cured}{\mathsf{Cured}}


Let $\mathsf{Adv}_\adv$ denote $\adv$'s advantage in winning the anonymity game, and let $(\mathsf{Adv}_\adv)|_{E}$ denote $\mathsf{Adv}_\adv$ conditioned on event $E$. 
Let $\Good$ be the event that each target recipient receives at least one challenge onion. 

\bigskip
\noindent \underline{Claim:} $\Good$ occurs with non-negligible probability, and $(\mathsf{Adv}_\adv)|_{\Good}$ is non-negligible, i.e., $\prob{\Good} \cdot (\mathsf{Adv}_\adv)|_{\Good}$. 

\bigskip 
\noindent 
$\bdv$ wins the mixing game if (i) $\bdv$ picks two valid valid onions (one for $R_i^0$ and the other for $R_i^1$), and (ii) $\adv$ wins the anonymity game. 
Thus, if the claim holds, $\bdv$ wins with non-negligible advantage since (i) occurs with non-negligible probability, and (ii) occurs with non-negligible advantage.

Thus, to prove that the reduction works, it suffices to show that the above claim holds. We do this below. 

Let an onion $O$ be a \emph{cured} onion if it can be processed into a challenge onion, i.e., letting $P$ denote the party who receives $O$ in the execution, $\proconion(\sk(P), O, P)$ produces a challenge onion $O'$. 

Let $\Zero$ be the 
event that there are cured onions for one of the target receivers. 
Given $\Zero$, the adversary $\adv$ has no advantage in winning the anonymity game: 

If there are no cured onions for either $R_i^0$ or $R_i^1$ during the execution, then $\adv$ doesn't observes any challenge onion/its destination. Since $\Pi$ is indifferent (and the underlying onion encryption scheme is secure),  
this implies that everything observable by $\adv$ is statistically the same on input $\sigma^0$ as it is on $\sigma^1$, and so $\adv$ cannot fare better than a random guess in winning the anonymity game. 

The adversary's advantage is also negligible in the case where there are cured onions for only one of the receivers. 
This is because a difference in the numbers of cured onions observed during the execution cannot relay any information regarding the input when $\Pi$ is indifferent and equalizes (and the underlying onion encryption scheme is secure). 


Let $\overline{\Zero}$ be the complement of $\Zero$. 
$\mathsf{Adv}_\adv$ can be expanded as: 
\begin{align}
\mathsf{Adv}_\adv &= \prob{\Zero} \cdot (\mathsf{Adv}_\adv)|_{\Zero} + \prob{\overline{\Zero}} \cdot (\mathsf{Adv}_\adv)|_{\overline{\Zero}} \nonumber
\intertext{From the above analysis, $(\mathsf{Adv}_\adv)|_{\Zero} = 0$. Thus,}
\mathsf{Adv}_\adv &= \prob{\overline{\Zero}} \cdot (\mathsf{Adv}_\adv)|_{\overline{\Zero}} , \label{eq:adv}
\end{align}
and so $\prob{\overline{\Zero}} \cdot (\mathsf{Adv}_\adv)|_{\overline{\Zero}}$ is non-negligible in the security parameter.  

Given $\overline{\Zero}$, $\adv$ can reliably drop all challenge onions for one target receiver (w.l.o.g., $R_i^0$) but leave at least one challenge onion for the other target receiver ($R_i^1$) only if $\adv$ could directly win the mixing game with non-negligible advantage. This implies our claim above. 
\end{proof}

%% file: proof-mix.tex
In this section, we prove Lemma~\ref{lem:mix}: $\PiBT$ mixes. 

To do this, we will prove that $\PiBT$ ``sufficiently shuffles'' the \comm\ onions, 
where a \comm\ onion is an onion that is formed by an honest party, which is not a checkpoint onion to be verified by an adversarial party. Formally, we define what this means using the following game. 

\paragraph{The game.}
Let $\mathcal{OE} = (\gen, \formonion, \proconion)$ be a secure onion encryption scheme. 
The mixing game $\mathsf{CommutableMixingGame}(\secparam, \Pi, \adv, r_1, r_2)$ is parametrized by the security parameter~$\secparam$, an onion routing protocol $\Pi$, an adversary $\adv$, and two round numbers $r_1$ and $r_2 \ge r_1$. 

First, the adversary $\adv$ and the challenger $\cdv$ set up the parties' keys and select the input $\sigma$ (exactly as we described for the original mixing game in Section~\ref{sec:mix}).  

Next, $\cdv$ interacts with $\adv$ in an execution of protocol $\Pi$ on input $\sigma$ with $\cdv$ acting as the honest parties adhering to the protocol and $\adv$ controlling the corrupted parties. Whenever the protocol~$\Pi$ specifies for an onion to be formed or processed, $\cdv$ runs the onion encryption scheme's onion-forming algorithm $\formonion$ or onion-processing algorithm $\proconion$. Whenever $\cdv$ forms a non-\comm\ onion (i.e., a checkpoint onion to be verified by an adversarial party), $\cdv$ provides $\adv$ with the input and output of the algorithm~$\formonion$: the message, the routing path, the keys associated with the parties on the path, the sequence of nonces, and the evolution of onion layers. 

Let $\mathsf{O}_{r_2}$ be the set of \comm\ onions received by the parties in round $r_2$. 
Let $\mathsf{O}_{r_1}$ be the set of (\comm) onions in round $r_1$ that ``evolve'' into an onion in $\mathsf{O}_{r_2}$; that is, an onion $O$ is in $\mathsf{O}_{r_1}$ iff a peeled version of $O$ is in $\mathsf{O}_{r_2}$. 

At the end of the execution, 
$\cdv$ provides $\adv$ with the following information: 
for each onion in $\mathsf{O}_{r_1}$, the onion's evolution from the first round to round $r_1$; and
for each onion in $\mathsf{O}_{r_2}$, the onion's evolution from round $r_2$ to the final round. 
Based on this auxiliary information and its view, $\adv$ chooses two onions $O_0, O_1 \in\mathsf{O}_{r_2}$. 
$\cdv$ picks a random bit $b \sample \bin$ and provides $\adv$ with the onion~$O'_b \in \mathsf{O}_{r_1}$ that evolves into $O_b$, and $\adv$ outputs a guess $b'$ for $b$.  

If $O_0$ and $O_1$ were formed by different senders and evolve into valid challenge onions (see the description of the original mixing game in Section~\ref{sec:mix} for a reminder of what a valid challenge onion is), and if $b'=b$, then $\adv$ wins with probability one. Otherwise, $\adv$ wins with probability one-half. 

We now define what it means for an onion routing protocol to mix \comm\ onions from round $r_1$ to round $r_2$. 

\begin{definition} [Mixing \comm\ onions from round $r_1$ to round $r_2$] \label{def:mix3} 
Let $\Pilong$ be an onion routing protocol; let $\adv$ be an adversary; and let $r_1$ and $r_2\ge r_1$ be round numbers. 
Let $E$ be an event in the probability space defined by the experiment, $\mathsf{CommutableMixingGame}(\secparam, \Pi, \adv, r_1, r_2)$. $\Pi$ mixes \comm\ onions from round $r_1$ to round $r_2$ conditioned on $E$ for $\adv$ if, given $E$, $\adv$ wins $\mathsf{CommutableMixingGame}(\secparam, \Pi, \adv, r_1, r_2)$ with negligible advantage, i.e.,
\mymathenv{
\left|\prob{\text{$\adv$ wins $\mathsf{CommutableMixingGame}(\secparam, \Pi, \adv, r_1, r_2)$} \,|\, E} - \frac{1}{2} \right| = \negl .
}

The protocol computationally (resp.\ statistically) mixes \comm\ onions from round $r_1$ to round $r_2$ if $\adv$ is computationally bounded (resp.\ unbounded). 
\end{definition}


We can relate Definition~\ref{def:mix3} to mixing (Definition~\ref{def:mix}) as follows:

\begin{lemma} \label{lem:comm2}
Let $\Pilong$ be a \regOR\ onion routing protocol; let $\adv$ be an adversary; and let $r_1$ and $r_2\ge r_1$ be round numbers. 
Let $E$ be an event in the probability space defined by the experiment, $\mathsf{CommutableMixingGame}(\secparam, \Pi, \adv, r_1, r_2)$. 
If $\Pi$ mixes \comm\ onions from round~$r_1$ to round $r_2$ conditioned on $E$ for $\adv$, then $\Pi$ mixes conditioned on $E$ for $\adv$.
\end{lemma}

\begin{proof}
Suppose that $\adv$ can ``break'' mixing, i.e., $\adv$ can win the mixing game $\mathsf{MixingGame}(\secparam, \Pi, \adv)$ with non-negligible advantage. Then, for any rounds $r_1, r_2 \ge r_1$, we can construct a reduction~$\bdv$ that can win $\mathsf{CommutableMixingGame}(\secparam, \Pi, \adv, r_1 , r_2)$ with non-negligible advantage as follows: 

$\adv$ chooses the set $\Bad$ of parties to corrupt. $\cdv$ generates the keys for the honest parties, and $\adv$ picks the keys for the corrupt parties. Then, the adversary chooses the target senders $\senders$, the target receivers $\receivers$, and the partial input vector $\tilde \sigma$, and $\cdv$ samples a random input vector $\sigma$ that ``completes'' $\tilde \sigma$. After interacting in a protocol run with $\cdv$ (via $\bdv$), $\adv$ outputs two onions $O_s$ and $O_{\bar{s}}$ received by parties in $\receivers$ and a party $P_s \in \senders$. $\bdv$ determines the onions $O_0, O_1 \in \mathsf{O}_{r_2}$ that ``evolve'' into $O_s$ and $O_{\bar{s}}$. ($\bdv$ can do this since $\Pi$ is \regOR\ and, therefore, round $r_2$ cannot occur after the final round in which $O_s$ and $O_{\bar s}$ are received.) $\bdv$ relays $(O_0, O_1)$ to $\cdv$, and $\cdv$ responds with the onion $O'_b \in \mathsf{O}_{r_1}$. If $O'_b$ was formed by $P_s$, then $\bdv$ guesses $b' = 0$; otherwise, $\bdv$ outputs a random bit for $b'$. 

The reduction works since, conditioned on $E$ and $\adv$'s challenge being valid, $\bdv$ wins half of the time that $\adv$ wins (when $b=0$). 
\end{proof}

We now prove Lemma~\ref{lem:mix} via Lemma~\ref{lem:comm2}. 

\begin{proof} 
\iftheory [Proof of Lemma~\ref{lem:mix}] 
\else [of Lemma~\ref{lem:mix}] 
\fi
Here, we prove that $\PiBT$ mixes. 
From Lemma~\ref{lem:comm2}, to prove that $\PiBT$ mixes, it suffices to show that the protocol mixes \comm\ onions from the the first round of the penultimate epoch, round~$r_1$, to the last round of the penultimate epoch, round~$r_2$. 

\underline{Case 1.} Let $E$ be the event that every honest party aborts by the first round the final epoch, round~$r_2 + 1$. Given $E$, $\PiBT$ mixes since w.o.p., no honestly formed onion will be delivered. (With overwhelming probability, the latter half of each honestly formed onion contains an honest party who has already aborted the run.)

\underline{Case 2.} Let $\overline{E}$ be the complement of $E$. That is, $\overline{E}$ is the event that there is an unaborted honest party at the first round of the final epoch, round $r_2 + 1$. Let us condition on $\overline{E}$. 
The adversary cannot drop more than a constant fraction of all \comm\ onions without also dropping a proportional number of checkpoint onions: 
if the adversary were to drop more than a constant fraction of all \comm\ onions, then, 
from known probability concentration bounds~\cite{HS05}, w.o.p., the adversary would drop close to a proportional number of checkpoint onions, which, in turn, would cause all honest parties to abort the run. 
Therefore, the average number of \comm\ onions routed to any party at a round $r \le r_2$ is superlogarithmic in the security parameter. 


Let $\mathsf{O}_{r_2}$ be the set of \comm\ onions at round $r_2$.  
Suppose that $\adv$ is given $\mathsf{O}_{r_2}$ and allowed to choose any two onions $O_0$ and $O_1$ in $\mathsf{O}_{r_2}$. 
Let $O'_0$ and $O'_1$ be onions at round $r_1$ that ``evolve'' into $O_0$ and $O_1$. Below, we prove the claim: even if $\adv$ were provided the set $\{O'_0, O'_1\}$ (in addition to her view and $\mathsf{O}_{r_2}$), 
the sequence of onion layers that preceded $O'_0$, and the sequence of onion layers that preceded $O'_1$, $\adv$ still wouldn't be able to guess which onion in $\{O'_0, O'_1\}$ is the peeled version of $O_0$ with better than negligible advantage. 

Proof of claim: 
For $r < r_2$, let $\mathsf{O}_{r}$ be the set of onions defined as follows: an onion $O_r \in \mathsf{O}_r$ iff there exists an onion $O_{r_2} \in \mathsf{O}_{r_2}$ such that we get $O_{r_2}$ by ``peeling off'' layers from $O_r$. 

For a round $r$ and a party $Q$, let $\volOnions(Q, r)$ denote the volume (i.e., number) of onions in $\mathsf{O}_{r}$ located at $Q$. 

Let $\tilde O_{r_2} \in \mathsf{O}_{r_2}$ be any \comm\ onion at round $r_2$, and let $P$ denote the location 
of $\tilde O_{r_1}$ (i.e., the party who receives $\tilde O_{r_1}$). 
Let $\mathcal{D}(Q, r)$ denote the (unbounded) adversary's ``best estimate'' that the onion $\tilde O_{r} \in\mathsf{O}_{r}$ at round $r$ from which $\tilde O_{r_2}$ evolved is at party $Q$, and define $g(r) \myeq \max_{Q} \frac{\volOnions(Q, r)}{|\mathsf{O}_{r}|} \mathcal{D}(Q, r) - \min_{Q} \frac{\volOnions(Q, r)}{|\mathsf{O}_{r}|} \mathcal{D}(Q, r)$. That is, $g(r)$ is the ``gap'' in probabilities at round $r$; it is the difference between the highest (normalized) probability at round $r$ and the lowest. Assume for a moment that there are no corrupt parties. Then, using Chernoff bounds, we can show that the gap at least halves with every round, i.e., $\frac{g(r)}{g(r+1)} \le \frac{1}{2}$; the proof is essentially the same argument made for proving an earlier result~\cite[Theorem 10]{ICALP:AndLysUpf18}. Since $r_2-r_1$ is superlogarithmic in~$\secpar$, this implies that $g(r_1)$ is negligible in~$\secpar$. In other words, $\tilde O_{r_2}$ cannot be traced back to $P$. This is still true when at most a constant fraction of the parties are corrupt; this is because w.o.p., any pair of \comm\ onions (formed by honest parties) both route to (possibly different) honest parties for at least a constant fraction of the rounds between $r_1$ and $r_2$. 

In conclusion, $\PiBT$ mixes either because it was aborted by all honest parties midway through the last epoch of the execution, or because the protocol sufficiently shuffles the \comm\ onions during the last epoch of the execution. The latter which implies that the protocol mixes conditioned on $\overline{E}$ from Lemma~\ref{lem:comm2}. 
\end{proof}

%% file: anon2.tex
First, we formally define what it means for the distinguisher to win the equalizing game then the adversary knows everything about non-\comm\ onions.

\paragraph{The game.}
\sloppy The game $\mathsf{CommutableEqualizingGame}(\secparam, \Pi, \adv, \ddv, \Sigma)$ is parametrized by the security parameter $\secparam$, an onion routing protocol $\Pi$, an adversary $\adv$, a distinguisher $\ddv$, and a set~$\Sigma$ of input vectors. 

The game starts exactly like the anonymity game (described in Section~\ref{sec:anon}): First, the adversary~$\adv$ and the challenger $\cdv$ set up the parties' keys. Then, $\adv$ selects two inputs $\sigma^0$ and $\sigma^1$ from $\Sigma$ (that are equivalent w.r.t.\ the set $\Bad$ of corrupted parties), and $\cdv$ picks a random bit $b \sample \bin$.

The parties interact in an execution of protocol $\Pi$ on input $\sigma^b$ with $\cdv$ acting as the honest parties adhering to the protocol and $\adv$ controlling the corrupted parties. Whenever the protocol~$\Pi$ specifies for an onion to be formed or processed, $\cdv$ runs the onion encryption scheme's onion-forming algorithm $\formonion$ or onion-processing algorithm $\proconion$. Whenever $\cdv$ forms a non-\comm\ onion (i.e., a checkpoint onion to be verified by an adversarial party), $\cdv$ provides $\adv$ with the input and output of the algorithm~$\formonion$: the message, the routing path, the keys associated with the parties on the path, the sequence of nonces, and the evolution of onion layers. 

At the end of the execution, $\cdv$ computes the statistics $\mathbf{v} = (v_1, \dots, v_N)$, where each $v_r$ represents the number of messages that party $P_r$ received during the execution. (See the description of the original equalizing game in Section~\ref{sec:equal} to recall how the statistics are computed for adversarial parties). 

$\cdv$ sends these statistics $\mathbf{v}$ to the distinguisher $\ddv$, who outputs a guess $b'$ for $b$ and wins the game if $b' = b$. We define equalizing from \comm\ onions as follows.

\begin{definition} [Equalizing from \comm\ onions] \label{def:equal2}
An onion routing protocol $\Pilong$ equalizes from \comm\ onions from the adversary class $\mathbb{A}$ w.r.t.\ the input set $\Sigma$ if for every adversary $\adv \in \mathbb{A}$ and for every distinguisher $\ddv$, $\ddv$ wins the game $\mathsf{CommutableEqualizingGame}(\secparam, \Pi, \adv, \ddv, \Sigma)$ with only negligible advantage, i.e.,
\mymathenv{\left|\prob{\text{$\ddv$ wins $\mathsf{CommutableEqualizingGame}(\secparam, \Pi, \adv, \ddv, \Sigma)$}} - \frac{1}{2} \right| = \negl .
}
The protocol computationally (resp.\ statistically) equalizes from \comm\ onions if the adversaries and the distinguishers are computationally bounded (resp.\ unbounded).
\end{definition}

If the distinguisher can win the equalizing game with non-negligible advantage, then she can also win the equalizing game from \comm\ onions (i.e., when the adversary is given auxiliary information). It follows that,

\begin{lemma} \label{lem:comm}
If the onion routing protocol $\Pilong$ equalizes from \comm\ onions from the adversary class $\mathbb{A}$, then $\Pi$ equalizes from $\mathbb{A}$.
\end{lemma}

%% file: proof-pitree.tex
Our proof of Lemma~\ref{lem:part1} essentially boils down to proving that the following undesirable events rarely happens: (1) For any honest party, the onions formed by the party do not travel together. (2) The first diagnostic fails to detect that the adversary dropped too many honest onions. Below, we show that events (1) and (2) can occur with only negligible probability. 

Recall that a ``\comm'' onion is either an honest merging onion or an honest checkpoint onion with a checkpoint for verification by an honest party.

Recall that $\numOnion$ is the number of merging onions formed by each honest party, and let $\hTree \myeq \log \numOnion+1$ denote the number of diagnostic rounds (or the number of epochs)\footnote{e.g., when $\numOnion=2$, there are two epochs, one corresponding to the leaf node and another to the root node of $G_{\triangle}$.} in a full unaborted execution of $\PiTree$.  

\begin{lemma} 
\label{lem:threshold}
In interacting with $\PiTree$, 
suppose that the adversary drops $\zeta$ fraction of all \comm\ onions before the first diagnostic such that $\zeta \frac{\numOnion}{\hTree} =\mathsf{polylog}(\secpar)$. 
Then, for all $0 < \delta \le 1$, w.o.p., each honest party~$I$ will notice at least $(1-\delta)(1-\corruptions) \zeta \frac{\numOnion}{\hTree}$ missing checkpoints at the first diagnostic. 
\end{lemma}

\begin{proof}
We recast this problem as a three-colored-balls problem with green balls, white balls and red balls. The different categories of balls correspond to different categories of onions (explained below). 

Fix an honest party $I$. 

Let $\mathcal{Z}$ be the set of all \comm\ checkpoint onions for verification by party~$I$ at the first diagnostic. (These correspond to the green onions/balls. All other \comm\ checkpoint onions belong to the set $\mathcal{Y}$ and are the white onions/balls.) Let $Z = |\mathcal{Z}|$.

If $F$ is a truly random function, 
\[
Z \sim \mathsf{Binomial}\left((1-\corruptions)N, q \right) . 
\]

Using Chernoff bound for Poisson trials, for any $0 < \delta' \le 1$:
\begin{align}
\prob{|Z - \expect{Z}| > \delta' \expect{Z}} &\le 2 \, \mathsf{exp}\left( -\mathsf{polylog}(\secpar) \right) = \negl . \label{eq:expect0} 
\end{align}
Thus, with overwhelming probability, 
$Z$ falls between $(1-\delta') \expect{Z} = (1-\delta')(1-\corruptions)\frac{\numOnion}{\hTree}$ and $(1+\delta') \expect{Z} = (1+\delta')(1-\corruptions)\frac{\numOnion}{\hTree}$. 

Let $\mathcal{X}$ be the set of all honest merging onions. (These correspond to the red onions/balls.) 

Let $\mathcal{I}$ be the set of all \comm\ onions, i.e., the set of all green, white and red onions/balls. 

Since the adversary cannot distinguish between any two onions in $\mathcal{I}$, the cumulative set $\mathcal{E} \subseteq \mathcal{I}$ of \comm\ onions that are \emph{eliminated} (or dropped) by the adversary by the first diagnostic is a random subset of the set $\mathcal{I}$ of all honest onions. 

Let $\zeta = \frac{|\mathcal{E}|}{|\mathcal{I}|}$ be the fraction of onions in $\mathcal{I}$ dropped by the adversary by the first diagnostic. 

Using a known concentration bound for the hypergeometric distribution~\cite{HS05}, when the expected number $\zeta Z$ of green balls in $\mathcal{E}$ is at least polylogarithmic in $\secpar$, i.e., $\zeta Z =\mathsf{polylog}(\secpar)$, the actual number $W$ of green balls in $\mathcal{E}$ is close to the expected value $\zeta Z$, i.e., for any $0 < \delta' \le 1$, $(1-\delta') \zeta Z \le W \le (1+\delta') \zeta Z$. Combining this with \eqref{eq:expect0} above, with overwhelming probability, the number of green balls in the random sample $\mathcal{E}$ falls between $(1-\delta')^2(1-\corruptions) \zeta \frac{\numOnion}{\hTree}$ and $(1-\delta')^2(1-\corruptions) \zeta \frac{ \numOnion}{\hTree}$. 

By choosing an appropriate $\delta$ such that $1-\delta \le (1-\delta')^2$ and $1+\delta \ge (1+\delta')^2$, we obtain our desired bound. 
\end{proof}

We now prove Lemma~\ref{lem:part1}. 

\begin{proof} 
\iftheory [Proof of Lemma~\ref{lem:part1}]
\else [of Lemma~\ref{lem:part1}]
\fi
Fix an input $\sigma$ and an honest party~$P$. 

Let $N$ be the number of parties, and 
let $\adv$ be the adversary who corrupts a random $\floor{\corruptions N}$ parties and at rounds $1$ and $2$, drops every droppable (delivered to a corrupted node) onion that could have been formed by party~$P$ (from the adversary's perspective).

Let $\numHOnions(r)$ be the number of \comm\ onions at round~$r$ that that could have been formed by party~$P$. 
In the first round, $P$ transmits $\numHOnions(1)$ onions. For any arbitrarily small $0<{\delta}'\le1$, with overwhelming probability, 
$
\numHOnions(1) \ge 2 (1-{\delta}') (1-\corruptions) \numOnion
$ 
because, with overwhelming probability, $P$ forms at least $(1-{\delta}') (1-\corruptions) \numOnion$ checkpoint onions with a checkpoint for an honest party (Chernoff bound) and $\numOnion$ merging onions. 

Let the \emph{span} at round~$r$, denoted $S(r)$, be the number of honest parties that receive an \comm\ onion that could have been formed by party~$P$ at round~$r$. From Chernoff bound, at least $(1-{\delta}')(1-\corruptions)$ of these $\numHOnions(1)$~onions go to honest parties. So, from Lemma~\ref{lem:bb} 
\ifdraft in Section~\ref{sec:first}, 
\else in Appendix~\ref{sec:first}, 
\fi 
\begin{align}
S(1) \ge 2(1-{\delta}')^2 (1-\corruptions)^2 \frac{\numOnion}{\log \secpar} . \label{eq:span}
\end{align}

Each of these $S(1)$ parties receives at least $2(1-{\delta}')(1-\corruptions)\numOnion$ \comm\ onions at round $1$ (Chernoff bound). Combining this with \eqref{eq:span}, there are at least $4(1-{\delta}')^3(1-\corruptions)^3 \frac{\numOnion^2}{\log \secpar}$ \comm\ onions that could have originated from party~$P$ at round 2; that is,   
\[
\numHOnions(2) \ge 4(1-{\delta}')^3 (1-\corruptions)^3 \frac{\numOnion^2}{\log \secpar} .
\]

At least $(1-{\delta}') \corruptions$ of these onions are routed to corrupted parties at round~$2$ (Chernoff bound); that is, the number of \comm\ onions from $\numHOnions(2)$ that go to corrupted parties is at least
\begin{align*}
4(1-{\delta}')^4 (1-\corruptions)^3 \corruptions \frac{\numOnion^2}{\log \secpar} 
&\ge 4(1-{\delta}')^4 (1-\corruptions)^3 \corruptions \frac{\chi^2}{\log \secpar}(\log \chi + 1)^2 \\
&\ge 4(1-{\delta}')^4 (1-\corruptions)^3 \corruptions (N \polylog \secpar)(\log \chi + 1)^2 \\
&\ge 4(1-{\delta}')^4 (1-\corruptions)^3 \corruptions (N \polylog \secpar) \hTree ,
\end{align*}
where $\chi \ge \sqrt{N \log^{1+\epsilon} \secpar}$, and $\hTree = \log \numOnion +1$. If all of them are dropped, the fraction $\zeta$ of the \comm\ onions that $\adv$ drops is at least 
\begin{align*}
\zeta 
&\ge \frac{ 4(1-{\delta}')^4 (1-\corruptions)^3 \corruptions (N \polylog \secpar) \hTree}{2(1+{\delta}')\numOnion N} \\
&= \left(\frac{4(1-{\delta}')^4 (1-\corruptions)^3 \corruptions}{2(1+{\delta}')}\right) \polylog \secpar \cdot \frac{\hTree}{\numOnion} ,
\end{align*}
because there are fewer than $2(1+{\delta}') \numOnion$ \comm\ onions in total (Chernoff bound).

Let $1-\delta =  \frac{(1-{\delta}')^5}{1+{\delta}'}$.   
From Lemma~\ref{lem:threshold}, each honest party~$Q$ will notice at least 
\begin{align*}
(1-{\delta}') \zeta \frac{\numOnion}{\hTree} 
&\ge \left(\frac{2(1-{\delta}')^5 (1-\corruptions)^3 \corruptions}{1+{\delta}'}\right) \polylog \secpar \cdot \frac{\hTree}{\numOnion}  \cdot \frac{\numOnion}{\hTree} \\
&= 2 (1-\delta)(1-\corruptions)^3\corruptions \polylog \secpar 
\end{align*}
missing checkpoints and will, therefore, abort the protocol. 

Any adversary that drops at least as many onions as $\adv$ will cause the honest parties to abort the protocol. 

For the adversary $\adv'$, let $V^{\tilde \adv, P}$ denote the number of party $P$'s merging onions that remain at the end of the first epoch in a run of $\PiTree$ on input $\sigma$, interacting with adversary $\adv'$. 
An adversary~$\adv'$ that drops at most as many onions as $\adv$ can only do worse than $\adv$; 
if $\adv'$ deviates from $\adv$ either by dropping fewer onions or waiting to drop onions, then 
\begin{align}
V^{\adv', P} \ge V^{\adv, P} . \label{eq:any-adv}
\end{align}

For any $0 < \delta \le 1$, at least $(1-\delta)(1-\corruptions)^2$ of party~$i$'s merging onions are randomly routed through only honest parties in rounds~$1$ and $2$ (Chernoff bound); it follows that
\begin{align}
V^{\adv, P} \ge (1-\delta)(1-\corruptions)^2 \numOnion . \label{eq:adv3}
\end{align}

Combining \eqref{eq:any-adv} and \eqref{eq:adv3}, we obtain our desired result. 
\end{proof}

%% file: proof-pibfly.tex
\subsubsection{Lemma~\ref{lem:singletons}} \label{app:singletons}

To prove Lemma~\ref{lem:equalizes}, we make use of Lemma~\ref{lem:singletons}, below. A consequence of Lemma~\ref{lem:singletons} is that w.o.p., the number of an honest party's \emph{mergeable pairs} at the start of any epoch is close to what is expected given the number of the party's merging onions at the start of the epoch. 

\begin{lemma} \label{lem:singletons}
Let $\mathcal{U}$ be a set of $2u$ balls paired into $u = \mathrm{polylog}(\secpar)$ distinct pairs of balls, and let $\mathcal{V}$ be a random subset of $\mathcal{U}$, such that $\nu = |\mathcal{V}|/|\mathcal{U}|$ is a constant factor. 
For any constant $\frac{2v-\nu}{2v-1} - 1 < \delta \le 1$, w.o.p.\ in $\secpar$, the number $W$ of paired balls in $\mathcal{V}$ is at least $(1-\delta) \nu |\mathcal{V}|$ and at most $(1+\delta) \nu |\mathcal{V}|$.
\end{lemma}

\begin{proof} 
Let $v = \frac{|\mathcal{V}|}{2}$. 
For every $i\in[u]$, let ${w}_i$ be one if both onions that comprise the $u^\textit{th}$ pair in $\mathcal{U}$ are in $\mathcal{V}$, and zero otherwise. 
\begin{align*}
\expect{{w}_i} = \prob{{w}_i = 1} = \frac{\binom{2u - 2}{2v-2}}{\binom{2u}{2v}} 
&= \frac{(2u-2)!}{(2v-2)!(2u-2v)!}\cdot\frac{(2v)!(2u-2v)!}{(2u)!} \\
&= \frac{2v(2v-1)}{2u(2u-1)} ,
\end{align*}
since there are $\binom{2u-2}{2v-2}$ ways to choose $2v-2$ balls from $2u-2$ balls; and likewise, there are $\binom{2u}{2v}$ ways of choosing $2v$ balls from $2u$ balls.

Let ${w}$ denote the number of pairs in $\mathcal{V}$
From the linearity of expectation, 
\begin{align*}
\expect{{w}} &= \sum_{i=1}^u \expect{{w}_i} = u \cdot \frac{2v(2v-1)}{2u(2u-1)} = \frac{2v-1}{2u-1} \cdot v .
\end{align*}

Recall that $\nu = \frac{v}{u} = \frac{|\mathcal{V}|}{|\mathcal{U}|}$. 
It follows that 
\begin{align*}
\expect{W} = \left( \frac{2v-1}{2u-1} \right) |\mathcal{V}| = \left( \frac{2v-1}{2v-\nu} \right) \left( \frac{\nu(2u-1)}{2u-1} \right) |\mathcal{V}| = \left( \frac{2v-1}{2v-\nu}\right) \nu |\mathcal{V}| .
\end{align*} 

For each $i\in[2v]$, let $Y_i$ be the $i^\textit{th}$ chosen ball in $\mathcal{V}$, and let 
\begin{align*}
Z_i = \expect{W | Y_1, Y_2, \dots, Y_i} .
\end{align*}
Then, $Z_0, Z_1, \dots, Z_{2v}$ is a Doob martingale by construction satisfying the Lipschitz condition with bound $1$. 
Thus, from the Azuma-Hoeffding inequality, for any $0 < \delta \le 1$, 
\begin{align*}
\prob{|W - \expect{W}| \ge \delta\expect{W}} 
\le 2\, \mathrm{exp} \left( - \frac{\delta^2\expect{W}^2}{\sum_{i=1}^{2v} 1}\right) 
&= 2\, \mathrm{exp} \left( - \frac{\delta^2(2v-1)^2}{(2u-1)^2}\cdot 2v \right) \\
&= 2\, \mathrm{exp} \left( - \bigsmallO{1} \cdot 2v \right) \\
&= \mathsf{negl}(\secpar) . 
\end{align*}

This completes our proof. 
\iftheory\else\qed\fi
\end{proof}

\subsubsection{Lemma~\ref{lem:zetafrac}} \label{app:zetafrac}

For all of the lemmas in this section: let $e$ be a round between the start of the equalizing phase, ``$\start$,'' and partway through the equalizing phase, ``$\partway$,'' and let $\kappa \le \frac{1}{2}$ be the upper bound on the corruption rate. 

Recall that an \emph{\comm} onion is either an honest merging onion or an honest checkpoint onion with a checkpoint for verification by an honest party, and that a \emph{singleton} is an onion that does not belong in any mergeable pair; it is either a checkpoint onion, or a merging onion without a ``mate.'' 

Let $\mathcal{I}_{e'}$ be the set of \comm\ singletons (onion evolutions) at epoch ${e'}$, and let $Y$ be the number of \comm\ checkpoint onions (onion evolutions) that are formed. 

Suppose that, for every epoch ${e'}$, the adversary drops $\alpha_{e'}$ fraction of the onions in $\mathcal{I}_{e'}$. Then, we expect that the adversary drops $\alpha_1$ of the $Y$ \comm\ checkpoint onions during epoch $1$, and another $\alpha_2$ of the remaining $(1-\alpha_1)Y$ \comm\ checkpoint onions during epoch $2$, and so on. 
Following this logic, \emph{by} epoch ${e'}$, we expect that $\expect{\zeta_{e'}}$ fraction of the $Y$ \comm\ checkpoint onions have been dropped, where $\expect{\zeta_{e'}}$ is defined recursively as follows: 
\begin{align}
\expect{\zeta_1} &= 0 \label{eq:base2} \\
\expect{\zeta_{e'}} &= \sum_{\tau = 1}^{{e'}-1} (1-\expect{\zeta_{\tau}}) \alpha_{\tau} , & {e'} \ge 2 . \label{eq:recur2}
\end{align}

From repeated applications of probability concentration bounds, we can show that if the adversary drops singletons during the equalizing phase, then 
(1) the adversary essentially drops a random sample from the remaining singletons (Lemma~\ref{lem:shuffle}), 
(2) w.o.p., the actual fraction~$\zeta$ of dropped \comm\ checkpoint onions is close to $\expect{\zeta}$ (Lemma~\ref{lem:threshold2}), and 
(3) w.o.p., the number of missing checkpoint onions at a party and round is strongly correlated with $\zeta$ (Lemma~\ref{lem:twocolor}). From these, it follows that, 

\begin{lemma} \label{lem:zetafrac}
In $\PiBfly$: 
Let $\alpha_{e}$ be the fraction of remaining \comm\ singletons that the adversary~$\adv \in \mathbb{A}_\corruptions$ drops during the ${e}^\textit{th}$ epoch, and let $\expect{\zeta_{e}}$ be as defined by \eqref{eq:base2} and \eqref{eq:recur2}.
\begin{enumerate}
\item[i] If $\expect{\zeta_{e}} \ge \frac{1}{2}$, then w.o.p., every honest party aborts by the ${e}^\textit{th}$ diagnostic. 
\item[ii] Conversely, if there is an unaborted honest party after the ${e}^\textit{th}$ diagnostic, then w.o.p., at least half of the \comm\ checkpoint onions remain in the system at the ${e}^\textit{th}$ diagnostic round.
\end{enumerate}
\end{lemma}

Below, we formally state and prove Lemmas~\ref{lem:shuffle}-\ref{lem:twocolor}, starting with Lemma~\ref{lem:shuffle}.

\begin{lemma} \label{lem:shuffle}
In $\PiBfly$: 
No matter how many and which singletons the adversary $\adv \in \mathbb{A}_\corruptions$ drops, $\adv$ essentially drops a random sample of the remaining singletons at the start of the epoch. 
\end{lemma}

We will prove Lemma~\ref{lem:shuffle} by showing that $\PiBfly$ sufficiently shuffles the \comm\ onions during the prior epoch. Before proceeding with the proof, below, we formally define what we mean by ``sufficiently shuffles'' using the following game. 

\paragraph{The game}
Let $\mathcal{OE} = (\gen, \formonion, \proconion)$ be a secure onion encryption scheme. 
The mixing game $\mathsf{CommutableShufflingGame}(\secparam, \Pi, \adv, r_1, r_2)$ is parametrized by the security parameter~$\secparam$, an onion routing protocol $\Pi$, an adversary $\adv$, and two round numbers $r_1$ and $r_2 \ge r_1$. 

First, the adversary $\adv$ chooses a subset $\Bad \subseteq \parties$ of the parties to corrupt and sends $\Bad$ to the challenger~$\cdv$. For each honest party in $\parties\setminus\Bad$, $\cdv$ generates a key pair for the party by running the onion encryption scheme's key generating algorithm $\gen$ and sends the public keys $\pk(\parties\setminus\Bad)$ of the honest parties to the adversary~$\adv$. $\adv$ picks the keys for the corrupted parties and sends the public keys portions $\pk(\Bad)$ to $\cdv$. 

Next, $\adv$ picks the input vector $\sigma$ and sends its choice to $\cdv$. $\cdv$ interacts with $\adv$ in an execution of protocol $\Pi$ on input $\sigma$ with $\cdv$ acting as the honest parties adhering to the protocol and $\adv$ controlling the corrupted parties. Whenever the protocol~$\Pi$ specifies for an onion to be formed or processed, $\cdv$ runs the onion encryption scheme's onion-forming algorithm $\formonion$ or onion-processing algorithm $\proconion$. Whenever $\cdv$ forms a non-\comm\ onion (i.e., a checkpoint onion to be verified by an adversarial party), $\cdv$ provides $\adv$ with the input and output of the algorithm~$\formonion$: the message, the routing path, the keys associated with the parties on the path, the sequence of nonces, and the evolution of onion layers. 

Let $\mathsf{O}_{r_2}$ be the set of \comm\ \emph{singletons} received by the parties in round $r_2$. 
Let $\mathsf{O}_{r_1}$ be the set of (\comm) onions in round $r_1$ that ``evolve'' into an onion in $\mathsf{O}_{r_2}$; that is, an onion $O$ is in $\mathsf{O}_{r_1}$ iff a peeled version of $O$ is in $\mathsf{O}_{r_2}$. 

At the end of the execution, 
$\cdv$ provides $\adv$ with the following information: 
for each onion in $\mathsf{O}_{r_1}$, the onion's evolution from the first round to round $r_1$; and
for each onion in $\mathsf{O}_{r_2}$, the onion's evolution from round $r_2$ to the final round. 
Based on this auxiliary information and its view, $\adv$ chooses two onions $O_0, O_1 \in\mathsf{O}_{r_2}$. 
$\cdv$ picks a random bit $b \sample \bin$ and provides $\adv$ with the onion~$O'_b \in \mathsf{O}_{r_1}$ that evolves into $O_b$, and $\adv$ outputs a guess $b'$ for $b$ and wins if $b' = b$.  

We now define what it means for an onion routing protocol to shuffle \comm\ singletons from round $r_1$ to round $r_2$. 

\begin{definition} [Shuffling \comm\ singletons from round $r_1$ to round $r_2$] \label{def:shuffle}
\sloppy An onion routing protocol $\Pilong$ shuffles \comm\ singletons from round $r_1$ to round $r_2$ conditioned for the adversary class $\mathbb{A}$ if for every adversary $\adv \in \mathbb{A}$, $\adv$ wins $\mathsf{CommutableShufflingGame}(\secparam, \Pi, \adv, r_1, r_2)$ with negligible advantage, i.e.,
\mymathenv{
\left|\prob{\text{$\adv$ wins $\mathsf{CommutableShufflingGame}(\secparam, \Pi, \adv, r_1, r_2)$}} - \frac{1}{2} \right| = \negl .
}

The protocol computationally (resp.\ statistically) shuffling \comm\ singletons from round $r_1$ to round $r_2$ if the adversaries in $\mathbb{A}$ are computationally bounded (resp.\ unbounded). 
\end{definition}

We now prove Lemma~\ref{lem:shuffle}. 

\begin{proof} [Proof of Lemma~\ref{lem:shuffle}]

To prove the lemma, it suffices to show that $\PiBfly$ shuffles singletons from prior to the $e^{\textit{th}}$ epoch. We do this by cases. 

\underline{Case 1.} In the first case, epoch $e$ is the first epoch of the equalizing phase. We show that $\PiBfly$ sufficiently shuffles the \comm\ singletons during the mixing phase; that is, w.o.p., the adversary cannot trace any \comm\ singleton $O$ at the end of the mixing phase back to its sender: 

Let $O_r$ denote the $r^\textit{th}$ layer of the evolution to which $O$ belongs. 

Let $P$ be the location of $O_1$, and let $\mathsf{binary}(P, i)$ be the $i^\textit{th}$ bit of the binary representation of $P$. Let the rounds that ``affect $\mathsf{binary}(P, i)$'' be those that correspond to the $i^\textit{th}$ stage of $B$; these are the rounds in epochs $i$, $n + i$, $2n+i$, \dots, $(z-1)n+i$. From the hypothesis, the number~$z$ of iterations of the butterfly network is polylogarithmic in the security parameter; thus, from Chernoff bounds, w.o.p., every honestly formed onion goes to at least one honest party $Q$ during a round $j$ that affects each $\mathsf{binary}(P, i)$. Moreover, $Q$ mixes the onion with other onions (from honest parties) unless the adversary drops most onions that were meant to shuffle between $Q$ and $Q$'s shuffling partner during the $j^\textit{th}$ epoch, in which case $Q$ will detect this from the diagnostic test at the end of the epoch and abort. (This follows from a known probability concentration bound for the hypergeometric distribution~\cite{HS05}.) $Q$ aborting the protocol run, in turn, will cause the network to be flooded with abort messages, and the remaining honest parties to eventually abort. (While at least half of the honest parties are unaborted, the number of aborted honest parties grows super-exponentially w.r.t.\ the number of rounds. This follows from recasting the problem as a martingale problem and applying the Azuma-Hoeffding inequality; see Lemma~\ref{lem:bb} in \emph{Supplementary materials}.) 

Thus, from the adversary's perspective, every bit of $\mathsf{binary}(P)$ is equally likely to be zero as it is one. 

\underline{Case 2.} When epoch $e$ is after the first epoch of the equalizing phase, $\PiBfly$ shuffles \comm\ singletons from the start of epoch $e-1$ to the end of epoch $e-1$ using an identical argument as our proof of Lemma~\ref{lem:mix}.
\end{proof}

This next lemma states that the actual fraction $\zeta_{e}$ is close to the expected, $\expect{\zeta_{e}}$. 

\begin{lemma} \label{lem:threshold2}
In $\PiBfly$: 
Let $\alpha_{e}$ be the fraction of remaining \comm\ singletons that the adversary drops during the ${e}^\textit{th}$ epoch, and let $\expect{\zeta_{e}}$ be as defined by \eqref{eq:base2} and \eqref{eq:recur2}. 
The fraction $\zeta_{e}$ of all \comm\ checkpoint onions that the adversary drops by the ${e}^\textit{th}$ epoch is close to $\expect{\zeta_{e}}$, i.e., for all $0 < \delta \le 1$, with overwhelming probability, 
\[
(1-\delta) \expect{\zeta_{e}} \le \zeta_{e} \le (1+\delta) \expect{\zeta_{e}} .
\]   
\end{lemma} 

\begin{proof}
The proof is by induction.

\underline{Base case (${e}=2$).} This follows from a known concentration bound~\cite{HS05} for the hypergeometric distribution. 

\underline{Inductive step (${e} > 2$).} Assume that $(1-\delta)  \expect{\zeta_{{e}-1}} \le \zeta_{{e}-1} \le (1+\delta)  \expect{\zeta_{{e}-1}}$. 

Let $\alpha'_{e}$ be the fraction of the (remaining) \comm\ checkpoint onions that the adversary drops during the ${e}^\textit{th}$ epoch. 
\begin{align}
\zeta_{e} &= \zeta_{{e}-1} + (1 - \zeta_{{e}-1}) \alpha'_{e} \nonumber \\
&\ge (1-\delta)  \expect{\zeta_{{e}-1}} + (1 - (1+\delta)  \expect{\zeta_{{e}-1}}) \alpha'_{e} \label{eq:hypoth} \\
&\ge (1-\delta)  \expect{\zeta_{{e}-1}} + (1-\delta) (1 - (1+\delta)  \expect{\zeta_{{e}-1}}) \alpha_{e} \label{eq:hypergeo} \\
&\ge (1-\delta)  \expect{\zeta_{{e}-1}} + (1-\delta) \alpha_{e} - (1 -\delta)  \expect{\zeta_{{e}-1}} \alpha_{e} \label{eq:squared} \\
&= (1-\delta)  \expect{\zeta_{e}} \nonumber ,
\end{align}
where \eqref{eq:hypoth} follows from the inductive hypothesis, \eqref{eq:hypergeo} follows from a known concentration bound~\cite{HS05} for the hypergeometric distribution, and \eqref{eq:squared} follows because $1-\delta^2 \ge 1-\delta$. 

We obtain the upper bound in a similar fashion.  
\end{proof}

If the adversary drops $\zeta_{e}$ fraction of the \comm\ checkpoint onions by the ${e}^\textit{th}$ diagnostic round, then every party would observe, on average, $(1-\corruptions) \zeta_{e} \frac{\numOnion}{\start+h}$ missing checkpoints at the ${e}^\textit{th}$ diagnostic, where $h \myeq \log x + 1$ denotes the number of epochs in the equalizing phase. 
We now prove that, with overwhelming probability, the actual number of missing checkpoint onions is close to this expected quantity. 

\begin{lemma} 
\label{lem:twocolor}
In $\PiBfly$: 
Suppose that the adversary drops at least a constant $0 \le \zeta_{e} \le 1$ fraction of all \comm\ checkpoint onions before the ${e}^\textit{th}$ diagnostic round (i.e., the ${e}y^\textit{th}$ round).  
If $F$ is a truly random function, then for all $0 < \delta \le 1$, with overwhelming probability, each party~$P_k$ will notice at least between $(1-\delta)(1-\corruptions) \zeta_{e} \frac{\numOnion}{\start+h}$ and $(1+\delta)(1-\corruptions) \zeta_{e} \frac{\numOnion}{\start+h}$ missing checkpoints at the ${e}^\textit{th}$ diagnostic round.
\end{lemma}

\begin{proof}
We recast this problem as a two-colored-balls problem. The different categories of balls correspond to different categories of onions (explained below). 

Fix a party $P_k$ and a diagnostic round ${{e} y}$. 

The green balls/onions, $\mathcal{Z}$, are all the \comm\ checkpoint onions for verification by party~$P_k$ at the ${e}^\textit{th}$ diagnostic; let $Z = |\mathcal{Z}|$. 

Let $\mathcal{Y} \supseteq \mathcal{Z}$ be all the \comm\ checkpoint onions, including those in $\mathcal{Z}$; and let $Y = |\mathcal{Y}|$. 
The white onions/balls are the onions in $\mathcal{Y}\setminus\mathcal{X}$; these are the \comm\ checkpoint onions \emph{not} for verification by $P_k$ at the ${e}^\textit{th}$ diagnostic. 

Since the onions in $\mathcal{Y}$ are \comm, if the adversary drops $\zeta$ fraction of them, the adversary \emph{eliminates} (or drops) a random sample $\mathcal{E} \subseteq \mathcal{Y}$ of size $\zeta Y$. 

Using a known concentration bound~\cite{HS05} for the hypergeometric distribution, when the expected number $\expect{W} = \zeta Z$ of green balls in $\mathcal{E}$ is at least polylogarithmic in the security parameter, with overwhelming probability, the actual number $W$ of green balls in $\mathcal{E}$ is close to $\expect{W}$, i.e., $W = (1\pm\delta')\expect{W}$. 

Let $X$ be the number of non-\comm\ checkpoint onions. If 
\begin{claim}
$X$, $Y$, and $Z$ are close to their respective expected values, i.e., for any $0 < \delta' \le 1$, with overwhelming probability, $X = (1 \pm \delta') \expect{X}$, $Y = (1 \pm \delta') \expect{Y}$, and $Z = (1 \pm \delta') \expect{Z}$; 
\end{claim}
then with overwhelming probability, at least $(1-\delta)(1-\corruptions) \zeta_{e} \frac{\numOnion}{\start+h}$ checkpoints onions will be missing for party $k$ at the ${e}^\textit{th}$ diagnostic. 


To complete the proof, we now prove the claim above: 

Let $\mathcal{Y}_1$ be the set of all (for all $i$'s) \comm\ checkpoint onions formed by party~$P_i$ for verification by party~$P_i$, excluding the (possible) onion formed by party $P_k$ for verification by party~$P_k$ at the ${e}^\textit{th}$ diagnostic.  Let $Y_1' = |\mathcal{Y}_1|$. 

Let $\mathcal{Y}_2$ be the set of all (for all $i$'s and all $j$'s) \comm\ checkpoint onions formed by party~$P_i$ for verification by party $P_j$, $j>i$, excluding any onion for verification by party $P_k$ at the ${e}^\textit{th}$ diagnostic as well as any onion \emph{formed by} party $P_k$ for verification at the ${e}^\textit{th}$ diagnostic. Let $Y_2' = |\mathcal{Y}_2|$. 

For every triple $(i, j, \tau)$ consisting of the index $\tau$ of a diagnostic round ${\tau y}$ and honest parties~$P_i$ and $P_j$, let $Y_{i\rightarrow j}^\tau$ be one if party~$P_i$ forms a checkpoint onion to be verified by party~$P_j$ at the $\tau^\textit{th}$ diagnostic (and zero, otherwise). 

Since $Y_{i\rightarrow j}^\tau =1 \iff Y_{j\rightarrow i}^\tau=1$ (i.e., party~$P_i$ creates an onion to be verified by party~$P_j$ at the $\tau^\textit{th}$ epoch iff party~$P_j$ creates a symmetric onion to be verified by party~$P_i$ at the $\tau^\textit{th}$ epoch), it follows that the total (over all $i$'s, all $j$'s, and all $\tau$'s) number of checkpoint onions formed by party~$P_i$ for party $P_j\not=P_i$ is $2\left(Y_2' + \sum_{i\not=k} Y_{i\rightarrow k}^{e}\right)$. 

Let $Y_1 = Y_1' + Y_{k\rightarrow k}^{e}$, and let $Y_2 = Y_2' + \sum_{i\not=k} Y_{i\rightarrow k}^{e}$. 
The total number $Y$ of \comm\ checkpoint onions is given by 
\begin{align}
Y 
&= (Y_1' + Y_{k\rightarrow k}^{e}) + 2\left(Y_2' + \sum_{i\not=k} Y_{i\rightarrow k}^{e}\right) 
= Y_1 + 2Y_2 . \label{eq:yequals}
\end{align}

If $F$ is a truly random function, the onions in $\mathcal{Z}\cup\mathcal{Y}_1\cup\mathcal{Y}_2$, are i.i.d.~Bernoulli random variables, each having probability $q = \frac{2\numOnion}{N(\start+\hTree)}$ of success. 
It follows that
\begin{align*}
Z &\sim \mathsf{Binomial}\left((1-\corruptions)N, q \right) , \\
Y_1 &\sim \mathsf{Binomial}\left((1-\corruptions)N(\start+\hTree), q \right) , \\
Y_2 &\sim \mathsf{Binomial}\left(\binom{(1-\corruptions)N}{2}(\start+\hTree), q \right) .
\end{align*}

Using Chernoff bound for Poisson trials, for any $0 < {\delta''} \le 1$:
\begin{align}
\prob{|Z - \expect{Z}| > {\delta''} \expect{Z}} &\le 2 \, \mathsf{exp}\left( - \mathsf{polylog} (\secpar)\right) = \mathsf{negl}(\secpar) \label{eq:expect} \\
\prob{|Y_1 - \expect{Y_1}| > {\delta''} \expect{Y_1}} &\le 2 \, \mathsf{exp}\left( - \mathsf{polylog} (\secpar)\right) = \mathsf{negl}(\secpar) \\
\prob{|Y_2 - \expect{Y_2}| > {\delta''} \expect{Y_1}} &\le 2 \, \mathsf{exp}\left( - \mathsf{polylog} (\secpar)\right) = \mathsf{negl}(\secpar) \label{eq:expect2}. 
\end{align}

Thus, with overwhelming probability, 
(i) $Z = (1\pm{\delta''}) \expect{Z} = (1-{\delta''})(1-\corruptions)\frac{2\numOnion}{\start+\hTree}$,  
(ii) $Y_1 = (1\pm{\delta''}) \expect{Y_1}$, and 
(iii) $Y_2 = (1\pm{\delta''}) \expect{Y_2}$. 


Facts (ii) and (iii) imply 
\begin{align}
Y 
&= (1\pm{\delta''}) (\expect{Y_1} + 2 \expect{Y_2}) \label{eq:expect3} \\
&= (1\pm{\delta''}) \left( (1-\corruptions)N(\start+\hTree) + 2 \left( \frac{(1-\corruptions)N((1-\corruptions)N-1)}{2}(\start+\hTree) \right) \right) q \label{eq:plug} \\
&= (1\pm{\delta''}) (1-\corruptions)^2N^2(\start+\hTree) \left(\frac{2\numOnion}{N(\start+\hTree)}\right) \nonumber \\
&= (1\pm {\delta''}) \cdot 2 (1-\corruptions)^2 \numOnion N , \nonumber 
\end{align}
where \eqref{eq:expect3} follows \eqref{eq:yequals} and \eqref{eq:expect}-\eqref{eq:expect2}, and \eqref{eq:plug} holds because 
\[
\expect{Y_1} = (1-\corruptions)N(\start+\hTree) q
\]
and
\[
\expect{Y_2} = \left( \frac{(1-\corruptions)N((1-\corruptions)N-1)}{2}(\start+\hTree) \right) q .
\]

Following a similar argument as above, we have $X = (1+\delta') \expect{X}$. This concludes are proof. 
\end{proof}

Combining Lemmas~\ref{lem:shuffle} and \ref{lem:twocolor} proves Lemma~\ref{lem:zetafrac}(ii). We now prove Lemma~\ref{lem:zetafrac}(i).

\begin{proof} [Proof of Lemma~\ref{lem:zetafrac}(i)]
If $\zeta_{{e}} \ge \frac{1}{2}$, then with overwhelming probability, every honest party aborts the protocol before the $({e}+1)^\textit{st}$ epoch: 

From Lemma~\ref{lem:shuffle}, the adversary essentially drops a random sample of the remaining singletons. 
From Lemma~\ref{lem:threshold2}, 
the actual fraction $\zeta_{e}$ of \comm\ checkpoint onions that have been dropped by the ${e}^\textit{th}$ epoch is close to the expected faction, $\expect{\zeta_{{e}}} \ge (1-\delta)\zeta$. 
From Lemma~\ref{lem:twocolor}, 
each party will notice close to the expected number of missing checkpoints: $(1-\delta) \expect{\zeta_{{e}}} \frac{\numOnion}{\start+h}$ (for an arbitrarily small $\delta$). 
\end{proof}

%% file: lowerproof.tex
\section{Proof that polylogarithmic onion cost is required}
\label{sec:lowerproof}

To prove the Theorem~\ref{thm:lower}, we make use of 
the following observation (Lemma~\ref{lem:critical}, below): 
If an onion routing protocol $\Pi$ is too efficient, then there exist many settings in which there exist parties $P_i$ and $P_j$ such that $P_i$ is neither a sender nor an intermediary node for recipient $P_j$. 

In a run of onion routing protocol~$\Pi$ interacting with adversary~$\adv$ on input~$\sigma$ and security parameter $\secparam$: 
\begin{itemize}
\item For an honest party $P_i$, let $\hops{\Pi, \adv}{i \rightarrow j \rightarrow k}{\secparam, \sigma}$ denote the number of onions created by party~$P_i$ and received by party~$P_j$ that will reach party~$P_k$ (if allowed to continue to $P_k$). 

\item For honest parties~$P_i$ and $P_j\not=P_i$, ``\emph{$P_i$ cannot affect $P_j$'s recipient}'' if 
\begin{align}
\expect{\hops{\Pi, \adv}{j\rightarrow i \rightarrow \recipient(j)}{\secparam, \sigma}} \le \frac{1}{2} , \label{eq:notaroute}
\end{align}
where $\recipient(j)$ is the index of the recipient $P_{\recipient(j)}$ for $P_j$. 
\end{itemize}

\begin{lemma} \label{lem:critical} 
Let $\sio$ denote the set of input vectors in the simple I/O setting. 

If the onion cost $\mathsf{OC}^{\Pi, \adv}(\secparam)$ of the onion routing protocol~$\Pilong$ interacting with the adversary~$\adv$ is sublinear in the number $N$ of parties, then there exists a set $\mathsf{Inputs} \subseteq \sio$, $|\mathsf{Inputs}| = \bigsmallO{|\sio|}$ s.t.\ for every $\sigma \in \mathsf{Inputs}$, there exists a set $\mathsf{Senders}_{\sigma} \subseteq \parties$, $|\mathsf{Senders}_{\sigma}| = \bigsmallO{N}$ s.t.\ for every party $P_i \in \mathsf{Senders}_{\sigma}$, 
\begin{enumerate}
\item[i.] $\mathbb{E}_{\$}\left[\outof{\Pi, \adv}{i}{\secparam, \sigma} \right] = \bigsmallO{1} \cdot \mathsf{OC}^{\Pi, \adv}(\secparam)$, and 
\item[ii.] there exists a party $P_{\sigma, i}$ such that $P_i$ cannot affect $P_{\sigma, i}$'s recipient (as defined in \eqref{eq:notaroute}). 
\end{enumerate}
\end{lemma}

\input{lemcritical} 

For the proof of Theorem~\ref{thm:lower}, below, we show that if the protocol is too efficient, then the adversary can ``isolate'' an honest party by blocking all network traffic originating from or passing through the party. 

\begin{proof} [Proof of Theorem~\ref{thm:lower}.]
Fix the corruption rate $\corruptions$. 
For a party~${P_i} \in \parties$, 
let $\adv_{i}$ be the adversary who corrupts a uniformly random set of $\floor{\corruptions N}$ parties
and, additionally, drops every onion that party ${P_i}$ transmits directly to a corrupted party. Otherwise, $\adv_{i}$ follows the protocol. 

Let $\adv$ be the adversary that chooses a random party ${P_i} \sample \parties$ to target and then follows $\adv_{i}$'s code. 

Assume that $\Pi$ is an onion routing protocol that is weakly robust and anonymous $\adv$, and the onion cost of $\Pi$ interacting with $\adv$ is $\bigO{f(\secpar)}$. 

W.l.o.g., assume  $\bigO{f(\secpar)} = \smallO{N}$. We assume this to be the case, since otherwise, there are known solutions with $\bigsmallO{N^2}$ communication complexity, e.g., using general purpose MPC.

From Lemma~\ref{lem:critical}, 
there exists an input $\sigma^0 \in \sio$, such that 
there exists a set $\mathsf{Senders}_{\sigma^0} \subseteq \parties$, $|\mathsf{Senders}_{\sigma^0}| = \bigsmallO{N}$ s.t.\ for every party $P_i \in \mathsf{Senders}_{\sigma^0}$, 
\begin{enumerate}
\item[i.] $\mathbb{E}_{\$}\left[\outof{\Pi, \adv}{i}{\secparam, \sigma^0} \right] = \bigO{f(\secpar)}$, and 
\item[ii.] there exists a party $P_{j}$ such that $P_i$ cannot affect the recipient of $P_{j}$. 
\end{enumerate}

We will now prove the following: 
In the event (with nonnegligible probability) that $\adv$ picks a party $P_i \in \mathsf{Senders}_{\sigma^0}$ to target, $\adv$ can distinguish the setting on input~$\sigma^0$ from the setting on input~$\sigma^1 = \mathsf{swap}(\sigma^0, P_i, P_{j})$ which is the same as $\sigma^0$ except that the inputs for parties~$P_i$ and $P_{j}$ are swapped. 

Let $\recipient$ be the recipient of $P_{j}$ in $\sigma^0$ (and also the recipient of $P_i$ in $\sigma^1$), and let $v_{\recipient}^b$ denote the number of messages that $\recipient$ receives in a protocol run of $\Pi$ interacting with adversary $\adv$ on input $\sigma^b$. 

Let $\mathsf{isolated}$ denote the event that $\adv$ manages to drop every onion that $P_i$ transmits. 
 
On input $\sigma^1$: 
Conditioned on $\mathsf{isolated}$, $\recipient$ never receives his message, i.e., 
\begin{align}
\prob{v_{\recipient}^1 |_{\mathsf{isolated}} = 0 } = 1 . \label{eq:contra1}
\end{align}

On input $\sigma^0$: 
Let $\mathsf{unaffectable}$ denote the event that $\hops{\Pi, \adv}{j\rightarrow i\rightarrow \recipient}{\sigma^0} = 0$. 
From (ii), 
$
\prob{\mathsf{unaffectable}} \le \frac{1}{2} .
$
Combined with weak robustness, it follows that $\recipient$ receives his message with nonnegligible probability, i.e., 
\begin{align}
\prob{v_{\recipient}^0 |_{\mathsf{isolated}} > 0 } = \nonnegl . \label{eq:contra2} 
\end{align}

If $\mathsf{isolated}$ occurs with nonnegligible probability on input $\sigma^0$: 
Then, from combining \eqref{eq:contra1} and \eqref{eq:contra2}, $\Pi$ doesn't equalize; and from Theorem~\ref{thm:implies}, $\Pi$ is not anonymous. 

To complete our proof, it suffices to prove that the probability of $\mathsf{isolated}$ is nonnegligible: 
From~(i), $\mathbb{E}_{\$}[\mathsf{out}_i^{\Pi, \adv}(\secparam, \sigma^0)] = \bigO{f(\secpar)}$. 
From Markov's inequality, there exists a constant $\alpha > 0$, such that $\mathsf{out}_i^{\Pi, \adv}(\secparam, \sigma^0) \le \alpha f(\secpar)$ with nonnegligible probability. 

Let $\mathsf{droppable}$ denote the event that $\mathsf{out}_i^{\Pi, \adv}(\secparam, \sigma^0) \le \alpha f(\secpar)$, and 
let $\mathsf{isolated}|_{\mathsf{droppable}}$ denote $\mathsf{isolated}$ conditioned on $\mathsf{droppable}$.
 
The probability of $\mathsf{isolated}|_{\mathsf{droppable}}$ is smallest when the location of each of the (at most) $\alpha f(\secpar)$ onions that $P_i$ transmits goes to a different location. This probability is bounded by the probability~$p$ that a random $(\alpha f(\secpar))$-size  sample from a set of $N$ balls, $\corruptions N$ of them which are green, are all green. When $\alpha f(\secpar) \le \beta \log \secpar \le \sqrt{N}$ for some positive constant $\beta$,
\begin{align*}
p 
\le \frac{\binom{\corruptions N}{\beta \log \secpar}}{\binom{N}{\beta \log \secpar}} 
&= (1 + \smallO{1}) \frac{(\corruptions N)^{\beta \log \secpar}}{(\beta \log \secpar)!} \cdot (1 + \smallO{1}) \frac{(\beta \log \secpar)!}{N^{\beta \log \secpar}} 
= \bigsmallO{\corruptions^{\beta\log\secpar}} ,
\end{align*}
which is nonnegligible in $\secpar$. 
Thus, $\prob{\mathsf{isolated}\wedge\mathsf{droppable}} = \prob{\mathsf{droppable}}\cdot\prob{\mathsf{isolated}|_{\mathsf{droppable}}}$ is nonnegligible in $\secpar$. It follows that $\mathsf{isolated}$ occurs with nonnegligible probability. 
\end{proof}

We can also prove the weaker result that if an onion routing protocol is \emph{robust} (Definition~\ref{def:robust}) and anonymous, then its onion cost is superlogarithmic (in the security parameter). The proof is a simpler contradiction showing that an onion routing protocol with logarithmic (in the security parameter) onion cost cannot be robust (rather than anonymous). 

To prove the lower bound, we used the fact that the adversary knows the number of messages received by each honest party in the protocol run.  However, the bound holds even when we exclude these statistics from the adversarial view.  We can prove the stronger result by using in place of Theorem~\ref{thm:implies}:  If an onion routing protocol is anonymous from adversaries who corrupt up to $\corruptions N + 1$ parties, then it essentially equalizes for adversaries who corrupt up to $\corruptions N $ parties. 

%% file: lemcritical.tex
\begin{proof} 
We can prove Lemma~\ref{lem:critical} by applying Markov's inequality several times.

Let $\mathsf{OC}^{\Pi, \adv}$ denote the onion cost of $\Pi$ interacting with $\adv$. 

From Markov's inequality, 
\[
\text{Pr}_{\secparam, \sigma} \left[ \mathbb{E}_{i, \$} \left[ \outof{\Pi, \adv}{i}{\secparam, \sigma} \right]  \ge 2 \mathsf{OC}^{\Pi, \adv} \right] \le \frac{1}{2} . 
\]
Thus, there exists a set $\mathsf{Inputs}$, $|\mathsf{Inputs}| = \bigsmallO{|\sio|}$ s.t.\ for every $\sigma \in \mathsf{Inputs}$,  
$
\mathbb{E}_{i, \$} \left[ \outof{\Pi, \adv}{i}{\secparam, \sigma} \right] <  2\mathsf{OC}^{\Pi, \adv} ,
$
Using Markov's inequality again, we have, for all $\sigma \in \mathsf{Inputs}$, 
\[
\text{Pr}_{i} \left[ \mathbb{E}_{\$} \left[ \outof{\Pi, \adv}{i}{\secparam, \sigma} \right] \ge 2 \mathbb{E}_{i, \$} \left[ \outof{\Pi, \adv}{i}{\secparam, \sigma} \right] \right] \le \frac{1}{2} .
\]
That is, there exists a set $\mathsf{Senders}_{\sigma} \subseteq [N]$, $|\mathsf{Senders}_{\sigma}| = \frac{N}{2}$ s.t.\ for sufficiently large $N$,
\begin{align}
&\mathbb{E}_{\$}\left[\outof{\Pi, \adv}{i}{\secparam, \sigma} \right] < 4 \mathsf{OC}^{\Pi, \adv} < \frac{N}{2} , &\forall P_i \in \mathsf{Senders}_{\sigma}. \label{eq:lazy}
\end{align}
This shows that (i) is satisfied. 

For every $P_i \in \mathsf{Senders}_{\sigma}$, there are at most $N-1$ distinct party $P_j \not=P_i$, such that 
\[
\expect{\hops{\Pi, \adv}{j \rightarrow i \rightarrow \recipient(j)}{\sigma}} \ge \frac{1}{2} ,
\]
where $\recipient(j)$ is the recipient of $P_j$ in $\sigma$. 
If this weren't the case, then the expected number of onions that party~$P_i$ transmits would be at least $\frac{N}{2}$, contradicting \eqref{eq:lazy}. 
Hence, we also satisfy (ii). 
\iftheory\else\qed\fi
\end{proof}